\documentclass[letterpaper,11pt]{article}
\usepackage{jheppub}
\pdfoutput=1
\usepackage[svgnames]{xcolor}
\usepackage{amsmath,amssymb,amsfonts,bm,amscd,mathtools}
\usepackage{graphicx}
\usepackage{multirow}
\usepackage{verbatim}
\usepackage{appendix}
\usepackage{fancybox}
\usepackage{array}
\usepackage{url}
\usepackage{float}
\usepackage{subfigure}
\usepackage{tikz}
\usepackage{dsfont}

\def\Tr{{\rm Tr \,}}

\def\CB{{\mathcal B}}

\def\CE{{\mathcal E}}
\def\CF{{\mathcal F}}
\def\CG{{\mathcal G}}
\def\CH{{\mathcal H}}

\def\CJ{{\mathcal J}}
\def\CK{{\mathcal K}}

\def\CM{{\mathcal M}}

\def\CO{{\mathcal O}}

\def\CQ{{\mathcal Q}}
\def\CR{{\mathcal R}}
\def\CS{{\mathcal S}}

\def\QQ{\mathcal{Q}}

\def\be{\begin{equation}}
\def\ee{\end{equation}}
\def\bea{\begin{eqnarray}}
\def\eea{\end{eqnarray}}

\newcommand{\cH}{{\mathcal H}}
\newcommand{\tq}{{\mathtt q}}

\newcommand{\ta}{{\mathtt a}}
\newcommand{\tb}{{\mathtt b}}
\newcommand{\tc}{{\mathtt c}}

\newcommand{\tp}{{\mathtt p}}

\newtheorem{theorem}{Theorem}

\title{Multi-invariants and bulk replica symmetry}
\preprint{YITP-24-130, LCTP-24-19}

\author[a]{Abhijit Gadde}
\author[b]{Jonathan Harper}
\author[a,c]{Vineeth Krishna}
\affiliation[a]{Department of Theoretical Physics \\ 
Tata Institute for Fundamental Research, Mumbai 400005, India}
\affiliation[b]{Center for Gravitational Physics and Quantum Information,\\ Yukawa Institute for Theoretical Physics, Kyoto University,\\ Kitashirakawa Oiwakecho, Sakyo-ku, Kyoto 606-8502, Japan}
\affiliation[c]{Leinweber Center for Theoretical Physics,\\University of Michigan, Ann Arbor, MI 48109, United States}
\emailAdd{abhijit@theory.tifr.res.in, jonathan.harper@yukawa.kyoto-u.ac.jp, vkt@umich.edu}

\abstract{In this paper, we analyze the question of replica symmetry in the bulk for multi-partite entanglement measures in the vacuum state of two dimensional holographic CFTs. We first define a class of multi-partite local unitary invariants, \emph{multi-invariants}, with a given replica symmetry that acts freely and transitively on the replicas. We look for a subclass of measures such that the dual bulk geometry also preserves replica symmetry. We obtain the most general solution to this problem if we require the bulk to preserve replica symmetry for general configurations of the regions.  
Orbifolding the bulk solution with the replica symmetry gives us a bulk geometry with a network of conical singularities. Our approach makes it clear that there are infinitely many infinitely large families of multi-invariants such that each family evaluates identically on the holographic state. 
Geometrically, these are equalities involving volumes of handlebodies, possibly of different genus, at particular points in the moduli space. 
In certain cases, we check our bulk computation with an explicit calculation in CFT. 
Finally we comment on the generalization to higher dimension.}

\begin{document}
\maketitle
\flushbottom

\section{Introduction and summary}

Quantum information theoretic ideas have been crucial for recent progress in understanding the black hole information paradox \cite{Hawking:1975vcx,Hawking:1976ra, Almheiri:2019hni, Penington:2019kki,Almheiri:2019qdq}. They have also been critical in understanding the locality of bulk to boundary encoding map in the AdS/CFT correspondence \cite{Almheiri:2014lwa,Dong:2016eik,Harlow:2016vwg,Penington:2019npb,Almheiri:2019psf,Cotler:2017erl}. Almost all of the new insights in this context have stemmed from study of bi-partite entanglement properties of the holographic quantum state. A holographic state is a state in quantum field theory that admits a geometric description via AdS/CFT correspondence. Apart from the tripartite measures of entanglement like Entanglement Negativity \cite{Vidal:2002zz,Kudler-Flam:2018qjo,Kusuki:2019zsp,Dong:2021clv,Dong:2024gud} and Reflected Entropy \cite{Dutta:2019gen, Hayden:2021gno, Akers:2021pvd, Akers:2022zxr, Akers:2024pgq, 2018NatPh..14..573U} (and it's multipartite generalizations \cite{Bao:2019zqc,Chu:2019etd,Yuan:2024yfg, 2018JHEP...10..152U, 2019JHEP...08..101H, 2021JHEP...03..116H}), the multi-partite entanglement properties of holographic states remain relatively less explored. This is partly because multi-partite entanglement is difficult to classify and quantify. Bi-partite entanglement admits an operational interpretation as a resource that can be distilled and consumed. While, it has been understood for a while that there are multiple types of multi-partite entanglement, their interpretation as a resource remains a mystery. Entanglement monotones make some progress towards classifying and quantifying multi-partite entanglement. Whatever are the measures of multi-partite entanglement, one thing is clear, they must be invariant under local unitary transformations because we expect that the entanglement properties of the state to be independent of the choice of basis. We call such quantities local unitary invariants or \emph{multi-invariants} for short.

In this paper, we focus on a class of multi-invariants that are constructed using polynomials of the state and its conjugate.\footnote{It is likely that the multi-invariants that we consider generate the ring of all multi-invariants. It would be good to show this rigorously.} Instead of imposing conditions that follow from quantum information theory such as monotonicity under local operations and classical communication, we will impose that these invariants, when evaluated for holographic states, admit a \emph{convenient} geometric dual description. Let us first explain what we mean by a holographic state and then by a convenient geometric dual description.

A CFT with large central charge and large gap is called a holographic CFT. Holographic states are the states in holographic CFT that are described by a classical gravity solution. 
The vacuum state of a holographic CFT is holographic. Other holographic states include states in holographic CFT constructed by acting on the vacuum with operators of large dimension. They are dual to AdS with propagating point particles. If the conformal dimension $\Delta\sim O(c)$, where $c$ is the central charge, then these particles can back-react on the geometry to create conical singularities. In this case also the state is holographic as it is described by a geometry albeit back-reacted by the heavy particle. The thermofield double state of the CFT is also holographic because it is described by a geometry, either thermal AdS or blackhole, depending on the temperature. 
We will be exclusively working with the vacuum state of a two dimensional holographic conformal field theory, but our ideas are general and could be applied to other holographic states. The vacuum state of a $2d$ CFT on a circle is prepared by the euclidean path integral on, say, the southern hemisphere of $S^2$.  Given a decomposition of the circular spatial slice  into $\tq$ regions, we can think of the state as a $\tq$-partite state by considering decomposition of the Hilbert space of the theory on $S^1$ into $\tq$ factors each associated to one of the $\tq$ regions on the circle. 
The norm of the state is computed by the partition function on $S^2$, which is obtained by gluing southern hemisphere i.e. bra with the northern hemisphere i.e. ket. The partition function of the holographic theory on any boundary manifold can be computed using AdS/CFT correspondence to be $e^{-S_{\rm grav}}$ where $S_{\rm grav}$ is the action of the dominant gravity solution that fills in the boundary. When the boundary is $S^2$, the dominant gravity solution is the Euclidean AdS$_3$, also denoted as $H^3$ for hyperbolic three space. 

A $\tq$-partite invariant of the $\tq$-partite state is constructed by taking $n_r$ copies of bras and $n_r$ copies of kets. Each of the copies is known as a replica and $n_r$ is called the replica number. The $n_r$ number of party $A$ regions on the bra circles are glued to $n_r$ number of party $A$ regions on the ket circles in some way. Similarly the party $B$ regions are also glued and so on. The gluing pattern of each party is independent and  together they define a multi-invariant. 
After the entire gluing process, we get a manifold that does not have any boundary components. It can, and in general will, have a higher genus and points with conical excesses. We call this the replicated manifold (associated to a given gluing pattern and hence to a given invariant). Because all the $A$ regions in the bra are glued to all the $A$ regions in the ket and so on, the partition function on the replicated manifold is invariant under local unitary transformations. Thanks to holography, the partition function is computed as $e^{-S_{\rm grav}}$ where $S_{\rm grav}$ is the gravitational action of the dominant bulk solution that fills in the replicated manifold. This is how one can compute a multi-invariant in a holographic theory. However this description is not very convenient. 

By convenient geometric description we mean that ($n_r$-th root of) the invariant should be computed from a geometry whose boundary is again the original $S^2$, but the bulk may not be $H^3$. An example of a familiar invariant that admits a convenient geometric description is the Renyi entropy ${\rm Tr} \rho_A^{n}$ where $\rho_A$ is the density matrix of associated to some region $A$. For simplicity, let us take this region to be connected. The Renyi entropy is calculated by $e^{-n S_{\rm grav}^{\rm orb}}$ where $S_{\rm grav}^{\rm orb}$ is the action of an orbifold geometry whose boundary is $S^2$ and the bulk consists of a conical singularity with a cone angle $2\pi/n$ around it (here $n_r=n$). We will be interested in constructing multi-invariants that admit such a description. We will soon see that the \emph{convenient} description is admitted if:
\begin{itemize}
    \item The invariant has a \emph{replica symmetry} acting freely and transitively on the replicas.
    \item The replica symmetry is preserved by the dominant bulk solution filling in the replicated manifold. 
\end{itemize}
We will explain these conditions in detail in the body of the paper. Let us outline them for now: As mentioned, the multi-invariants that we consider are constructed by taking $n_r$ copies of the bra and $n_r$ copies of the ket. The gluing can be done in a way that preserves some symmetry. This symmetry is known as the replica symmetry. The replicated manifold enjoys the action of this symmetry. If the replica symmetry acts freely and transitively on the replicas then the quotienting of the replicated manifold produces the original un-replicated manifold, namely $S^2$. This satisfies the first condition for having a convenient description. Realizing the second condition is much more non-trivial. The dominant bulk geometry filling in the replicated manifold may not have the replica symmetry that the boundary enjoys. In case it does, the bulk geometry can be orbifolded. As a result, the boundary goes back to $S^2$ as discussed and we get a bulk solution that fills it in. As the action of the replica symmetry on the bulk solution may have fixed points, the orbifolded geometry may have conical  singularities. The fixed point loci are generically co-dimension $2$ and are fixed by group elements with some finite order $n$. After orbifolding, these  loci give rise to co-dimension $2$ conical singularity with an angle $2\pi/n$ around them. As we will explain in the body of the paper, these conical singularity can form a tri-valent graph subject to certain conditions on the vertices. The multi-invariant is then $e^{-n_r S_{\rm grav}^{\rm orb}}$. This is because, due to orbifolding, the action of the gravity solution filling in the replicated manifold is $S_{\rm grav}=n_r S_{\rm grav}^{\rm orb}$. We define the normalized multi-invariant to be $n_r$-th root of the polynomial multi-invariant. Then the normalized invariant is simply given by $e^{-S_{\rm grav}^{\rm orb}}$. The condition that the replica symmetry of the replicated manifold can be extended into the bulk is highly non-trivial and requires a careful analysis. This paper is dedicated to this analysis for the case of two dimensional conformal field theories. If this condition is obeyed then we say that the multi-invariant preserves bulk replica symmetry. 

The answer to whether a given multi-invariant preserves bulk replica symmetry depends not only on the invariant, but also on the party region decomposition of the circular spatial slice. It is possible that for certain configurations of party regions the bulk replica symmetry is preserved and for certain other configurations  it is not. We find a large class of multi-invariants that preserve bulk replica symmetry for \emph{some} configuration of regions. We also find the most general multi-invariants that preserves the bulk replica symmetry for \emph{all} configurations of regions. Interestingly, the invariants that preserve replica symmetry for all region configurations  are associated with finite Coxeter groups. In fact, they have finite Coxeter groups as their extended replica symmetry groups. The extended replica symmetry group is a certain ${\mathbb Z}_2$ extension of the replica symmetry group. We will explain this in detail in the body of the paper. 

The theory of Kleinian groups, in particular their construction using the so-called Klein-Maskit combination theorems, plays an important role in classifying the invariants that preserve bulk replica symmetry. Kleinian groups are discrete subgroups of $PSL(2,{\mathbb C})$ or $SO(3,1)$ which is the isometry group of $H^3$. 
It turns out that the orbifold geometries that are dual to the normalized invariants are quotients of $H^3$ by virtually free Kleinian groups.\footnote{Virtually free groups are the groups that have a normal subgroup which is free. More on this later.} Alternatively, each of these orbifolds is also obtained by quotienting handlebodies of various genera by a finite subgroup. This variety in obtaining a given orbifold by quotienting handlebodies of different genera has an interesting consequence. It implies that there are families, in fact infinitely large, of normalized multi-invariants that are identical for the vacuum state of the $2d$ holographic CFT! Moreover, infinitely many such families can be constructed. 

The rest of the paper is organized as follows. In section \ref{graphical}, we introduce the general theory of polynomial multi-invariants, focusing especially on the replica symmetry and its ${\mathbb Z}_2$ extension. In section \ref{qft-inv}, we discuss the construction of  symmetric invariants in conformal field theories, pointing out its alternate formulation as a correlation function of twist operators. We also make more precise the idea of computing the invariants in holographic theories with the help of an  orbifold. In section \ref{bulk-replica}, we discuss the bulk solutions that have non-trivial isometry groups. We review the construction of hyperbolic handlebodies as quotients by the Schottky group and outline the main idea of constructing symmetric handlebodies by quotienting $H^3$ by virtually free Kleinian groups. The importance of studying Kleinian groups is thus established. 
Section \ref{klein-geom} is dedicated to the study of Kleinian groups. First we review the action of conformal isometry group $PSL(2,{\mathbb C})$ in two dimensions and its extension into the bulk $H^3$ as isometry of $H^3$. We then discuss finite Kleinian groups and their action on the boundary. We also discuss their action on $H^3$ and the associated orbifolds. There are only a few finite Kleinian groups. In order to produce a general set of bulk replica symmetry preserving invariants, it is important to work with infinite Kleinian groups. We study their construction using Klein-Maskit combination. Algebraically, this combination is what is known as the amalgamation of  groups. We study the action of the resulting Kleinian group on the $S^2$ as conformal isometry and on $H^3$ as isometry, explicitly, in a number of cases. In section \ref{replica-solution}, we compile our analysis of Kleinian groups and apply it to find a general solution to the bulk replica symmetry problem i.e. to find the general multi-invariant preserving bulk replica symmetry. We also discuss the case of certain special multi-invariants that do not preserve replica symmetry. We show that they do admit a bulk solutions that preserves the bulk replica symmetry, however they are sub-leading. These sub-leading solutions turn out to be Euclidean wormholes with two boundary components with conical singularities between them. In section \ref{cft-comp}, we check our prediction from the bulk analysis from direct CFT computation in a number of cases. This involves computing Liouville action for genus $0$ covering maps. We find a perfect agreement with our bulk result.  Finally we end with outlook in section \ref{outlook}. The appendix \ref{l-uni-app} reviews computation of the Liouville action associated to a covering map, while \ref{app:reps} lists representations, group elements and twist operator monodromies for the multi-invariants presented.

\section{Multi-invariants and replica symmetry}\label{graphical}
Consider a quantum state $|\Psi\rangle\in {\cal H}$ where ${\cal H}$ admits the factorization $\CH=\CH_1\otimes \ldots\otimes \CH_{\tq}$. Each factor $\CH_\ta$ is called a party and the state $|\Psi\rangle$ is called a $\tq$-partite state. We are interested in characterizing such states up to unitary transformations that act on individual factors $\CH_{\ta}$. 
This is naturally done by constructing functions of the state that are invariant under local unitary transformations. We will concern ourselves only with the functions that are ``monomials'' in the wavefunction and its conjugate.
Let \(\lvert e^{i_\tp}_\ta\rangle\) be the basis of the party $\ta$  Hilbert space \(\cH_\ta\) (\(i_\ta = 1,\ldots,d_\ta\) and \(\ta = 1,\ldots, \tq\)). We can expand the state \(\lvert \Psi \rangle\) in these basis,
\begin{equation}
    \lvert \Psi \rangle = \sum_{i_1=1}^{d_1} \ldots \sum_{i_\tq=1}^{d_\tq} \psi_{i_1\ldots i_\tq} ~\lvert e^{i_1}_1\rangle \otimes \ldots \otimes \lvert e^{i_\tq}_\tq \rangle.
\end{equation}
The components \(\psi_{i_1\ldots i_\tq}\) are collectively called the wavefunction of the state. It transforms in the fundamental representation under the action of a unitary transformation on any of the individual \(\tq\) parties. Local unitary invariants are constructed by taking multiple copies of the wavefunction $\psi_{i_1\ldots i_\tq}$ and its conjugate $\bar\psi^{j_1\ldots j_\tq}$ and contracting all the fundamental indices with the anti-fundamental indices in any way possible. As long as no index remains un-contracted, the resulting quantity is a local unitary invariant. We call it a multi-invariant $\CE$.  Note that, to generate a multi-invariant we must take an equal number of $\psi$'s and $\bar \psi$'s. We will refer to each copy of $\psi$ as a replica and the total number of $\psi$'s as the replica number $n_r$. As the invariant is homogenous in $\psi$'s and $\bar \psi$'s, it is morally a monomial. In particular, it obeys 
\begin{align}
    \CE(|\psi_1\rangle\otimes |\psi_2\rangle)= \CE(|\psi_1\rangle) \cdot \CE(|\psi_1\rangle).
\end{align}
Here $|\psi_1\rangle$ and $|\psi_2\rangle$ are $\tq$-partite states. The product state $|\psi_1\rangle\otimes |\psi_2\rangle$ is also thought of as a $\tq$-partite state where each of its party is a tensor product of corresponding parties of $|\psi_1\rangle$ and $|\psi_2\rangle$.
The set of all possible index contraction patterns gives rise to invariants whose number grows super-exponentially with the replica number. Below we will describe how to characterize the index contractions. 

We assign each $\psi$ (and $\bar \psi$) a replica index that takes values from $1$ to $n_r$. Now we contract the party $1$ fundamental indices of all $\psi$'s with the party $1$ anti-fundamental indices of all $\bar \psi$'s. This involves assigning a permutation element $g_1\in S_{n_r}$ to party $1$. Similarly, the index contractions of party $2$ is described by another permutation element $g_2\in S_{n_r}$ and so on. Concretely, the multi-invariant corresponding to the choice $(g_1,\ldots, g_\tq)$ permutation elements is given as
\begin{align}\label{inv-def}
    \CE(g_1,\ldots, g_\tq)&=  (\psi_{i^{(1)}_1\ldots i^{(1)}_\tq} \ldots \psi_{i^{(n_r)}_1 \ldots  i^{(n_r)}_\tq}) (\bar \psi^{j^{(1)}_1\ldots j^{(1)}_\tq} \ldots \bar \psi^{j^{(n_r)}_1 \ldots j^{(n_r)}_\tq}) \delta^{\vec i_1}_{g_1 \cdot \vec j_1} \ldots \delta^{\vec i_\tq}_{g_\tq \cdot \vec j_\tq}\notag\\
    {\rm where}, \quad \delta^{\vec i_\tp}_{g_\tp \cdot \vec j_\tp} &\equiv \delta^{i_\tp^{(1)}}_{j_\tp^{(g_\tp \cdot 1)}} \ldots \delta^{i_\tp^{(n_r)}}_{j_\tp^{(g_\tp \cdot n_r)}}.
\end{align}
This notation is a little cumbersome to process, so let us give an example of the familiar bi-partite invariant ${\rm Tr} \rho^n$, where $\rho$ is reduced density matrix on one of the parties, in this notation. A choice of $g_1$ and $g_2$ that gives ${\rm Tr}\rho^n$ is,
\begin{align}
    g_1=e, \qquad g_2=(1,2, \ldots, n).
\end{align}
Here we have picked $n_r=n$ and  $e$ stands for the identity element of $S_n$ and $g_2$ permutes the $n$ replica cyclically. Written explicitly as in \eqref{inv-def}, it is
\begin{align}
    {\rm Tr} \rho^n= (\psi_{i_1^{(1)}i_2^{(1)}}\ldots \psi_{i_1^{(n)}i_2^{(n)}})(\bar \psi^{j_1^{(1)}j_2^{(1)}}\ldots \bar \psi^{j_1^{(n)}j_2^{(n)}})\, (\delta^{i_1^{(1)}}_{j_1^{(1)}}\ldots \delta^{i_1^{(n)}}_{j_1^{(n)}}) (\delta^{i_2^{(1)}}_{j_2^{(2)}}\ldots \delta^{i_2^{(n-1)}}_{j_2^{(n)}}\delta^{i_2^{(n)}}_{j_2^{(1)}})
\end{align} 
As we will see soon, it is useful to pick the set $g_\ta$ such that it generates a subgroup of $S_n$. 
In defining a general $\tq$-partite invariant $\CE$ using the equation \eqref{inv-def}, the assignment of $g_\ta$ is not unique. We have the redundancy,
\begin{itemize}
    \item Left multiplication: $\qquad $ $\CE(g_1,\ldots, g_\tq) = \CE(g \cdot g_1,\ldots, g \cdot g_\tq),\qquad {\rm for}\quad g\in S_{n_r}$.
\end{itemize}
This indicates the freedom in labeling $\psi$'s once the labeling of $\bar \psi$'s is fixed. An element $h\in S_{n_r}$ is known as a replica symmetry element if it obeys
\begin{align}\label{left-red}
    h\cdot (g_1,\ldots, g_\tq)\cdot h^{-1}=(g \cdot g_1,\ldots, g \cdot g_\tq), \qquad \qquad {\rm for}\,\,{\rm some}\quad g\in S_{n_r}.
\end{align}
The replica symmetry elements form a group that we call the replica symmetry group ${\cal R}$ or simply replica symmetry. 
It is often useful to fix the left-multiplication redundancy by setting one of the permutation elements, say $g_1=e$. This is done by choosing $g=g_1^{-1}$ so the ``gauge fixed'' tuple is $(1, \hat g_2\equiv g_1^{-1}g_2, \ldots, \hat g_\tq\equiv g_1^{-1}g_\tq)$.  After this gauge fixing, the replica symmetry can be defined as the commutant of  $\hat g_\ta$ for $\ta=2, \ldots, \tq$. We also define the normalized version of the invariant $\CE$ as
\begin{align}
    \tilde \CE\equiv \CE^{1/{n_r}}, 
\end{align}
where ${n_r}$ is the number of replicas. It is the normalized version that we will always be ultimately interested in, but it is convenient to discuss the computation of the un-normalized version $\CE$ because it is a polynomial in the state $\psi$ (and its conjugate $\bar \psi$). 

\begin{figure}[H]
    \begin{center}
        \includegraphics[]{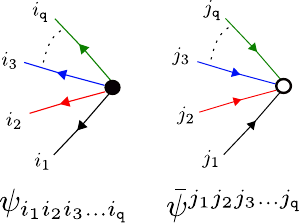}
    \end{center}
    \caption{Graphical notation for the wavefunction (Black vertex with colored edges and arrows directed away from the vertex) and it's conjugate (White vertex with colored edges and the arrows directed towards the vertex).}
    \label{psigraph}
\end{figure}

\begin{figure}[H]
    \begin{center}
        \includegraphics[]{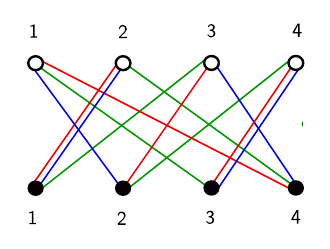}
    \end{center}
    \caption{An example of a polynomial multi-invariant for three parties. Every vertex has three edges (each of different color i.e. different party) incident on it. The three permutation that define the multi-invariant are - $g_1 = (1~3)(2~4)$, $g_2 = (1~2~3~4)$ and $g_3=(1~2)(3~4)$. The associated edges are drawn in green, red and blue respectively.}
    \label{kempegraph}
\end{figure}

It is convenient to visualize these index contractions with the help of a graph. Denote $\psi$'s and $\bar \psi$'s with black and white vertices respectively. Let each vertex have $\tq$ edges of fixed, but distinct colors incident on it (see Fig. \ref{psigraph}). We take the edge of color $\ta$ to correspond to the index of party $\ta$. A graph with this property - which, in particular, is a bi-partite graph - is constructed using the above method and hence corresponds to a local unitary invariant (an example is shown in Fig. \ref{kempegraph}). The replica symmetry corresponds to the automorphism of the associated graph that preserves its bi-partite structure.

\subsection{Symmetric invariants}\label{sym-inv}
We are particularly interested in invariants that have a freely and transitively acting replica symmetry group. In this section, we will see how to construct an invariant of a $\tq$-partite state given a finite group $\cal G$ and its $\tq$ number of  generators $g_\ta$. Let us take the group $\cal G$ - thought of as a set - as the set of replicas.  Index its elements from $1$ to $n_r=|\cal G|$. The action of any element $g\in \cal G$ on the element $g'$ giving $g'\cdot g$ i.e. the right-action, can be realized as a permutation in $S_{|\cal G|}$. An invariant $\CE(g_1,\ldots, g_\tq)$ is constructed by thinking of $g_\ta$ as the corresponding permutations in $S_{|\cal G|}$. The advantage of this construction is that the left-action of $\cal G$, thought of as a subgroup of $S_{|\cal G|}$, commutes with the right-action and acts freely and transitively on the replicas. It can be used to construct the replica symmetry of $\CE$ in the following way. 
The replica symmetry group ${\cal R}$ is not $\cal G$ because of the left-multiplication ``gauge freedom'' in defining the invariant using the permutation tuple. The actual replica symmetry is obtained from $\cal G$ by quotienting it by simultaneous left-multiplication on the generators $g_\ta$. As a result, the replica symmetry is generated by $\sigma_{\ta\tb}\equiv g_\ta^{-1}g_\tb$. This is because the combination $\sigma_{\ta\tb}$ is invariant under a simultaneous left-multiplication. Note that the group generated by $\sigma_{\ta\tb}$ is the same as the group generated by the gauge fixed generators $\hat g_\ta$. To see this, note
\begin{align}
    \sigma_{\ta\tb}\equiv g_\ta^{-1}g_\tb=\hat g_{\ta}^{-1} \hat g_{\tb},\qquad {\rm and}\qquad \hat g_{\ta}= \sigma_{1\ta}.
\end{align}
In case ${\cal R}$ is a proper subgroup of ${\cal G}$, it does not act on the replica set in a transitive fashion. We consider a single orbit of ${\cal R}$ and treat it as the new replica set. On this set, ${\cal R}$ acts freely and transitively. The invariant is uniquely specified by specifying the replica group $\cal R$ and the gauge fixed generators $\hat g_{a}$. We simply take the replica set to be ${\cal R}$ and permutation tuple to be $(1,\hat g_1, \ldots, \hat g_\tq)$. Using this construction, an invariant with any given replica symmetry group $\CR$ can be constructed. In addition to $\CR$, specification of its generating set is also required.

There is another, more symmetric, way to present the replica symmetry group starting with $\cal G$ and its generators $g_\ta$. We extend $\cal G$ by ${\mathbb Z}_2$ by considering a new element $p$ that squares to identity $e$. Let $r_\ta\equiv  p\cdot g_\ta$ obey $r_\ta^2=e$. Let the group generated by $r_\ta$ be $\hat {\cal R}$.  Because it is a group presented with generators $r_\ta$ obeying $r_\ta^2=e$, it is a quotient of the Coxeter group. See
the beginning of section \ref{coxeter} for the definition of a Coxeter group.  The replica symmetry group ${\cal R}$ is a subgroup of index $2$ of $\hat {\cal R}$. Consider the sign homomorphism $\epsilon: \hat {\cal R} \to {\mathbb Z}_2$ defined as $\epsilon(r_\ta)=-1$. The replica symmetry group ${\cal R}$ is then the ${\rm Ker}\,\epsilon$. In other words, it contains the words made with even number of letters $r_\ta$. To see that it is indeed the case, observe
\begin{align}
    r_\ta r_\tb=(p\cdot g_\ta)\cdot (p\cdot g_\tb)= (p\cdot g_\ta)^{-1}\cdot (p\cdot g_\tb)= g_\ta^{-1} g_\tb= \sigma_{\ta\tb},
\end{align}
and that the subgroup of even number of letters is precisely generated by $r_\ta r_\tb$. Given the group ${\hat \CR}$ and its generators $r_\ta$, the generators $\hat g_\ta$ of the replica symmetry group are constructed as $r_\ta r_1$. We call the group ${\hat \CR}$, the extended replica symmetry group. This is because it extends the replica symmetry group by the ``reflection'' generator $p$ that maps bra to ket and vice versa.

The symmetric invariants played an important role in \cite{Gadde:2022cqi,Gadde:2023zzj} where $\CG$ was taken to be an abelian group and $g_\ta$, its independent generators\footnote{This includes a class of multi-invariants, the $n$th Renyi \emph{multi-entropy} on $\tq$ parties, which have abelian group symmetry $\mathbb{Z}_n^{q-1}$ and $g_\ta$ are chosen such that the length of the cycles is $n$. See also \cite{Penington:2022dhr, 2024ScPP...16..125H, 2024JHEP...02..025G, 2024arXiv241008284L, Yuan:2024yfg} for recent progress in this direction.}. In \cite{Gadde:2024jfi}, a class of multi-partite pure state entanglement monotones - local unitary invariants that are monotonic under local operations and classical communication - was constructed using graph theoretical methods. The monotonicity property was reduced ``edge-convexity'' of a graph which in turn required the graph to be ``edge-reflecting''. It was argued in \cite{Gadde:2024jfi} that the graph is edge-reflecting if and only if it corresponds to a symmetric invariant. Although a special class of edge-reflecting graphs were shown to be edge-convex, the question of edge-convexity of general edge-reflecting graphs was left open. It is possible that the symmetric invariants discussed here are significant from quantum information theory point of view in that they give rise to monotones under local operations and classical communications. It would be interesting to explore this direction further.  Additionally, unlike some of the other measures of multi-partite entanglement, such as entanglement of purification, the symmetric invariants are easy to compute for general quantum states. They are only polynomial in the wavefunction coefficients and their complex conjugates and thus can be calculated just given the state of the system and operation of partial trace.

\subsection{Graphical presentation}
We associated three types of groups to an invariant:
\begin{itemize}
    \item $G$: The group used to construct the index contractions with the tuple $(g_1,\ldots, g_\tq)$.
    \item $\hat {\cal R}$: The extended replica symmetry group. The group whose generators $r_\ta$ which square to $1$.
    \item $\cal R$: The replica symmetry group, index $2$ subgroup of $\hat {\cal R}$ defined as ${\rm Ker}\, \epsilon$, where $\epsilon$ is the sign homomorphism. This is generated by $\sigma_{\ta\tb}=g_\ta^{-1}g_{\tb}$ or equivalently by $r_\ta r_\tb$.
\end{itemize}
We can understand all of them graphically in terms of their Cayley diagrams. 
The Cayley diagram is associated to a group $H$ and its generators $h_\ta$'s. It is a graph whose vertex set is $H$. A directed edge of color $\ta$ is drawn from vertex $i$ to vertex $j$ if and only if $i=  j\cdot h_\ta$. The resulting graph has $H$ as its automorphism group, where the automorphism acts on the vertices by left-action. This automorphism is free and transitive because the action of a group on itself by left-multiplication is free and transitive.  

\begin{figure}[H]
    \begin{center}
        \includegraphics[scale=0.5]{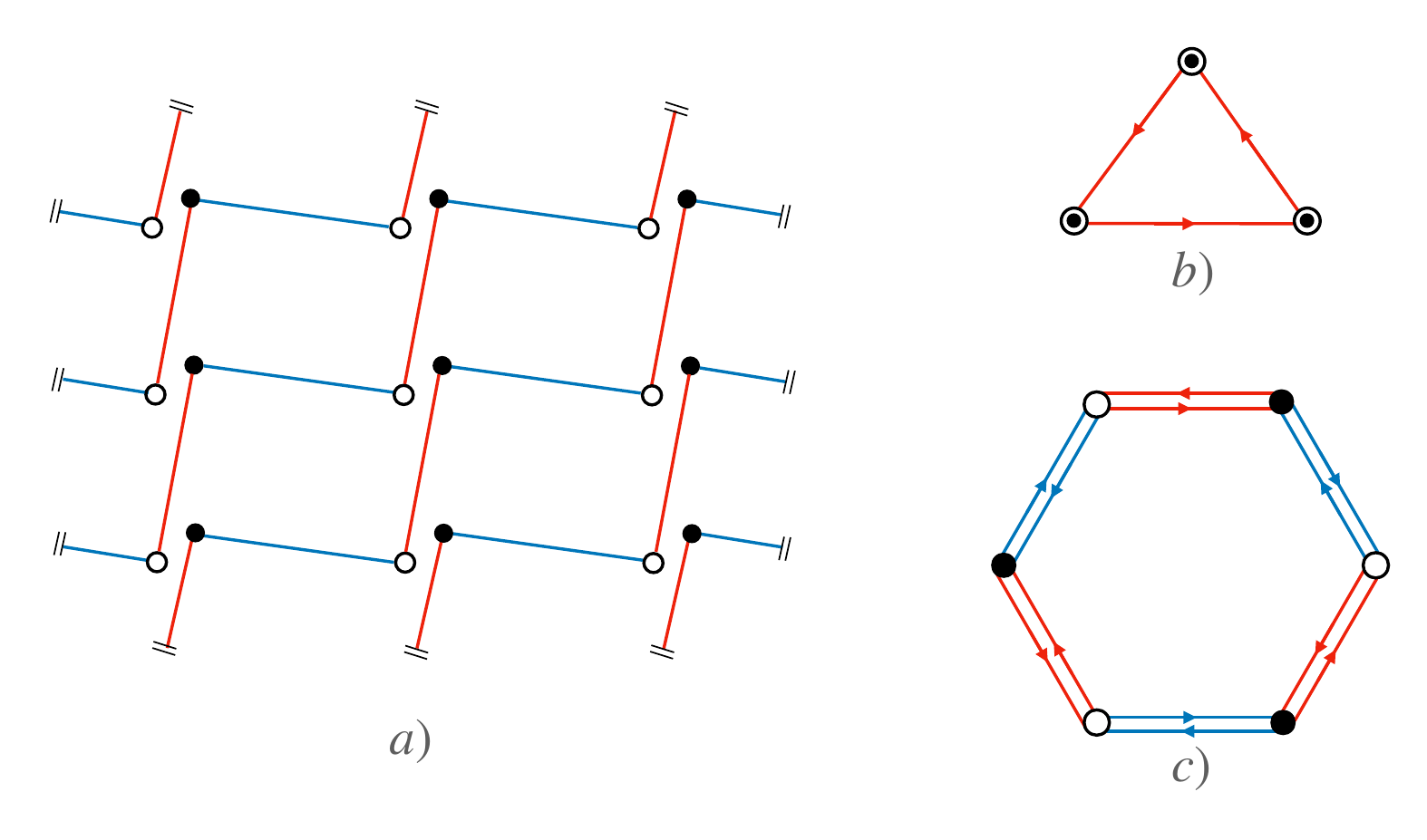}
    \end{center}
    \caption{Three graphical presentations of a multi-invariant. The figures $a), b)$ and $c)$ emphasize the symmetries $G,\CR$ and $\hat \CR$ respectively. Figure $a)$ shows the construction of the multi-invariant using $G={\mathbb Z}_3\times {\mathbb Z}_3$. It is disconnected, consisting of three identical connected components. The other two figures describe a single connected component. 
    Figures $b)$ and $c)$ are Cayley diagrams of $\CR={\mathbb Z}_3$ and $\hat \CR= \mathbb{D}_6$, the dihedral group respectively.}
    \label{q233}
\end{figure}

Cayley diagrams of the replica symmetry ${\cal R}$ and of $\hat {\cal R}$ form different but equivalent graphical presentation of a given invariant. Consider the example $G={\mathbb Z}_3\times {\mathbb Z}_3$ and $g_1$ and $g_2$ being the two independent ${\mathbb Z}_3$ generators. The invariant corresponding to this choice is graphically presented in figure \ref{q233}. The replica symmetry is obtained by gauge fixing $\hat g_1=1$, then $\hat g_2= g_1^{-1}g_2$ which is also a ${\mathbb Z}_3$ element. To see its Cayley diagram, we contract the edge of color $1$ (represented as red in the figure). We get three copies of Cayley diagram formed by the edge of color $2$ (represented as blue in the figure). 
Because ${\cal R}$ is a proper subset of ${\cal G}$, we take only one copy. It is the Cayley diagram of ${\cal R}={\mathbb Z}_3$ as expected.

The Cayley diagram for ${\hat \CR}$ is obtained by going back to the diagram describing the invariant and making all the arrows bi-directional. These correspond to generators $r_\ta$ of ${\cal R}$ that square to $1$. In this case also, the graph is disconnected into three identical copies. Each copy  consists of six vertices. It is the Cayley diagram of the ${\mathbb Z}_2$ extension $\mathbb{D}_{6}$ of the replica symmetry ${\mathbb Z}_3$. Considering paths of even length, we recover the Cayley diagram of ${\cal R}$. As we have discussed earlier, the graphs labeling the invariant are bi-partite, white vertices corresponding to bras and black vertices corresponding to kets. The group $\hat {\cal R}$ is the automorphism of this graph (with bi-directional) edges that does not necessarily preserve the bi-partite structure while ${\cal R}$ is the automorphism group that preserves the bi-partite structure of the graph. The maps that don't preserve the bi-partite structure are ``orientation reversing'' in that they map bras to kets and vice versa. The comment about them being orientation reversing will make sense when we discuss the invariants in the context of quantum field theory.
There it will be more convenient to work with the extended replica symmetry group ${\hat \CR}$. 

\begin{figure}[H]
    \begin{center}
        \includegraphics[scale=0.5]{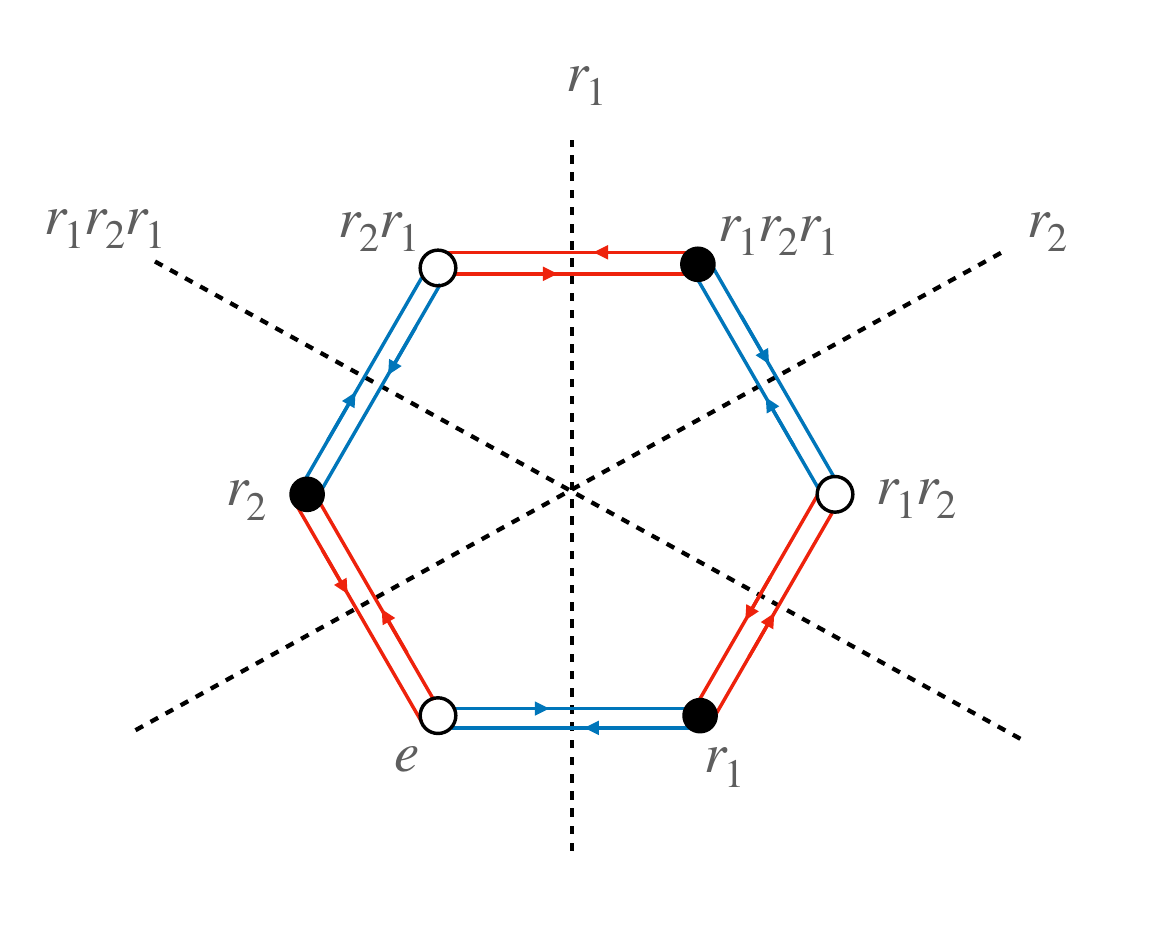}
    \end{center}
    \caption{The Cayley graph of $\hat \CR=\mathbb{D}_6$. An arrow of color red (blue) is drawn from vertex $i$ to $j$ if an only if $i=r_1\cdot j$ ($i=r_2\cdot j$). The automorphisms of the graph are reflections across the dotted lines. They correspond to left-multiplication by generators.}
    \label{left-right}
\end{figure}

This example also gives us an opportunity to highlight the difference between the generators $r_\ta$ used to construct the Cayley diagram of ${\hat \CR}$ and its automorphisms, also generated by $r_\ta$. The Cayley graph is constructed by drawing an arrow of color $\ta$ from vertex $i$ to $j$ if an only if $i=j\cdot r_\ta$ i.e. if and only if $i$ is obtained from $j$ using \emph{right-multiplication} by the generator. This allows us to label all the vertices of the Cayley graph by the group elements. For $\hat \CR=\mathbb{D}_6$, this is done in figure \ref{left-right}.
On the other hand, as explained in section \ref{sym-inv}, the automorphisms of the Cayley graph is obtained by \emph{left-multiplication} by the generators. This is illustrated in figure \ref{left-right} by reflections across the dotted lines. For example, the vertical dotted line represents the reflection that corresponds to left-multiplication by $r_1$ as can be straightforwardly checked. The other dotted lines are labeled by the associated  generators, realizing reflections by left-multiplication. In section \ref{qft-inv} and onwards, when we discuss multi-invariants in quantum field theory,  we will often talk about the automorphism of the replicated manifold and that of the associated bulk geometries. In that context, the distinction between the right-multiplication defining the multi-invariant and the left-multiplication realizing the automorphisms is worth keeping in mind.

\subsection{Coxeter invariants}\label{coxeter}

A \emph{Coxeter group}, along with its canonical presentation, is defined as the abstract group generated by reflections i.e. with generators $r_\ta$ obeying $r_\ta^2=e$. It is also assumed that the order $m_{\ta\tb}$ of $r_\ta\cdot r_\tb$ is finite and that there are no other relations on the generators $r_\ta$:
\be
\langle r_1,\cdots r_{\tq}|(r_ir_j)^{m_{ij}}=e\rangle, \quad m_{ii}=1, \quad m_{ij}=m_{ji}\geq2
\ee
Coxeter classified the matricies $m_{\ta\tb}$ for which the Coxeter group is finite, these are classified by Dynkin diagrams. Each vertex of the Dynkin diagram represents a generator $r_\ta$. If vertices $\ta, \tb$ don't share an edge, then $m_{\ta\tb}=2$. If they share an unlabeled edge then $m_{\ta\tb}=3$ and if they share an edge with label $m$ then $m_{\ta\tb}=m$. For finite Coxeter groups, except in the case of a single edge labeled $n$, an edge is either unlabeled or has at most a single label of either $4$ or $5$. See figure \ref{fin-cox} for the Dynkin diagram presentation of finite Coxeter groups: 
\begin{figure}[h]
    \begin{center}
        \includegraphics[scale=0.22]{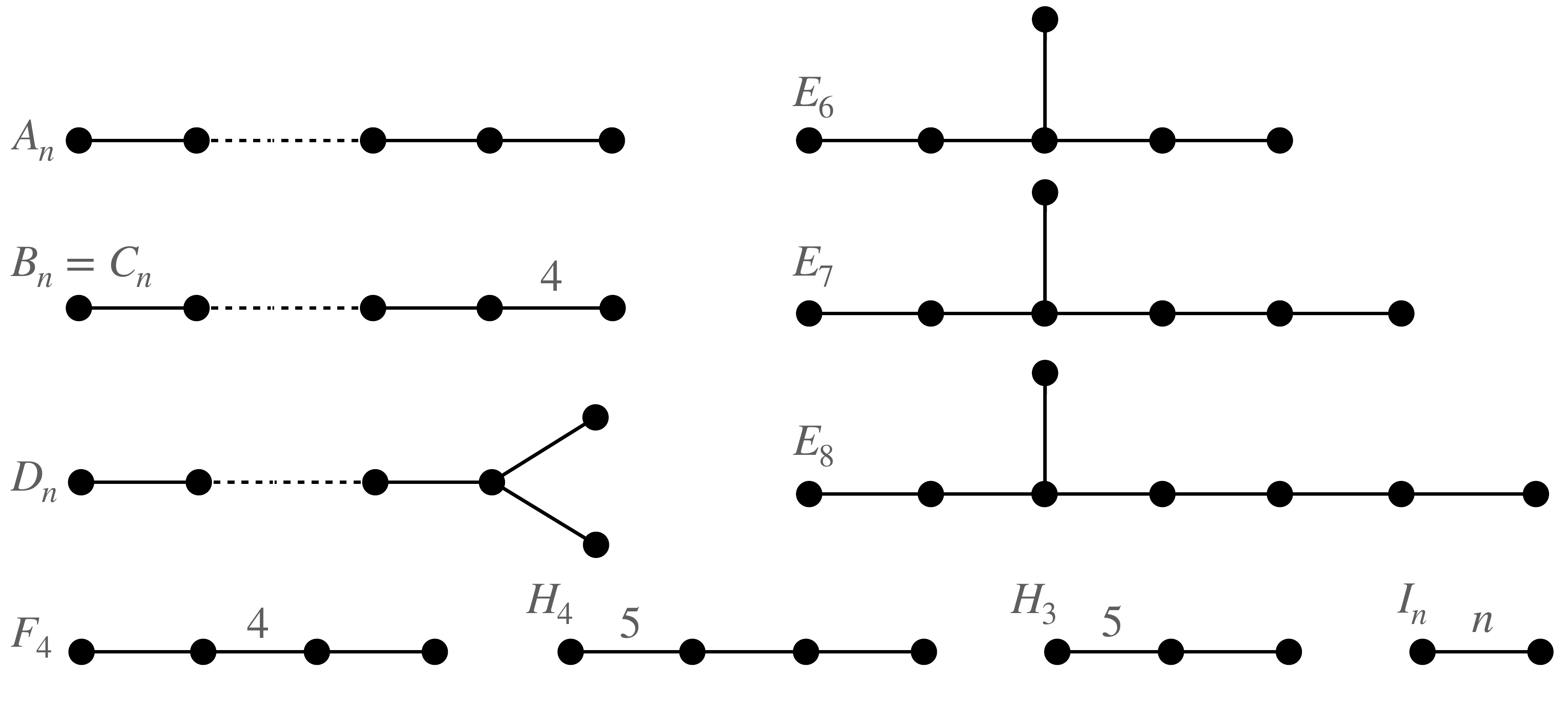}
    \end{center}
    \caption{Classification of finite Coxeter groups by Dynkin diagram. Common names of the associated Lie algebras are also indicated. Note that $A_n$ is the symmetric group $\mathbb{S}_{n+1}$ and $I_n$ is the dihedral group $\mathbb{D}_{2n}$.}
    \label{fin-cox}
\end{figure}

Every Coxeter group has a normal index 2 alternating subgroup which consists of those elements which are constructed from the product of an even number of generators \cite{brenti2007alternatingsubgroupscoxetergroups}. This subgroup can be constructed as follows: Of the generators we select any one generator say $r_1$ and construct the product $\sigma_i=\sigma_{i1}=r_i r_1$ to rewrite the group presentation in terms of $\sigma_2\cdots \sigma_{\tq}$ and $r_1$. Importantly we have
\be
r_i^2=r_ir_1^2r_ir_1^2=(\sigma_ir_1)^2=e
\ee
which together with the group presentation indicates that $G$ can be written as the semi-direct product
\be\label{eq:coxrsig}
\langle \sigma_2,\cdots \sigma_{\tq},r_1|r_1^2=\sigma_i^{m_{i1}}=(\sigma_i\sigma_j^{-1})^{m_{ij}}=e,\;(\sigma_ir_1)^2=e\rangle
\ee
In particular the alternating subgroup is given by the quotient by $\mathbb{Z}_2$ (heuristically this can be though of as setting $r_1=0$)
\be\label{eq:coxalternating}
\langle \sigma_2,\cdots \sigma_{\tq}|\sigma_i^{m_{i1}}=(\sigma_i\sigma_j^{-1})^{m_{ij}}=e\rangle.
\ee

Coxeter invariants will play an important role in this paper. In fact, an important result of the paper is that the $\tq$-partite Coxeter invariant - corresponding to Dynkin diagram with $\tq$ nodes - preserves the bulk replica symmetry if the circular spatial slice of the vacuum state is decomposed into $\tq$ regions, each corresponding to a party. 

\section{Invariants in quantum field theory}\label{qft-inv}
For any local unitary invariant $\CE$, specified either with the choice of permutation tuple $(g_1,\ldots, g_\tq)$ or as a bi-partite edge-colored graph, we can formulate its computation in a quantum field theory as a path integral on a manifold $\CM_{\CE}$ that depends on $\CE$. For concreteness, we will fix the state to be the vacuum state of a $2d$ quantum field theory $\QQ$. The bra is prepared by a path integral over the southern hemisphere. Its dual, the ket, is prepared by a path integral over the northern hemisphere. The norm of the state is computed by gluing the two hemispheres and doing the path integral over the resulting sphere. 

We think of the vacuum state as a multi-partite state by associating Hilbert space factors to various regions of the equatorial circle. Let the number of distinct disconnected regions be $s$. Note that  $s$ need not be equal to the number of parties $\tq$ because multiple disconnected regions may be identified with the same party. It is often convenient to treat each disconnected region as a separate party i.e. $s=\tq$, but it is not really necessary. 
An invariant $\CE(g_1,\ldots, g_\tq)$ is obtained by taking $n_r$ copies of bras and kets. The region associated to party $\ta$ in each of the bras i.e. southern hemispheres is glued to the region associated to party $\ta$ of the kets i.e. northern hemispheres according to the permutation $g_\ta$. After all the gluing, the resulting manifold does not have any boundaries, but does have $s$ ramification points i.e. points of conical excess at points that separate the regions associated to different parties. We will call this manifold associated to the invariant $\cal E$ as $\CM_{\cal E}$. As remarked earlier, the manifold that corresponds to the norm is $S^2$. The following discussion is general and allows for the manifold corresponding to the norm of the state to be other than $S^2$. For example, if we take the state in question to be the thermofield double state then the manifold that computes the norm is the torus. To keep the discussion general, we take the manifold that computes the norm to be $\CM$. The only condition that $\CM$ needs to satisfy is that it needs to be symmetric under the reflection that maps the bra part (southern hemisphere when $\CM=S^2$) to the ket part (northern hemisphere when $\CM=S^2$) and vice versa. This endows $\CM$ with an orientation reversing ${\mathbb Z}_2$ isometry. The locus along which the bra part of $\CM$ is glued to the ket part is fixed by this. 
We will often resort to the case of $\CM=S^2$ for concrete computations. 
 
The invariant $\CE$ is computed by a path integral on $\CM_{\CE}$ with the normalization that sets the partition function on $\CM$ to $1$. This  normalization ensures that the norm of the vacuum state is $1$. In other words, 
\begin{align}
    \CE=\frac{Z_{\CM_{\CE}}}{Z_{\CM}^{n_r}},\qquad{\rm i.e.}\qquad  \tilde \CE= \frac{(Z_{\CM_{\CE}})^{1/{n_r}}}{Z_{\CM}}
\end{align}
where $n_r$ is the number of replicas in $\CE$.


\subsection{In conformal field theory}\label{cft-sec}
If the theory is conformal, the expression for $\CE$ can be simplified further. This is because, in two dimensions - thanks to the uniformization theorem - the manifold $\CM_\CE$ that has points with conical excess is conformally equivalent to a smooth or ``uniformized'' manifold, say $\CM^{\rm uni}_\CE$ of the same topology. The partition function $Z_{\CM_\CE}$ is related to that of the uniformized  manifold as
\begin{align}\label{uni-Z}
    Z_{\CM_\CE} &=  e^{-S_L^{(c)}[\phi]} \, Z_{\CM^{\rm uni}_\CE},\notag\\
    S_L^{(c)}[\phi]&\equiv \frac{c}{96\pi }\int d^2 z \sqrt{g}(\partial_\mu\phi \partial^\mu \phi+ 2R\phi).
\end{align}
Here $\phi$ is the Weyl factor relating the metric on $\CM_\CE$ and on $\CM_{\CE}^{\rm uni}$. More precisely, 
\begin{align}
    g_{\CM} = e^{\phi}g_{\CM^{\rm uni}}.
\end{align}
and $S_L^{(c)}[\phi]$ is the Liouville action with central charge $c$ - same as that of the CFT - evaluated on $\CM_{\CE}^{\rm uni}$. Equation \eqref{uni-Z} can be used to compute $\CE$ if $e^{-S_L[\phi]}$ and $Z_{\CM^{\rm uni}_\CE}$ can be computed separately. It turns out that $\phi$ can be computed using the so-called covering map. It is the holomorphic map that maps $\CM_\CE^{\rm uni}$ to $\CM$. In section \ref{cft-comp}, we will see how to compute such a covering map  in the case when $\CM$ and $\CM_\CE^{\rm uni}$ are $S^2$. The contribution $e^{-S_L[\phi]}$ from the Weyl factor is universal in that it does not depend on the details of the CFT. It depends only on the CFT through its central charge $c$ as the Liouville action depends on $c$. The other piece, $Z_{\CM^{\rm uni}_\CE}$, on the other hand does depend on the details of the theory. The only exception is when $\CM_\CE$ has genus $0$. In that case $\CM_\CE^{\rm uni}$ is the round sphere and hence $Z_{\CM^{\rm uni}_\CE}$ is normalized to $1$. Computation of genus $0$ symmetric invariants is done in section \ref{cft-comp}. 

If the theory is holographic, then $Z_{\CM^{\rm uni}_\CE}$ can be computed from the action of the dominant gravity solution that ``fills in'' the boundary $\CM^{\rm uni}_\CE$. It is widely believed that the hyperbolic manifold of minimum regularized volume (which is the Einstein-Hilbert action for hyperbolic three-manifolds) that fills in a Riemann surface is a handlebody. See \cite{2007arXiv0710.2129Y, 2008JHEP...09..120Y} for more discussion on this point. We will assume that this is true.
We will use this method to compute certain invariants for which the genus of $\CM_\CE$ is $1$ in section \ref{cft-comp}. 

There is an alternate, but equivalent way of formulating $\CE$ as a correlation function of twist operators. Consider the theory $\CQ^{\otimes n_r}$ that is $n_r$ copies of the original theory $\CQ$. It has an obvious global symmetry $S_{n_r}$ which permutes the $n_r$ copies. As a result, the theory has twist operators each of which is labeled by an element $\sigma$ of $S_{n_r}$. A twist operator $\CO_\sigma$ is point-like and implements a monodromy on the fields of $\CQ^{\otimes {n_r}}$ by the permutation $\sigma\in S_{n_r}$ around it. 
They are inserted at the points that separate adjacent party regions. 
As the number of disconnected regions in $s$, the number of twist operators needed to separate them is also $s$.  
If a twist operator is separating regions of party $\ta$ and party $\tb$ then the twist operator will be of type $\sigma_{\ta\tb}=g_\ta^{-1}g_\tb$. As remarked earlier, for symmetric invariants, the permutations of all twist operators generate the replica symmetry group. 
With this definition of the twist operator, $\CE$ is simply the correlation function of the twist operators. More precisely,
\begin{align}
    \CE=\langle\CO_{\sigma_1}(x_1)\ldots \CO_{\sigma_s}(x_s) \rangle_{\CM}/ Z_{\CM}^{n_r}.
\end{align}
The subscript of the correlation function emphasizes that it is evaluated on $\CM$. The formulation of $\CE$ as a correlation function of twist operators has another advantage. Just like entanglement entropy all our invariants $\CE$ and their normalized versions $\tilde \CE$ are UV divergent. Thinking of them as a correlation function of twist operators offers a natural way of regularizing this divergence. We simply canonically normalize all the twist operators inserted at finite points such that their two point function at unit separation is $1$. Let us denote the twist operators at finite points  normalized in this way as $\CO_{\sigma}^{\rm norm}(x)$. 
To emphasize the difference in normalization, we will denote the invariant obtained this way as $\CE^{\rm reg}$.
\begin{align}
    \CE^{\rm reg}=\langle\CO_{\sigma_1}(x_1)\ldots \CO_{\sigma_s}(x_s) \rangle_{\CM}/\prod_{i=1}^s(\CO_{\sigma_i}(0)\CO_{{\sigma_i}^{-1}}(1))^\frac12 =\langle\CO_{\sigma_1}^{\rm norm}(x_1)\ldots \CO_{\sigma_s}^{\rm norm}(x_s) \rangle_{\CM}. 
\end{align}
We will often be interested in inserting one of the twist operators at $\infty$. In this case, even the regularized invariant $\CE^{\rm reg}$ becomes $0$. We define the twist operators at $\infty$ as the limit
\begin{align}
    \CO_\sigma^{\rm norm}(\infty) = \lim_{x\to \infty} \CO_\sigma(x)^{\rm norm}\, |x|^{2\Delta_\sigma}.
\end{align}
As a result $\CO_\sigma(x) \CO_{\sigma^{-1}}(\infty)=1$. 

If a twist operator is separating regions corresponding to party $\ta$ and $\tb$, then it is of the type $\sigma_{\ta\tb}= g_\ta^{-1} g_{\tb}$. Let $p_k(\sigma)$ be the number of $k$ cycles in the permutation element $\sigma$. The Euler characteristic  of $\CM_\CE$ can be computed from the cycle structure of all the twist operators using the Riemann-Hurwitz formula
\begin{align}\label{riem-hur-gen}
    \chi_{\CM_\CE} = {n_r} \, \chi_{\CM} - \sum_{\CO_\sigma} \sum_{k}p_k(\sigma) (k-1).
\end{align}
The $\sum_{\CO_\sigma}$ is over all the twist operators and $\chi_\CM$ is the Euler characteristic of $\CM$. In the entirety of the paper, we will encounter twist operators which have all the cycles of equal length\footnote{This is a consequence of Cayley's theorem for finite groups which states that every group $G$ of order $|G|=n$ is isomorphic to a subgroup of $\mathbb{S}_n$. This is done by constructing the ``regular representation" which can be regarded as a set of permutations of $\mathbb{S}_n$. In particular the regular representation has a group action which acts freely and transitively and has the property that each permutation consists of cycles all of the same length where the length is the order of that element. When choosing representations for the replica symmetry we will always work with the regular representation.}. We will only focus on twist operators of such type.  Letting the length of the cycle of twist operator $\CO_\sigma$ to be $k_\sigma$, the number of cycles in $\sigma$ is ${n_r}/k_\sigma$. The Riemann-Hurwitz formula, specialized to this case is
\begin{align}\label{riem-hur}
    \chi_{\CM_\CE} = {n_r} \, \Big(\chi_{\CM} - \sum_{\CO_\sigma}\Big(1-\frac{1}{k_\sigma}\Big)\Big).
\end{align}

An important feature to note about $\CM_\CE$ is that it inherits the replica symmetry $\CR$ of $\CE$ as discrete isometry. Moreover, if the replica symmetry acts in a free and transitive fashion on the replicas then the orbifolding $\CM_\CE$ by $\CR$ gives the original manifold $\CM$. This orbifolding is especially powerful when the quantum field theory that we are working with is a holographic CFT (see section \ref{hol-replica}). This is what makes our formalism of symmetric invariant useful in computing them in holographic CFTs.

\subsection{In holography: using bulk replica symmetry}\label{hol-replica}
Another way to use holography to compute $Z_{\CM^{\rm uni}_\CE}$ is to make use of the replica symmetry. As remarked earlier, the replica symmetry of $\CE$ becomes the discrete isometry of $\CM_\CE$ and as $\CM_\CE^{\rm uni}$ is conformally equivalent to $\CM_\CE$, the replica symmetry also acts on $\CM_\CE^{\rm uni}$ a discrete conformal isometry. We will also assume that the replica symmetry acts freely and transitively on the replicated boundary $\CM_\CE$. 

If the replica symmetry extends to the dominant bulk solution $\CB_\CE^{\rm uni}$ as a discrete isometry of $\CB_\CE^{\rm uni}$, then we can orbifold $\CB_\CE^{\rm uni}$ with this symmetry. The orbifold $\tilde \CB_\CE^{\rm uni}$ has singular loci corresponding to the fixed points of $\CR$. If the locus is fixed under the action of element $g$ of $\CR$, then the associated conical singular locus in $\tilde \CB_\CE^{\rm uni}$ has a cone angle $2\pi/m$ where $m$ is the order of $g$ i.e. $m$ is the smallest integer such that $g^m=1$. The boundary of $\tilde \CB_\CE^{\rm uni}$, called $\tilde \CM_\CE^{\rm uni}$, has conical singularities but is conformally equivalent to $\CM$.  

The orbifold $\tilde \CB_\CE^{\rm uni}$ and its boundary $\tilde \CM_\CE^{\rm uni}$ also inherit the orientation reversing ${\mathbb Z}_2$ isometry of $\CM$. If we extend the symmetry $\CR$ of $\CB_\CE^{\rm uni}$ by this ${\mathbb Z}_2$, we get the extended replica symmetry group $\hat \CR$ which contains elements that are both orientation preserving and reversing. As we have seen in section \ref{sym-inv}, $\hat \CR$ is generated by reflections $r_\ta$. If we orbifold $\CM_{\CE}^{\rm uni}$ by $\hat \CR$, we get only the bra (or ket) part of $\tilde \CM_{\CE}^{\rm uni}$. This has a boundary, along which it is glued to the ket part. The boundary is a union of intervals, each of which is fixed under some reflection $r_\ta$. The Hilbert space associated with this interval is identified with party $\ta$. This reverse engineers the invariant $\CE$ completely in terms of $\hat \CR$ and the reflection generators $r_\ta$'s along with the definition of the party regions. 

It is often useful to orbifold $\CB_\CE$, the bulk solution that fills in $\CM_\CE$ rather than its uniformized version $\CB_\CE^{\rm uni}$. These two orbifolded manifolds differ only in the Weyl factor of the boundary. The boundary of the former is the original $\CM$ while the boundary of the latter is $\tilde \CM_{\CE}^{\rm uni}$ which is only Weyl equivalent to $\CM$. Orbifolding of $\CB_\CE$ has the advantage that the boundary of the orbifold  $\tilde \CB_\CE$ is $\CM$ and not $\tilde \CM_{\CE}^{\rm uni}$. As a result, we don't have to worry about the uniformization factor $e^{-S[\phi]}$ coming from a Weyl rescaling. The normalized invariant is written straightforwardly in terms of the gravitational action for the orbifold solution $\hat \CB_\CE$ as,
\begin{align}\label{norm-gravity}
    \tilde \CE= e^{-S_{\rm grav}[\hat \CB_\CE]+S_{\rm grav}[\CB]}.
\end{align}
Here $\CB$ is the bulk solution that fills in the original manifold $\CM$ and $S_{\rm grav}$ is the gravitation action for the given solution. This is the key formula that will help us prove interesting equalities between various $\tilde \CE$'s. We will do so by showing that $\hat \CB_{\CE}$ is the same orbifold for infinitely large families  of symmetric  invariants. 
To characterize the action of replica symmetry group, it is convenient to work with the uniformized version $\CM_{\CE}^{\rm uni}$ as it allows for an effective application of the theory of Kleinian group as we will see shortly.

\section{Bulk replica symmetry}\label{bulk-replica}
Now we are in a position to characterize handlebodies which have a non-trivial isometry group. We reverse engineer the invariants $\CE$ that preserve the bulk replica symmetry from this characterization. 
The only novel concept needed for this characterization is the way of constructing handlebodies as quotients of the hyperbolic ball $H^3$ by \emph{Schottky groups}. Let us describe this briefly. 

\subsection{Handlebodies as Schottky quotients}
Take $g$ pairs of circles $\{C_i, C_i'\}$ on $S^2$ such that  interiors of all the circles are disjoint. For a given pair $i$, consider the conformal transformation $L_i$ that maps interior of one to the exterior of the other. The conformal generators $L_i$ do not obey any relations so they generate a free group on $g$ generators. This group is called the Schottky group $\CS$. Quotienting $S^2$ by Schottky generators effectively identifies the circles in each pair creating $g$ handles of a genus $g$ surface. The fundamental domain of the quotient is $S^2$ with the interior of all the $2g$ circles removed. This procedure is summarized in the figure \ref{schottky-fig}:
\begin{figure}[h]
    \begin{center}
        \includegraphics[scale=0.20]{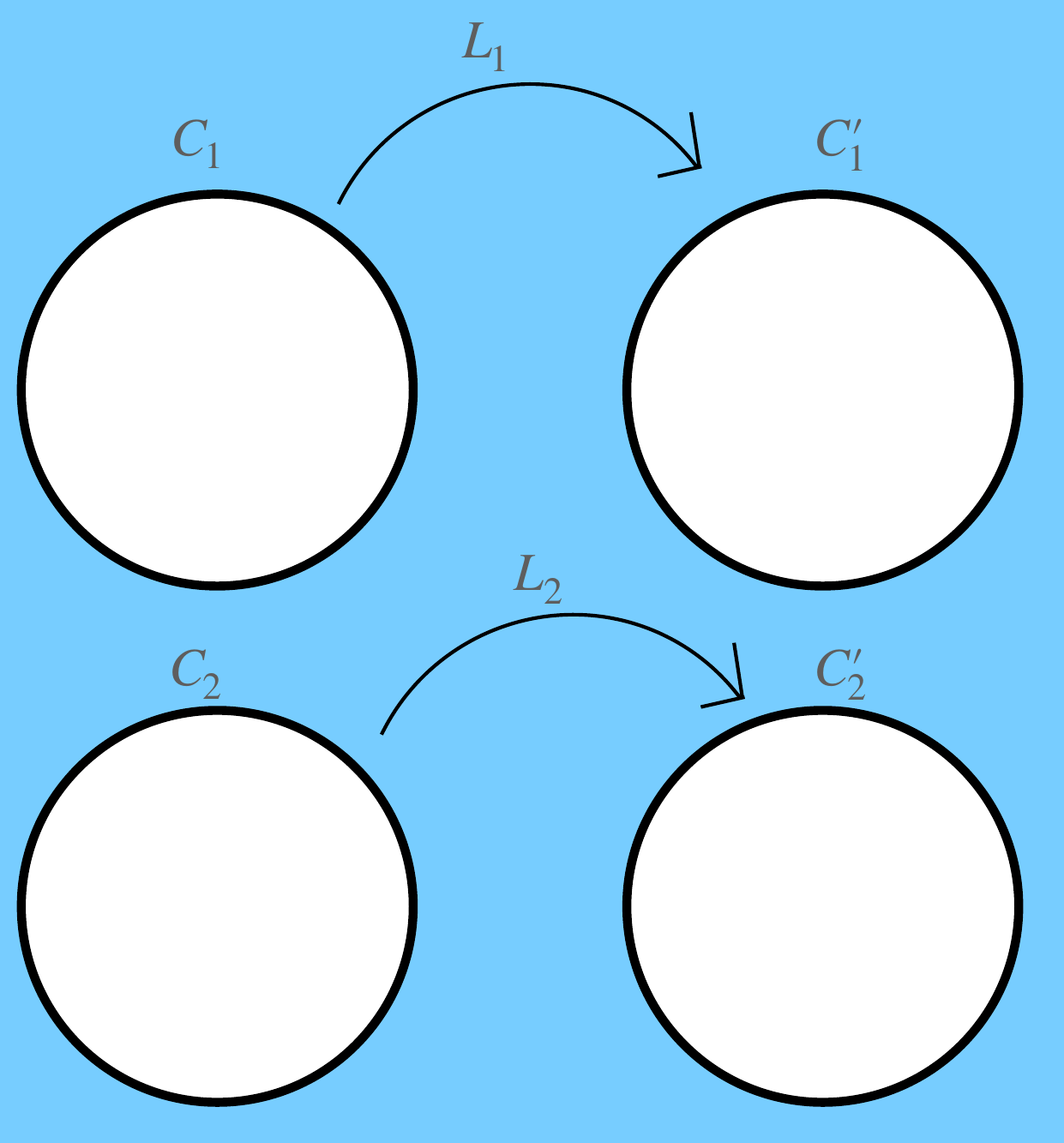}
        \qquad 
        \includegraphics[scale=0.30]{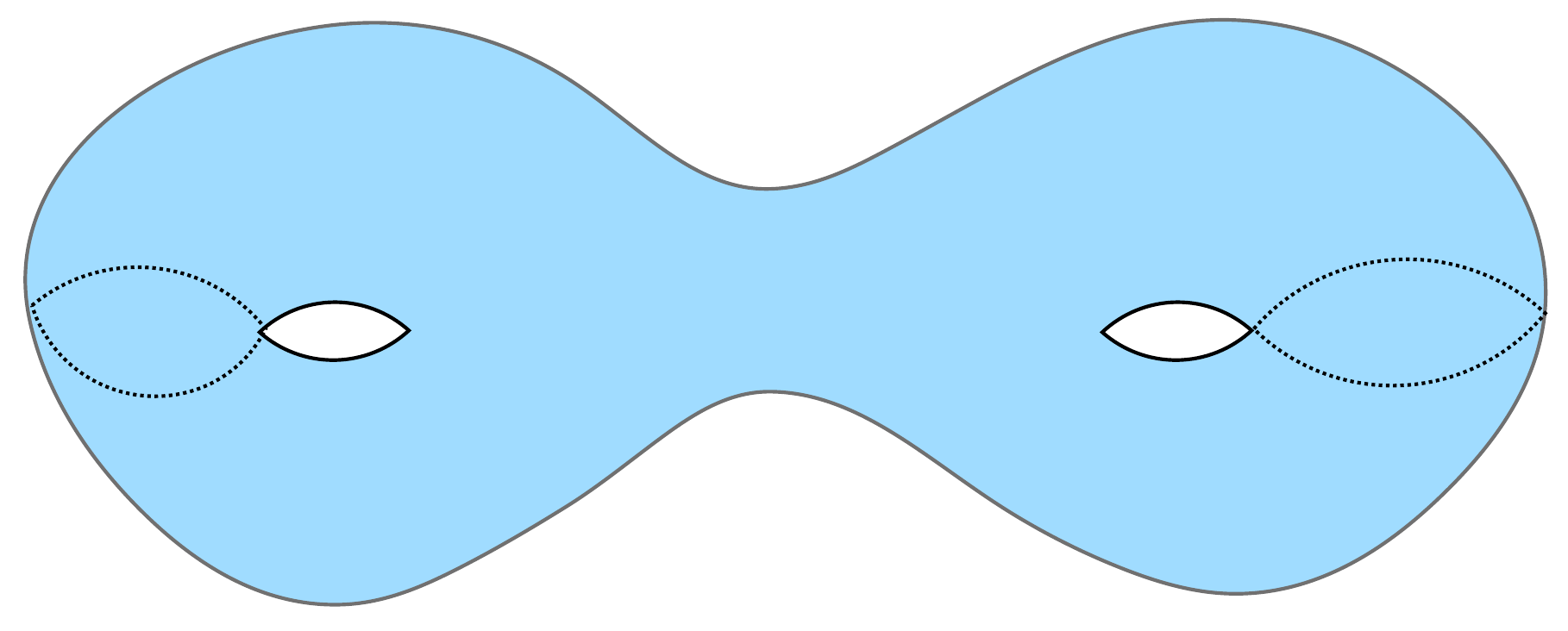}
    \end{center}
    \caption{The first figure shows two pairs of circles $\{C_1, C_1'\}$ and $\{C_2, C_2'\}$ on a sphere. The exterior of these circles is colored blue. The second figure shows the genus $2$ Riemann surface obtained by identifying $\{C_i, C_i'\}$ by conformal transformations $L_i$. The fundamental domain for this quotient is precisely the exterior of all the circles.}
    \label{schottky-fig}
\end{figure} 

The advantage of thinking of  the Riemann surface as a Schottky quotient is that the Schottky representation uniquely constructs a handlebody with hyperbolic metric whose boundary is the Riemann surface. 
If we think of the $S^2$ on which the Schottky group acts as a boundary of hyperbolic ball $H^3$ then the action of Schottky group can be extended uniquely  to the bulk in a way that preserves the hyperbolic structure. Recall that the conformal transformations on $S^2$ act as isometries on $H^3$. We simply extend the action of Schottky generators thought of as conformal transformations on the sphere to the isometric action on $H^3$. As a result, it preserves the hyperbolic metric. The quotient effectively removes the hemispherical ``scoops'' from $H^3$ and identifies the resulting boundary pairwise. The pair of circles on the boundary that are mapped to each other by Schottky generators become cycles that are contractible in the handlebody. 
In this way, a handlebody is uniquely associated to a Schottky representation of the Riemann surface. As $H^3$ is simply connected, the fundamental group of the quotient genus $g$-Handlebody is a free group on $g$ generators namely, the Schottky group, as expected. We summarize with the following theorem:
\begin{theorem}[\cite{Maskit1987-ee, Marden2016-sb}]
    The following are equivalent:
\begin{enumerate}
\itemsep-.1em
    \item $\CS$ is a Schottky group of rank $g$.
    \item $\CS$ is freely generated by $g$ loxodromic conformal isometries.
    \item Let $C_1,C_1'\cdots C_g,C_g'$ be $2g$ disjoint simple closed curves\footnote{For this paper we will always take these to be circles. Such Schottky groups are said to be ``classical". Any handlebody admitting an anti-conformal involution may be constructed from the quotient of $H^3$ by a classical Schottky group \cite{reto_hid}.}. which bound the region $D\subset \hat{\mathbb{C}}$ and $L_1,\cdots L_g$ the set of conformal isometries with $L_i(C_i)=C'_i$ and $L_i(D)\cap D =\emptyset$. $\CS$ is a Kleinian group generated by $L_1,\cdots L_g$ with fundamental region $D$.
    \item The quotient $H^3/\CS$ is a handlebody of genus $g$.
\end{enumerate}
\end{theorem}

\subsection{Symmetric handlebodies}\label{sym-handle}
The bulk solution $\CB_{\CE}^{\rm uni}$ is a handlebody whose boundary is a Riemann surface $\CM_{\CE}^{\rm uni}$. The genus $g$ of $\CM_{\CE}^{\rm uni}$ is computed with the Riemann-Hurwitz formula \eqref{riem-hur}. It is useful to think of the handlebody as a Schottky quotient of the $3$-ball $H^3$ where the action of the Schottky group is realized as its isometry or equivalently as a conformal isometry of the boundary $S^2$. Refer to the beginning of section \ref{klein-geom} for a quick review of the conformal isometries of $S^2$ and their extension into the bulk $H^3$ as isometries.
Algebraically, a Schottky group is a free group on $g$ generators where $g$ is the genus. 
The handlebody  $\CB_{\CE}^{\rm uni}$ obtained after the quotient by the Schottky group  is assumed to have a further action by a discrete isometry group. This is possible if we can find a discrete subgroup $\CK$, possibly infinite, of the isometry group of $H^3$ which has a free group $\CS$ with finite index as a normal subgroup. The finite index condition simply means that its quotient $\CK/\CS$ will be a finite set. Because the free subgroup $\CS$ is also normal, the quotient $\CK/\CS$ is in fact a finite group. 
A discrete subgroup of the isometry group of $H^3$ is called a \emph{Kleinian group}\footnote{The classic reference is \cite{Maskit1987-ee}. See also \cite{Marden2016-sb} for a modern treatment in the context of hyperbolic 3-manifolds.}. A group which has a finite index normal subgroup that is free is called a virtually free group. With this nomenclature, what we are looking for are virtually free Kleinian groups. They completely characterize handlebodies with a non-trivial symmetry group. Let us explain this now.  

Identify the free subgroup $\CS$ with the Schottky group. The number of generators of $\CS$ is the genus of the resulting handlebody after quotienting $H^3$ by $\CS$. The quotient handlebody $H^3/\CS$ is identified with $\CB_\CE$. The quotient group $\CK/\CS$ acts on the handlebody as a discrete group of isometries. Because $\CS$ is of finite index in $\CK$, $\CK/\CS$ is a finite group and is identified with the replica symmetry group $\CR$. The original manifold $\CM$ that computes the norm of the state is then Weyl equivalent to the boundary of $\CB_\CE/\CR=(H^3/\CS)/(\CK/\CS)=H^3/\CK$. As remarked earlier, $\CM$ further enjoys the action of a ${\mathbb Z}_2$ reflection. This means that the Kleinian group $\CK$ can be extended to an \emph{extended Kleinian group} $\hat \CK$ acting on $H^3$ by this orientation reversing ${\mathbb Z}_2$. It is such that $\hat \CK/\CS= \hat \CR$. 
As explained in section \ref{hol-replica}, looking at the boundary regions of $\CM=\tilde \CM_\CE/{\hat R}= (\partial H^3)/{\hat K}$ that are fixed under reflections $r_\ta$, determines the multi-invariant $\CE$ along with the definition of party regions completely. 

Let us summarize the prescription for reverse engineering replica symmetry preserving multi-invariants:
\begin{itemize}
    \item Construct a virtually free extended Kleinian group $\hat \CK$. 
    \item Let $\CS$ be a finite index free subgroup of $\hat \CK$. Require that the quotient $\hat \CK/\CS$ be generated by reflections $r_\ta$. 
    \item The quotient $(\partial H^3)/\hat \CK$ is a surface with boundaries. The boundary is divided into regions according to the reflection element $r_\ta$ that fixes it. 
\end{itemize}
The bra is prepared by the CFT path integral on $(\partial H^3)/\hat \CK$ and the region associated to the reflection $r_\ta$ is defined as party $\ta$. The group $\hat \CK/\CS$ is identified with the extended replica symmetry group $\hat R$. 

\begin{figure}[h]
    \begin{center}
        \includegraphics[scale=0.2]{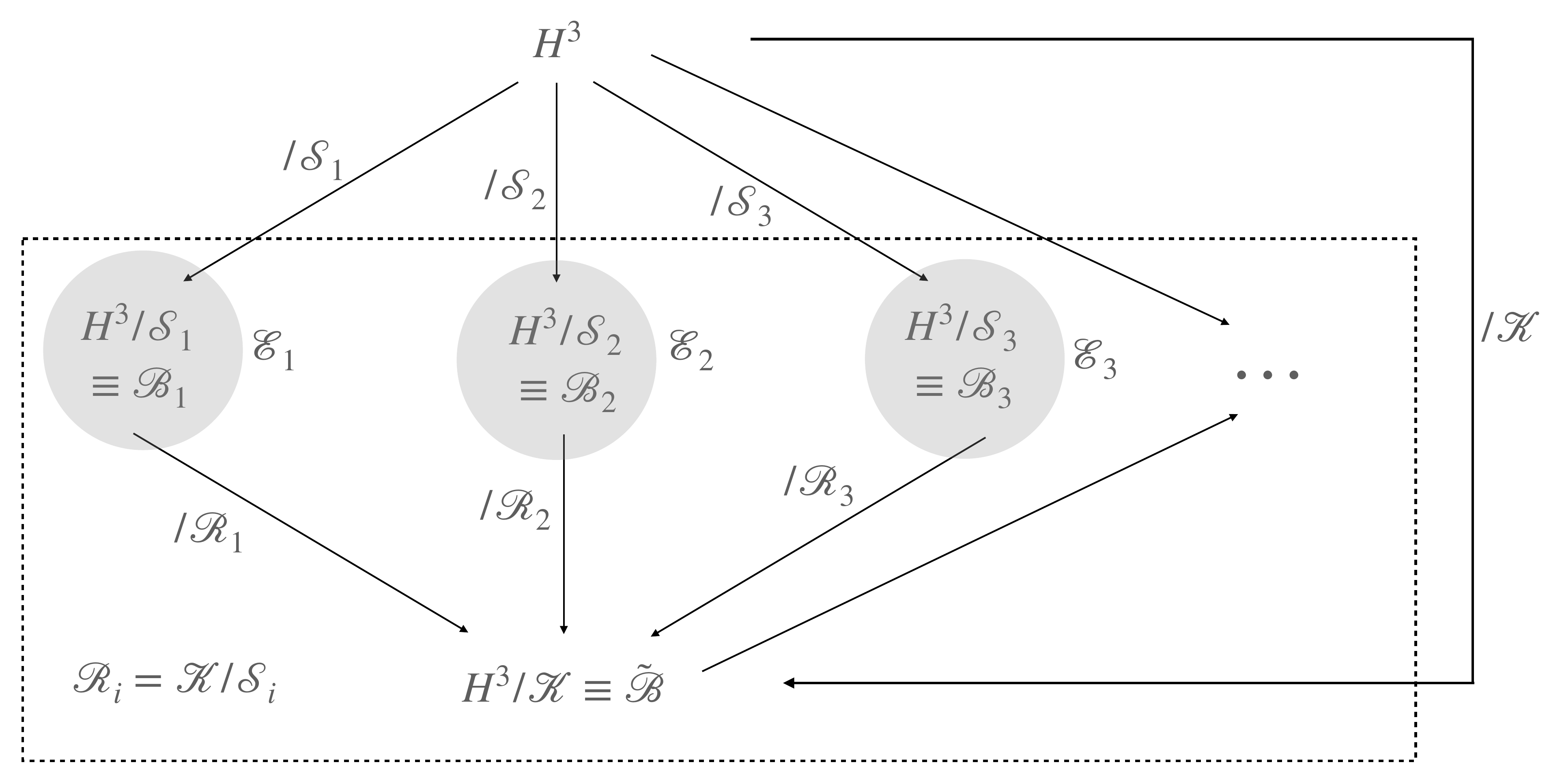}
    \end{center}
    \caption{
        The figure shows that the final orbifold geometry is obtained by quotienting $H^3$ by a Kleinian group $\CK$. However, this quotient can be done in steps. Consider some free normal subgroup $\CS_i$ of $\CK$. Quotienting $H^3$ by $\CS_i$ first gives a genus $g_i$ handlebody where $g_i$ is the number of generators of $\CS_i$. This handlebody is identified with $\CB_i$, the bulk geometry filling in the replicated manifold for the invariant $\CE_i$.  
        The quotient group $\CK/\CS_i$ acts on $\CB_i$. This is the replica symmetry group of ${\CE_i}$. This shows that for all $i$, the orbifold geometry evaluating the invariant $\CE_i$ is the same, namely $H^3/\CK$.}
    \label{key-idea}
\end{figure}

Our prescription for reverse engineering bulk replica symmetry preserving invariants has a remarkable physical consequence. The state whose multi-invariant is evaluated is fixed completely by choice of the extended Kleinian group $\hat \CK$. But the invariant itself depends on the choice of the free normal subgroup $S$ in $\CK$. This is clear because the number of generators of $\CS$ is the genus of $\CM_\CE$. Also, the replica symmetry is $\hat \CK/\CS$ which certainly depends on $\CS$. However, the invariant is evaluated by the gravitational action on $H^3/{\CK}$ which doesn't depend on the choice of $\CS$. This implies that there are multiple normalized invariants $\tilde \CE$ for a given state with the given definition of the regions  that are identical! The key idea described above is summarized in figure \ref{key-idea}.
What is more is that our analysis depends only on the symmetry properties of $H^3$ and only uses the classical nature of the bulk theory which enables us to compute the CFT partition function from the action of the bulk saddle point. 

One may be tempted to conclude that it is robust against higher derivative corrections as long as higher derivative corrected theory admits $H^3$ as a solution. This is indeed so, however, three dimensional gravity does not have any dynamical degrees of freedom and hence does not have any non-trivial S-matrix. As the space of gravitational actions up to field redefinitions is the same as the space of S-matrix, we do not expect any non-trivial higher derivative corrections.  

\subsection{Index of the free subgroup}
As we will be working with virtually free groups, say $\CK$ and their free subgroups, say $\CS$, it is useful to recall a mathematical result relating the number of generators $g$ of  $\CS$ i.e. the rank of $\CS$ and its index $I$ in $\CK$ i.e. the cardinality of the coset $\CK/\CS$. This is given by \cite{10.1112/plms/s3-59.2.373}
\begin{align}\label{free-index}
    I=\frac{1}{\chi(\CK)} (1-g),
\end{align}
where $\chi(\CK)$ is called the Euler characteristic of $\CK$. In order to avoid confusion between this and the Euler characteristic of Riemann surfaces that appear elsewhere in the paper, we will call $\chi(\CK)$ the characteristic of $\CK$. If $\CK$ itself is a free group $\CF$ of $n$ generators then $\chi(\CF)=1-n$. In this case, the above result becomes the  well-known result of Nielsen and Schreier. The virtually free groups that we are interested in are Kleinian groups. Their characteristic is usually a negative rational number. Also the free subgroups of these that we are interested in are normal, so the index of $\CS$ in $\CK$ is the order of the quotient group $\CK/\CS$ which is the replica symmetry group. As the replica symmetry group acts freely and transitively on the replicas, it is also equal to the number of replicas.
We will give a formula for calculating the characteristics for virtually free Kleinian groups obtained as ``amalgams'' of two finite Kleinian groups is given  in section \ref{amalgam-sec} in equation \eqref{two-amalgam}. A more general formula that is valid for amalgamations of multiple finite Kleinian groups is given in  \ref{general-tree} in equation \eqref{bulk-char}. The rank of the Schottky group is also the equal to the genus of the Riemann surface $\CM_\CE$.  

The discussion so far has been quite general, abstract and algebraic. But as the problem pertains to Kleinian groups, it is inherently geometrical. Now we will outline a prescription to construct Kleinian and extended Kleinian groups and in section \ref{amalgam-sec}, we will carry it out explicitly for a number examples. 

\section{Geometry of Kleinian groups}\label{klein-geom}

Let us start by a closer look at the conformal group acting on the two-sphere $S^2$. This group is $SO(3,1)\equiv PSL(2,{\mathbb C})$. It is most convenient to represent its action as M\"{o}bius transformations on the extended complex plane $\hat {\mathbb C}$ which are the set of orientation preserving conformal isometries
\begin{align}
A(z)=\frac{az+b}{cz+d}\quad \Leftrightarrow \quad \begin{pmatrix}
a & \quad b \\
c & \quad d 
\end{pmatrix},\quad ad-bc=1
\end{align}
where the parameters $a,b,c,d$ are complex. Composition of M\"{o}bius transformations corresponds to the multiplication of $2\times 2$ matrices. 
The conformal group consists of isometries of the complex plane namely rotations $z\to e^{i \theta}z$ and translations $z\to z+a$, in addition to the conformal transformations that include scaling $z\to \lambda z$ and special conformal transformations. From a mathematical point view it is useful to classify M\"{o}bius transformations by examining their fixed points
\begin{align}
    A(z)=z\quad \Longrightarrow\quad  z= \frac{a-d\pm \sqrt{(a+d)^2-4}}{2c}.
\end{align}
The group elements are classified according to the sign of the discriminant $(a+d)^2-4 ={\rm Tr}(A)^2 - 4$. If $\lambda$ in the loxodromic element is real then it is called a hyperbolic element\footnote{Some authors instead use the terms ``hyperbolic" and ``pure hyperbolic" in place of ``loxodromic" and ``hyperbolic" respectively.}.
\begin{table}[H]
    \centering
    \begin{tabular}{|l|l|l|l|}
    \hline
        Type & $\Tr(A)^2-4$ & Motion  & Representative \\ \hline\hline
        elliptic & $<0$  & rotation & $e^{i\theta}z$ \\ \hline
        parabolic & $=0$  & translation & $z+1$ \\ \hline
        loxodromic & $>0$  & scale/(scale+rotation) & $\lambda z, \quad \lambda\in {\mathbb C}$\\ \hline
    \end{tabular}
    \caption{Classification of Mobius transformations}
\end{table}
\noindent
As ${\rm Tr}A$ is invariant under conjugation $g \cdot A\cdot  g^{-1}$ for some $g\in PSL(2,{\mathbb C})$, the type of the element does not change under conjugation. The last column of the table gives the representative element in each conjugacy class. A M\"{o}bius transformation is parabolic if and only if it has only one fixed point in the extended complex plane, while elliptic and loxodromic elements have precisely two fixed points.
The only elements that can have finite order are elliptical transformations. It is conjugate to the transformation $z\to z\, e^{2\pi i/p}$. In this paper, finite order elements will play an important role.

In this paper, we are interested not only in the Kleinian groups, but also  in extended Kleinian groups. These are subgroups of $PSL(2,{\mathbb C})$ extended by an orientation reversing conformal isometry. We can take the generator of this transformation to be reflection across the $x$ axis i.e. complex conjugation $r_x$ defined as $r_x(z)= \bar z$. A general orientation reversing element is obtained by composing complex conjugation with a general  M\"{o}bius transformation.
\begin{align}
    A(z)=\frac{a\bar z+b}{c\bar z+d}.
\end{align}
A orientation reversing conformal isometry is sometimes called an anti-conformal isometry. Special among these are the ones which square to $1$ which are called reflections. 
It is useful to note that the fixed point locus of a reflection that is conjugate to complex-conjugation $r_x$ is a circle. In particular, if the conjugation is  by conformal transformation $A(z)$, then the fixed point locus of $A\cdot  r_x \cdot A^{-1}$ is the image of the $x$-axis under $A$. In  fact, a reflection $r_f$ is uniquely determined by its fixed point locus $f$. This is because if we conjugate $r_f$ by a conformal transformation $A$ that keeps $f$ fixed set-wise then $A \cdot r_f\cdot A^{-1}=r_f$. This is readily seen by first taking $r_f=r_x$ and $A(z)=z+a$ with $a\in {\mathbb R}$ and then conjugating this to prove for general $r_f$. This justifies labeling of a reflection by its fixed point locus.

\subsection{Extension to $H^3$}

The conformal isometries of $S^2$ are straightforwardly extended to $H^3$ as isometries. Parametrizing $H^3$ as the upper half space with the boundary identified with the extended complex plane as $(z,\bar z, t)$, the metric on $H^3$ is given by
\begin{align}
    ds^2= \frac{1}{t^2}(dt^2+dz d\bar z).
\end{align}
A general M\"{o}bius transformation is extended to $H^3$ as,
\begin{align}
    z\longrightarrow z'=\frac{(az+b)\overline{(cz+d)}+a\bar{c}t^2}{|cz+d|^2+|c|^2t^2},\quad t\longrightarrow t'=\frac{t}{|cz+d|^2+|c|^2t^2}. 
\end{align}
Complex conjugation is extended to $H^3$ simply as $(z,\bar z)\to (\bar z, z)$ and $t\to t$. Let us emphasis that a conformal transformation of $S^2$  completely determines the isometric transformation of $H^3$. The group of conformal isometries of $S^2$ and those of $H^3$ are isomorphic to each other. It is useful to note that the circular fixed locus $L$ of a reflecting   anti-conformal transformation extends into the bulk as a hemisphere whose boundary is $L$. When $L$ is circle with a point at infinity as in the case of complex conjugation $r_x$, the fixed point locus in the bulk becomes a plane. Now we are set to discuss construction of Kleinian and extended Kleinian groups concretely.

\subsection{Finite Kleinian groups}\label{fin-klein}
It is useful to first construct the Kleinian groups that are finite\footnote{The finite groups are a subset of the \emph{elementary} Kleinian groups. These are completely classified based upon the number of hyperbolic and parabolic fixed points: either 0, 1 parabolic, or 2 hyperbolic. All other Kleinian groups have an infinite number of hyperbolic fixed points. For examples of explicit constructions see \cite{eleklein_sato}.}. They turn out to be finite subgroups of isometries of the sphere. As the group is finite, every element of the group must have finite order and the only elements of the conformal group that have finite order are elliptical i.e. rotations. So a finite subgroup of conformal transformations must consist entirely of rotations and hence must be a subgroup of $SO(3)$, the group of isometries of the sphere. 
\begin{figure}[h]
    \begin{center}
         \includegraphics[scale=0.25]{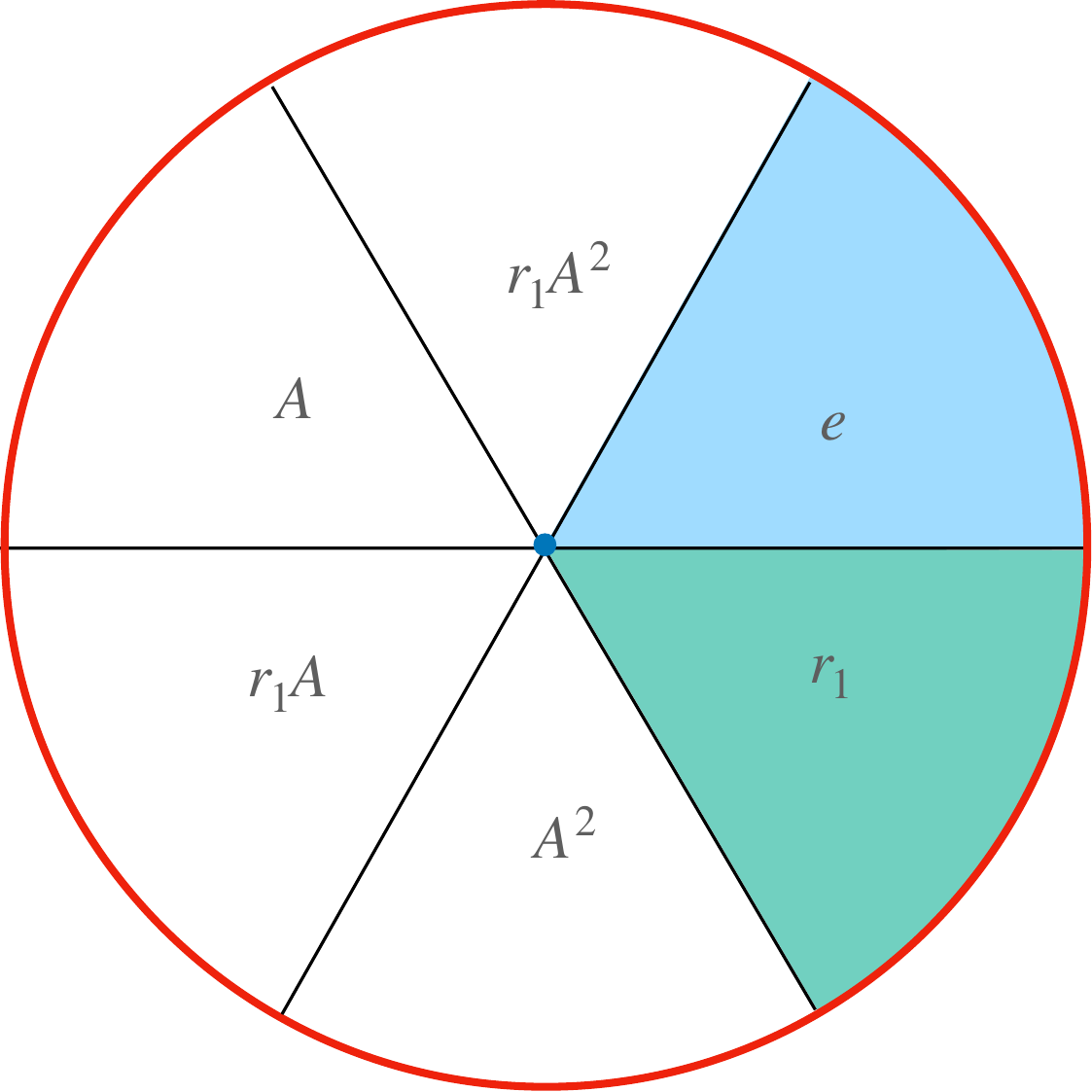}
        \qquad 
        \includegraphics[scale=0.25]{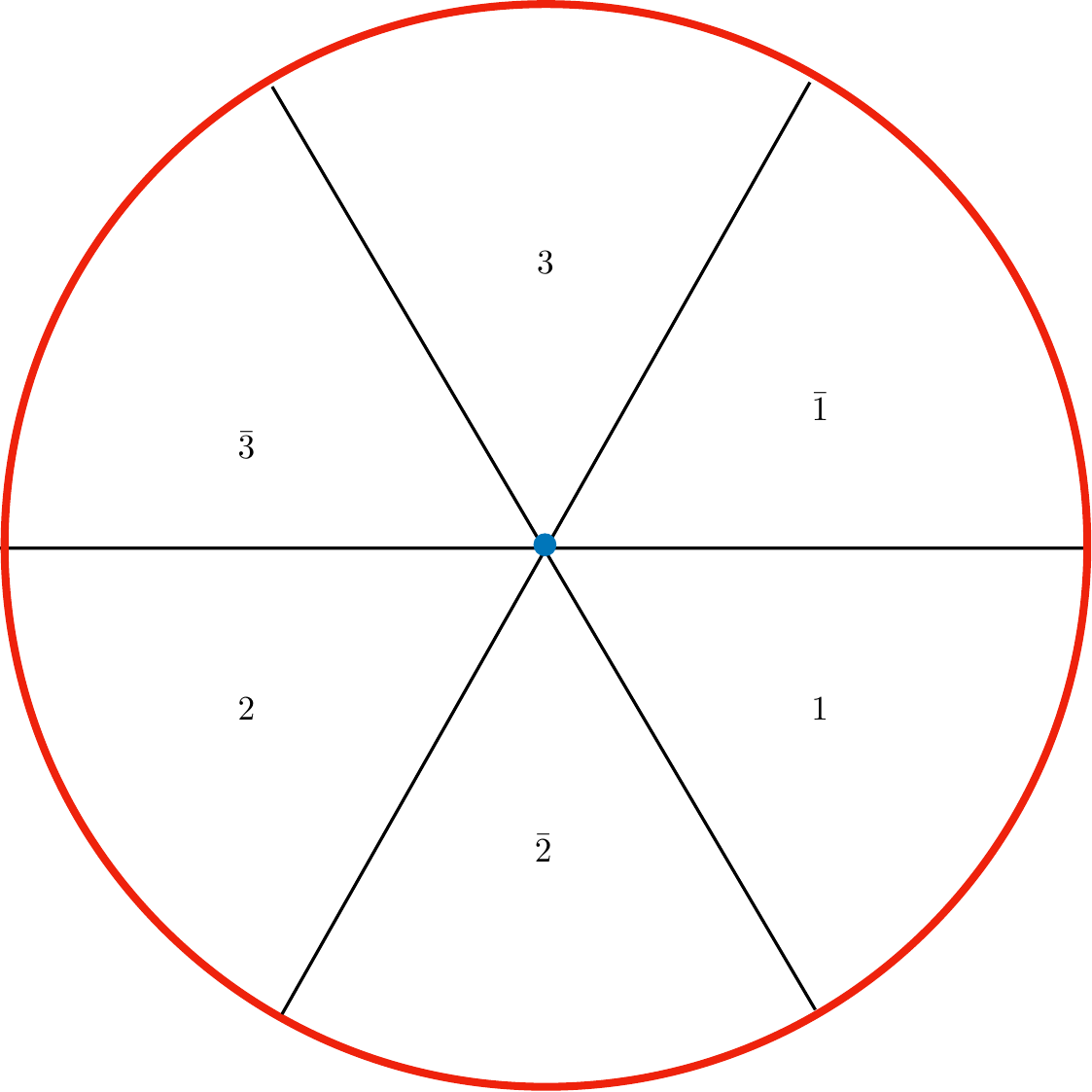}
        \qquad      
        \includegraphics[scale=0.35]{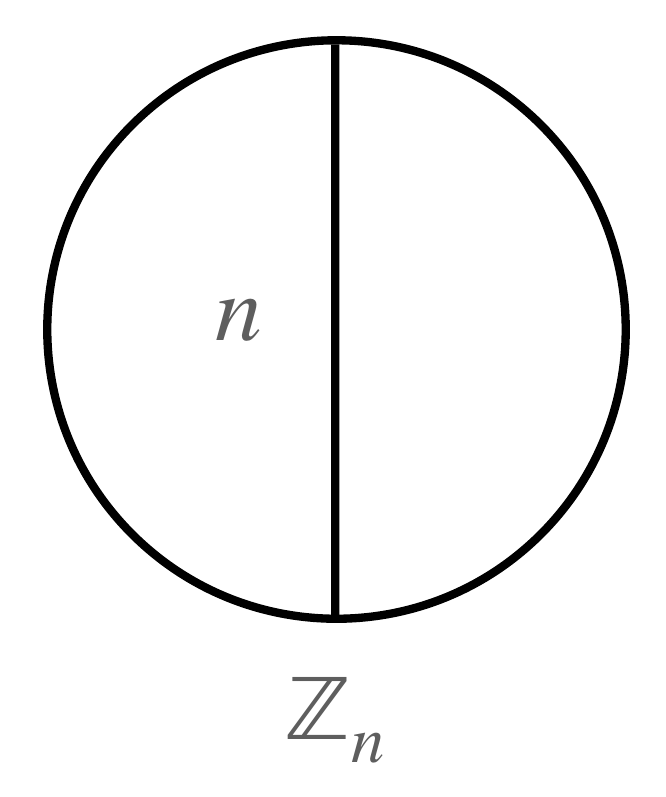}
    \end{center}
    \caption{The two reflections $r_1(z)=\bar{z},r_2(z)=e^{\frac{2 \pi i}{6}}\bar{z}$ generate an extended Kleinian group $\hat\CK=\langle r_1^2=r_2^2=e, \; (r_2r_1)^3=e \rangle$ isomorphic to ${\mathbb Z}_6$. By considering compositions of these reflections the boundary $\hat{\mathbb{C}}$  is divided into a number of regions forming a tessellation by bi-gons with angles $\frac{\pi}{3}$. As a consequence the bulk $H^3$ is also divided into the same number of cells whose boundary on $\hat{\mathbb{C}}$ is one of these regions. In the first figure we have shown the tessellation as well as each region labeled by a unique group element. They are labeled in terms of $r_1$ and the conformal isometry $A=r_2r_1=e^{\frac{2 \pi i}{3}}z$. The blue region labeled with the identity $e$ is the fundamental region of the extended Kleinian group $\hat\CK$ while the union of the blue and teal region labeled $r_1$ are the fundamental region of the Kleinian group $\CK=\langle A^3=e\rangle \sim \mathbb{Z}_3$. The labels correspond to the image of the fundamental region under the listed transformation. In the tessellation each vertex will be a unique elliptical fixed point. Here there are two: the point at zero in blue and the point at infinity represented as the red circle. They have been labeled by different colors because they can not be mapped onto each other by the action of $\CK$. In order to determine the multi-invariant we consider an explicit permutation representation of the replica symmetry $\mathbb{Z}_3 :\langle a^3=e\rangle$ where $a=(123)$. Working with the same tessellation we choose a group element for each side of the fundamental region which will correspond to parties in the boundary CFT: $g_O=e,\quad g_A=a^{2}$. Now labeling the fundamental region $\bar{1}$ we look at the adjacent regions to $O,A$ and label them with unbarred numbers according to the action of $g_O$ and $g_A$: 1 and 3 respectively. The result of continuing this procedure is shown in the second figure. The regions with unbarred labels represent copies of the bra and the regions with barred labels represent copies of the ket. Using the Euclidean path integral this can be translated into an explicit multi-invariant constructed from $g_O$ and $g_A$. Here the result is the familiar quantity  ${\rm Tr}\rho_A^n$ with $n=3$. The third figure shows the singular locus obtained by orbifolding $H^3$ by the replica symmetry group ${\mathbb Z}_n$. This arises because the bulk geodesic connecting the elliptical fixed points at 0 and $\infty$ is fixed in $H^3$ by the action of $\CK$. The cone angle around the singularity is $2\pi/n$. This is the familiar ``heavy'' cosmic brane solution dual to the $n$-th Renyi entropy.}
    \label{zn-orb-fig}
\end{figure}
The simplest of such groups is ${\mathbb Z}_n$. The action of ${\mathbb Z}_n$ has two fixed points, say the north pole and south pole. Its action can be extended by a reflection to ${\mathbb Z}_{2n}$. The fundamental region of the ${\mathbb Z}_{2n}$ is bounded by two great semi-circles i.e. meridians with an angle $\pi/n$ between them. As a result, the invariant corresponding to this action is a bi-partite one. In fact, as the replica symmetry group is ${\mathbb Z}_n$ the invariant is the familiar ${\rm Tr}\rho^n$ used to define the $n$-th Renyi entropy. The tessellation of the sphere by the fundamental regions of the extended replica symmetry group ${\mathbb Z}_6$ is shown in figure \ref{zn-orb-fig}. The replica symmetry group in this case is ${\mathbb Z}_3$. In the same figure we have also shown the singular locus obtained by orbifolding $H^3$ by the replica symmetry group ${\mathbb Z}_n$

\begin{figure}[t]
    \begin{center}
        \includegraphics[scale=0.28]{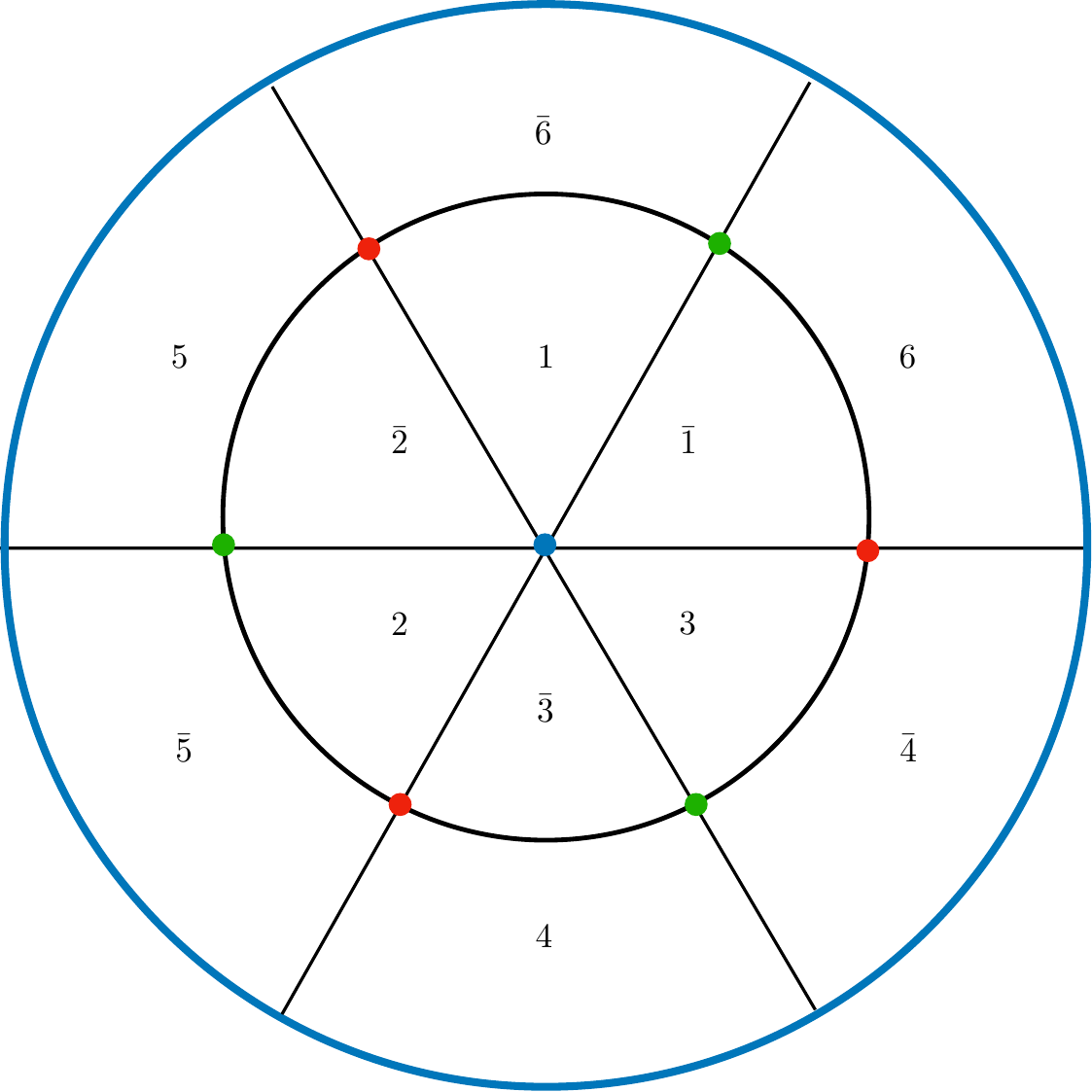}\,\,
        \includegraphics[scale=0.20]{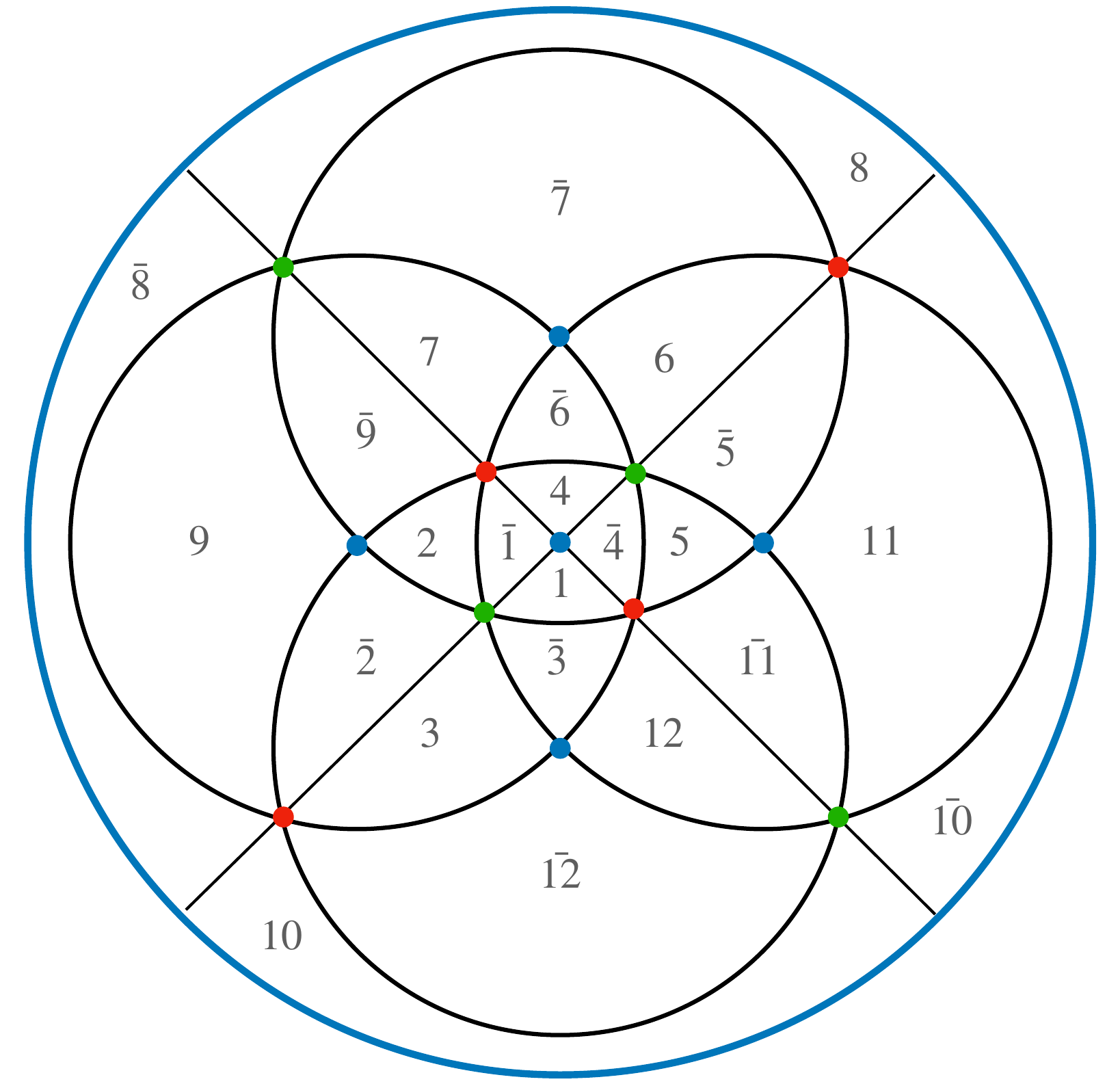}\,\,
        \includegraphics[scale=0.20]{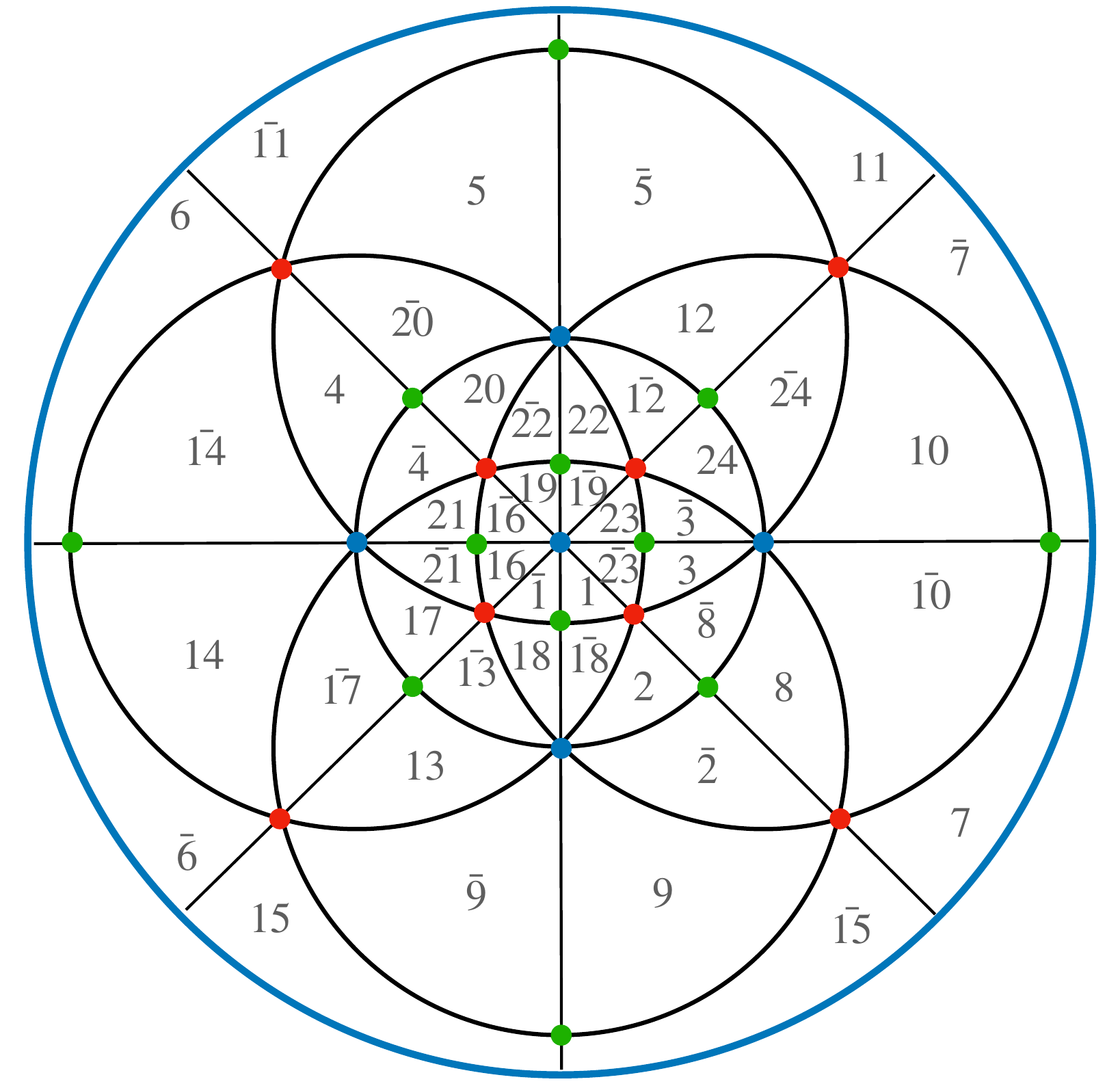}\quad
        \includegraphics[scale=0.26]{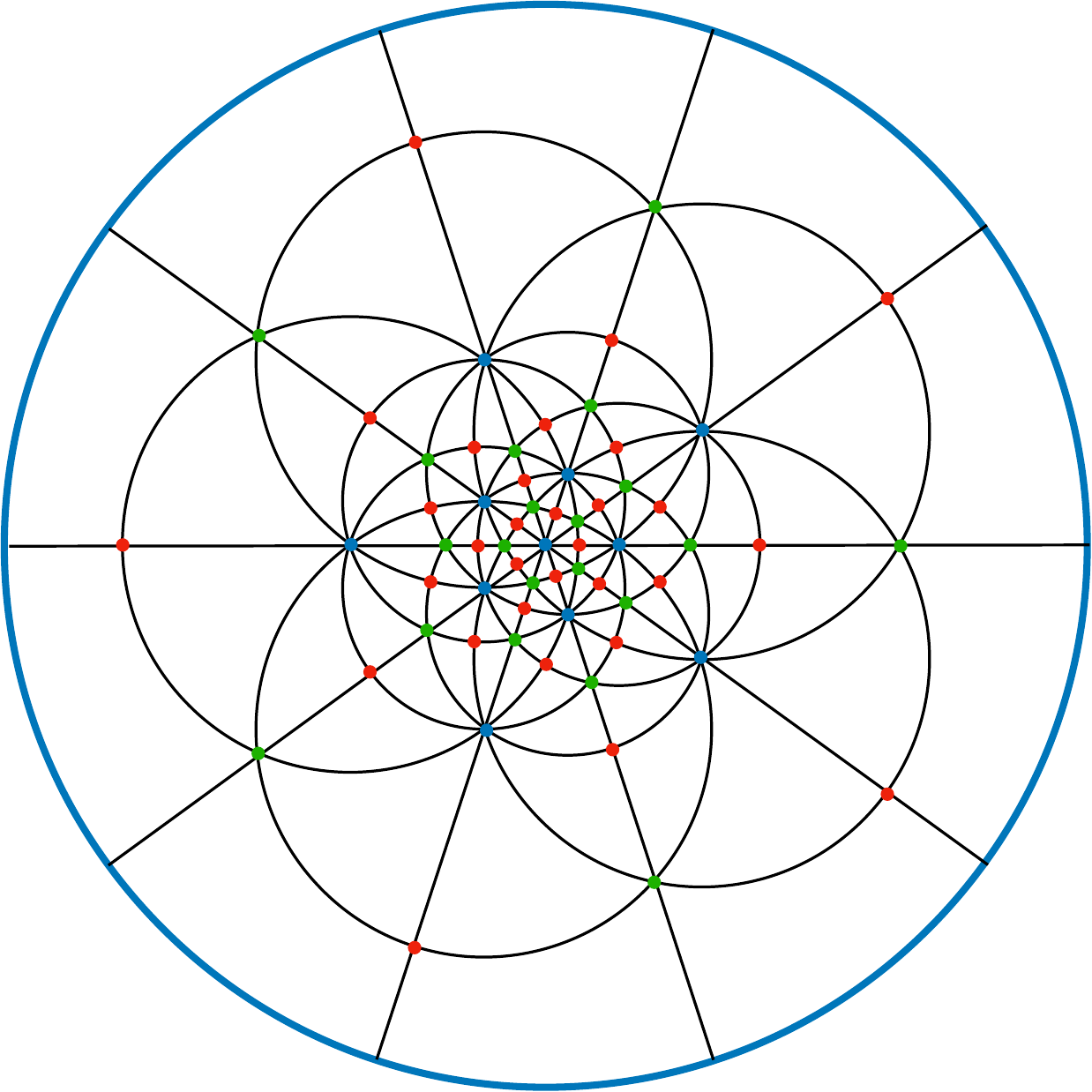}
    \end{center}
    \caption{The first figure shows the stereographic projection of the tesselation of the sphere by triangular fundamental regions of the Coxeter group $(2,2,3)$. The outer blue circle is mapped to the point at infinity. The regions with unbarred labels represent copies of the bra and the regions with barred labels represent copies of  the ket. The second, third and fourth figures represent the action on the sphere by the Coxeter groups $(2,3,3), (2,3,4)$ and $(2,3,5)$ respectively. We have not label the regions in the last figure to avoid clutter.}
    \label{fin-cox-fig}
\end{figure}

\begin{figure}[t]
    \begin{center}
        \includegraphics[scale=0.35]{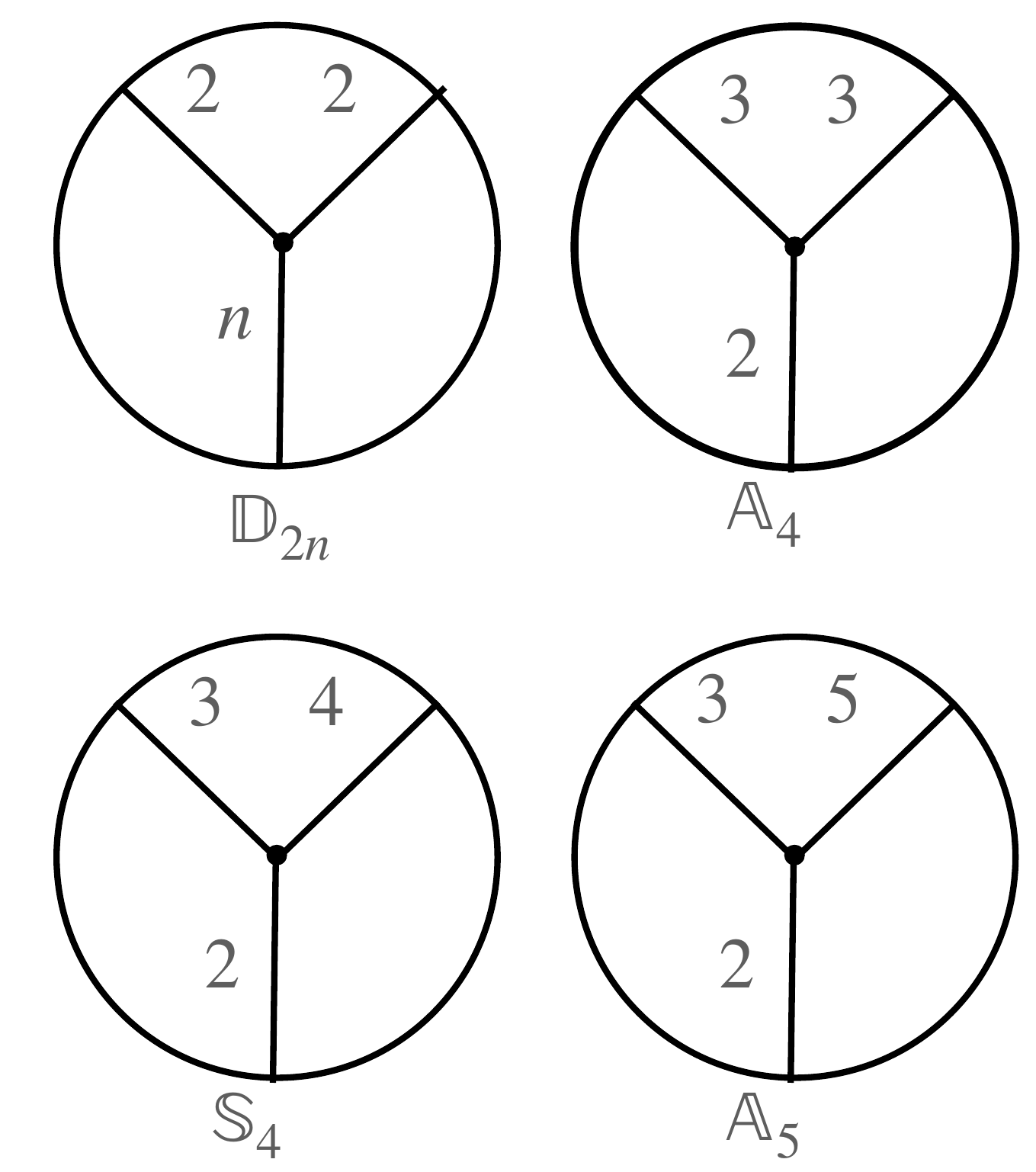}
    \end{center}
    \caption{The singular locus for the tri-partite invariant corresponding to the extended replica symmetry group being the Coxeter groups $(2,2,n), (2,3,3), (2,3,4)$ and $(2,3,5)$ respectively. The label $n$ on the edge indicates that the conical angle around it $2\pi/n$. These solution correspond to tri-valent junctions of heavy cosmic branes with differing tensions. 
    We have also included the names of the replica symmetry group in each of these cases. }
    \label{fin-cox-orb}
\end{figure}

More nontrivial subgroups of  $SO(3)$ have more than two fixed points and are classified by the symmetry groups of platonic solids. If we extend this replica symmetry group by reflections, the resulting groups $\hat \CR$ are easy to characterize (In this case, $\hat \CR=\hat \CK$ because the replicated manifold is a genus zero surface and hence $\CS$ is trivial). They are simply finite Coxeter groups with three generators. Defining $m_{\ta\tb}$ to be the order of the element $r_\ta r_\tb$, these Coxeter groups are given by the following choices of the tuple $(m_{12}, m_{23}, m_{31})$.
\begin{align}\label{spherical-tuples}
    (m_{12}, m_{23}, m_{31}) = \quad (2,2,n), \quad (2,3,3),\quad (2,3,4), \quad (2,3,5).
\end{align}
The groups given by the tuples $(2,3,3), (2,3,4)$ and $(2,3,5)$ are the extended (i.e. including orientation reversal) symmetry groups of a regular tetrahedron, a cube (or a regular octahedron) and a regular dodecahedron (or a regular icosahedron) respectively. The replica symmetry groups for these Coxeter groups i.e. the orientation preserving subgroups are the dihedral group $\mathbb{D}_{2n}$, the alternating group $\mathbb{A}_4$ (group of even permutations of $4$ objects), the symmetric group $\mathbb{S}_4$ and $\mathbb{A}_5$ respectively. The finite Kleinian groups and their extended versions are sometimes called spherical and extended spherical groups respectively. We summarize this discussion in the table below. The manifolds $\CM_{\CE}^{\rm uni}$ and the orbifolds ${\tilde B}_{\CE}$ for these invariants are given in figure \ref{fin-cox-fig}\footnote{For the remainder of the multi-invariants presented we have relegated the specific details regarding the choice of representation for the replica symmetry, group elements for each region, and monodromies of twist operators to appendix \ref{app:reps}.} and figure \ref{fin-cox-orb} respectively. 
\begin{table}[h]\label{tbl:g0mt}
    \centering
    \begin{tabular}{|l|l|l|l|}
    \hline
        Lengths & Group $G$ & Order $|G|=n_r$ & Symmetry \\ \hline
        $(n,n)$ & $\mathbb{Z}_n$ & $n$ & Cyclic \\ \hline
        $(2,2,n)$ & $\mathbb{D}_{2n}$ & $2n$ & Dihedral \\ \hline
        $(2,3,3)$ & $\mathbb{A}_4$ & 12 & Tetrahedral \\ \hline
        $(2,3,4)$ & $\mathbb{S}_4$ & 24 & Octohedral \\ \hline
         $(2,3,5)$ & $\mathbb{A}_5$ & 60 & Icosohedral \\ \hline
    \end{tabular}
    \caption{The finite Kleinian groups}
\end{table}


The boundary of the orbifold ${\tilde B}_{\CE}$ in each case is a sphere and it has three segments of conical singularities with cone angle $2\pi/m_{12}, 2\pi/m_{23}$ and $2\pi/m_{31}$ which meet at a trivalent junction in the bulk. The locus with conical angle $2\pi/m_{\ta\tb}$ originates from the loci in the bulk solution $\CB_{\CE}^{\rm uni}$ that are invariant under the ${\mathbb Z}_{m_{\ta\tb}}$ subgroup. Recall that this subgroup is generated by $r_\ta r_\tb$. We can associate the labels to the segments of conical singularity in the following way. First find the slice of $\tilde \CB_{\CE}$ that is fixed by the ${\mathbb Z}_2$ reflection that maps bra to ket and vice versa. The conical singular loci decompose this slice into chambers in one to one correspondence with the boundary regions. Let us label the chamber which has boundary region $\ta$ as a part of its boundary as $\ta$. 
The singular locus that separates chambers with labels $\ta$ and $\tb$ has the label $m_{\ta\tb}$ and has a cone angle $2\pi/m_{\ta\tb}$ around it.
We compute the associated regularized invariants $\CE^{\rm reg}$ for all the spherical groups by computing the bulk action for the orbifolds in figure \ref{fin-cox-orb} in section \ref{bulk-orb-compute}. 
In each of these cases, the quotient $(\partial H^3)/\hat \CK$ is a spherical triangle with the edges of the triangle fixed by a each of the three reflections respectively. 
This gives rise to a tri-partite invariant for the vacuum state. 
The angles between the two edges $\ta$ and $\tb$ of the triangle is $\pi/m_{m_{\ta\tb}}$.

\subsection{Intermezzo: Cone manifolds}
In the case of ${\mathbb Z}_n$ orbifold, the singular locus consists of a single segment with a cone angle $2\pi/n$ around it. This orbifold geometry computes ${\rm Tr} \rho^n$. To compute the entanglement entropy, one needs to analytically continue the action of these geometries in $n$. What helps in this regard is that the orbifold geometries themselves can be analytically continued in $n$. The geometries thus obtained are known as cone-manifolds in mathematics literature. They have the property that they have a co-dimension $2$ singular locus with a cone angle $2 \pi \alpha$ around it, where $0\leq \alpha\leq 1$. The manifold is hyperbolic everywhere else, even arbitrary close to the conical singularity.

It turns out that the orbifolds $H^3/\hat \CK$ where $\CK$ is any finite Kleinian group also admit such an ``analytic continuation''. That is, there are solutions consisting of three segments of conical singularity with continuous cone angles $2 \pi\alpha, 2 \pi\beta$ and $2 \pi\gamma$, such that $0\leq\alpha,\beta, \gamma \leq 1 $, meeting at a trivalent junction such that the geometry is hyperbolic everywhere else. However, the parameters $\alpha, \beta$ and $\gamma$ need to satisfy a joint condition. Consider a small sphere surrounding the conical junction. It has three conical singularities with cone angles $2 \pi\alpha, 2 \pi\beta$ and $2 \pi\gamma$. We can think of this sphere as being obtained by gluing two identical spherical triangles, one at the top and the other at the bottom. The conical singularities are the three vertices of both of the triangles. The angles of the triangle are $ \pi\alpha,  \pi\beta$ and $ \pi\gamma$. In order for this triangle to be spherical, the sum of its angles must be strictly greater than $\pi$ \cite{cooper2000three}. We have
\begin{align}\label{angle-ineq}
    \alpha+\beta+\gamma > 1.
\end{align}
For the special case of cone-manifolds that are orbifolds, the conical singularity must result from the fixed point locus of the action of certain group element. If the order of the element is $n$, then the cone angle around the singularity is $2\pi/n$. In other words, for orbifolds, the parameters $\alpha, \beta$ and $\gamma$ must take the form $1/n$ for some integer $n$. Indeed the spherical groups discussed above can be obtained from this reasoning. The only solution set to the equation 
\begin{align}
    \frac{1}{m_{12}}+\frac{1}{m_{23}}+\frac{1}{m_{31}} > 1, \qquad {m_{\ta\tb}\in {\mathbb Z}_+}.
\end{align}
are the tuples listed in equation \eqref{spherical-tuples}! 

The point we would like to highlight  is that a cone-manifold exists for arbitrary cone-angle parameters $(\alpha, \beta, \gamma)$ as long as they satisfy the inequality \eqref{angle-ineq}. This has an interesting implications for the discussion of bulk duals of multi-entropy \cite{Gadde:2022cqi} and holographic probe measures \cite{Gadde:2023zzj}. Our discussion shows that the analytic continuations defined \cite{Gadde:2022cqi, Gadde:2023zzj} exist at least for holographic state for small values of the cone angle. In particular, it is valid in the probe limit when the angle of the conical singularity approaches $2\pi$. For the multi-entropy, the parameters $(\alpha, \beta, \gamma)$ are taken equal and $\alpha$ is taken close to $1$ and for general holographic probe measures, $(\alpha, \beta, \gamma)$ all are taken close to $1$ with their ratios kept fixed. This results in ``probe-branes'' with varying tension. Now we return to the discussion of orbifolds.

\subsection{Dihedral group}\label{dihedral1}
Let us discuss in more detail the case where the extended Kleinian group is the Coxeter group given by the tuple $(2,2,n)$. The Kleinian group corresponding to this is the dihedral group $\mathbb{D}_{2n}$. We will pick three anti-conformal maps $r_1, r_2$ and $r_3$ such that the orders of $r_2 r_1$, $r_3 r_2$ and $r_3 r_1$ are $2,2$ and $n$ respectively. It is easy to check that the following anti-conformal maps do the job 
\begin{align}\label{choice-ref}
    r_1(z)=\bar z,  \qquad r_2(z)= \frac{1}{\bar z}, \qquad r_3 (z)= e^{2 \pi i/n} \bar z. 
\end{align}
\begin{figure}[H]
    \begin{center}
        \includegraphics[scale=0.35]{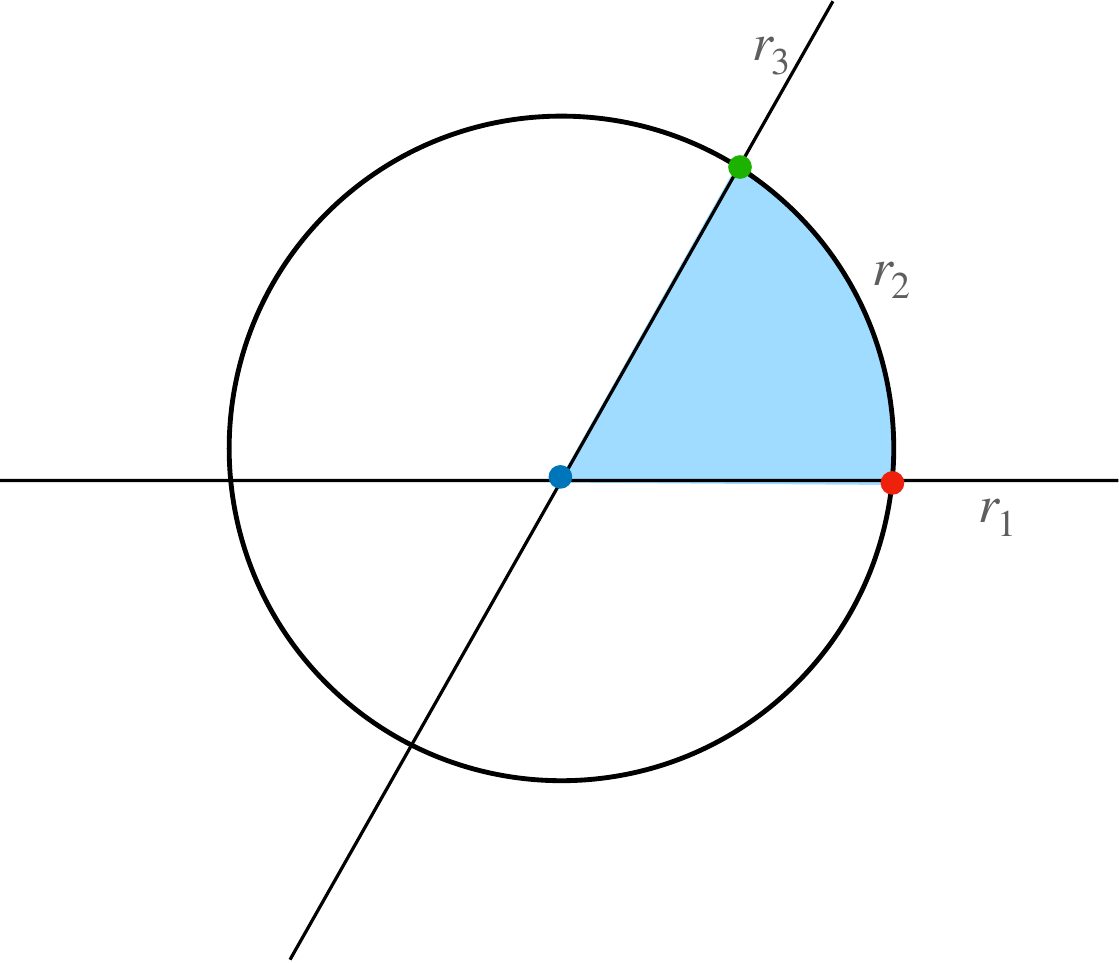}
    \end{center}
    \caption{The fundamental domain of the extended Kleinian group that is a three node Coxeter group with the tuple $(2,2,3)$.}
    \label{dihedral-fig}
\end{figure}
These can be constructed by following considerations. The fundamental domain of the group is a triangle with each side being a circular arc (or a straight line). Each side is fixed by one of the reflections. Let us denote the side fixed by $r_\ta$ as $s_\ta$. The angle between $s_\ta$ and $s_\tb$ is $\pi  /{m_{\ta \tb}}$, where $m_{\ta \tb}$ is the order of $g_\ta g_\tb$. For the case at hand, we would like to construct a triangle with angles $\pi/2, \pi/2$ and $\pi/{n}$ respectively. A convenient choice of such a triangle is shown in figure \ref{dihedral-fig}. 

The reflection element $r_\ta$ associated to the side $s_\ta$ of the triangle are constructed by conjugating $r_x: z\to \bar z$ by a conformal transformation $A_\ta$ that maps $x$-axis to the circle whose arc is $s_\ta$. The side $s_1$ is a part of the $x$-axis itself, so $A_1(z)=z$. The side $s_3$ is part of the $x$-axis rotated by $\pi/n$, so $A_3(z)=e^{\pi i /z}$. The side $s_3$ is an arc of the unit circle. The conformal transformation that maps $x$-axis to unit circle is $A_2(z)=(z-i)/(z+i)$. Conjugating $r_x$ by these conformal maps, we obtain the reflections in equation \eqref{choice-ref}. The extended Kleinian group is then given by
\be
\hat\CK=\langle r_1^2=r^2_2=r_3^2=e,\;(r_2r_1)^2=(r_3r_2)^2=(r_3r_1)^n=e\rangle
\ee
which is shown in figure \ref{fig:dCT}:

\begin{figure}[h]
    \begin{center}
        \includegraphics[scale=0.25]{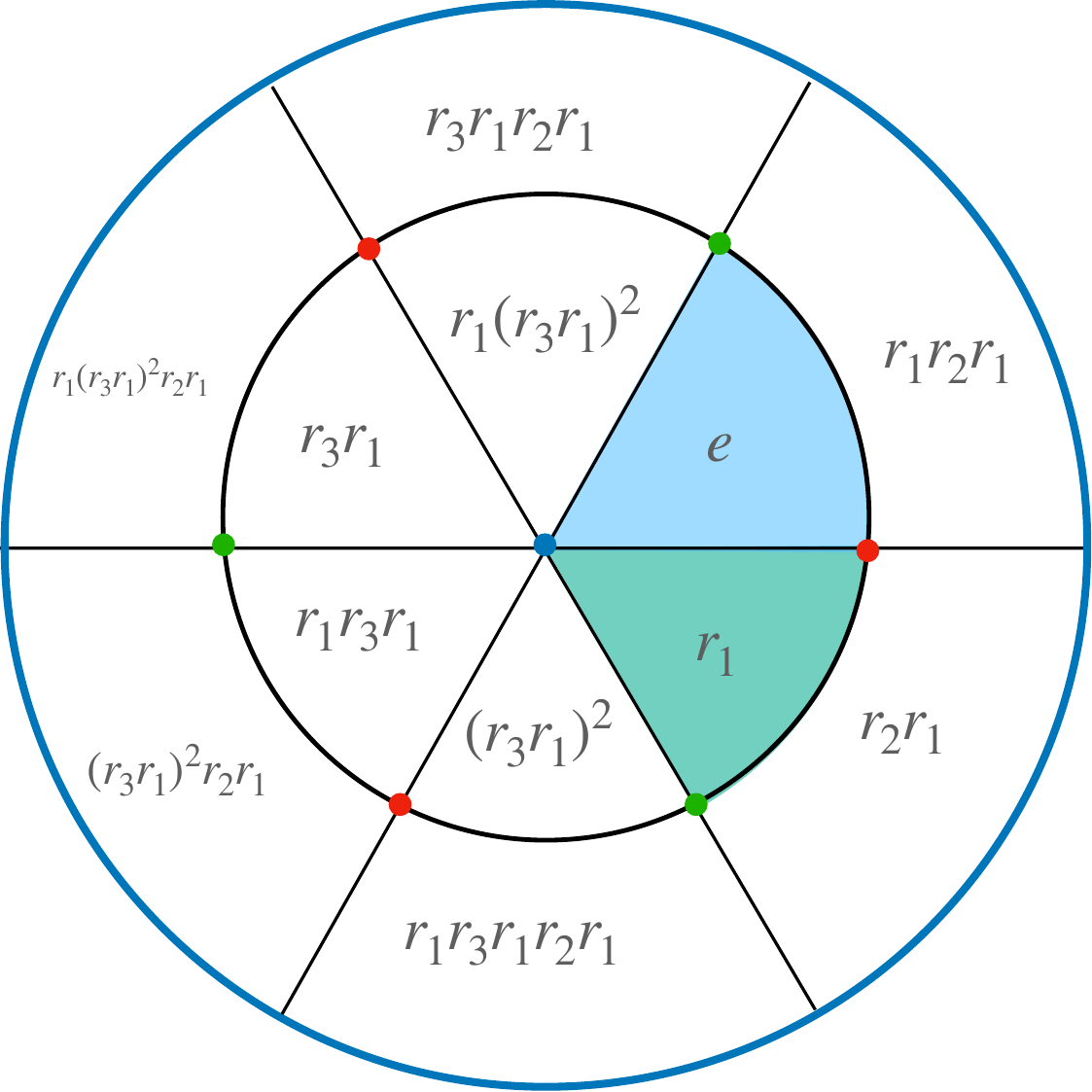}\quad 
        \includegraphics[scale=0.25]{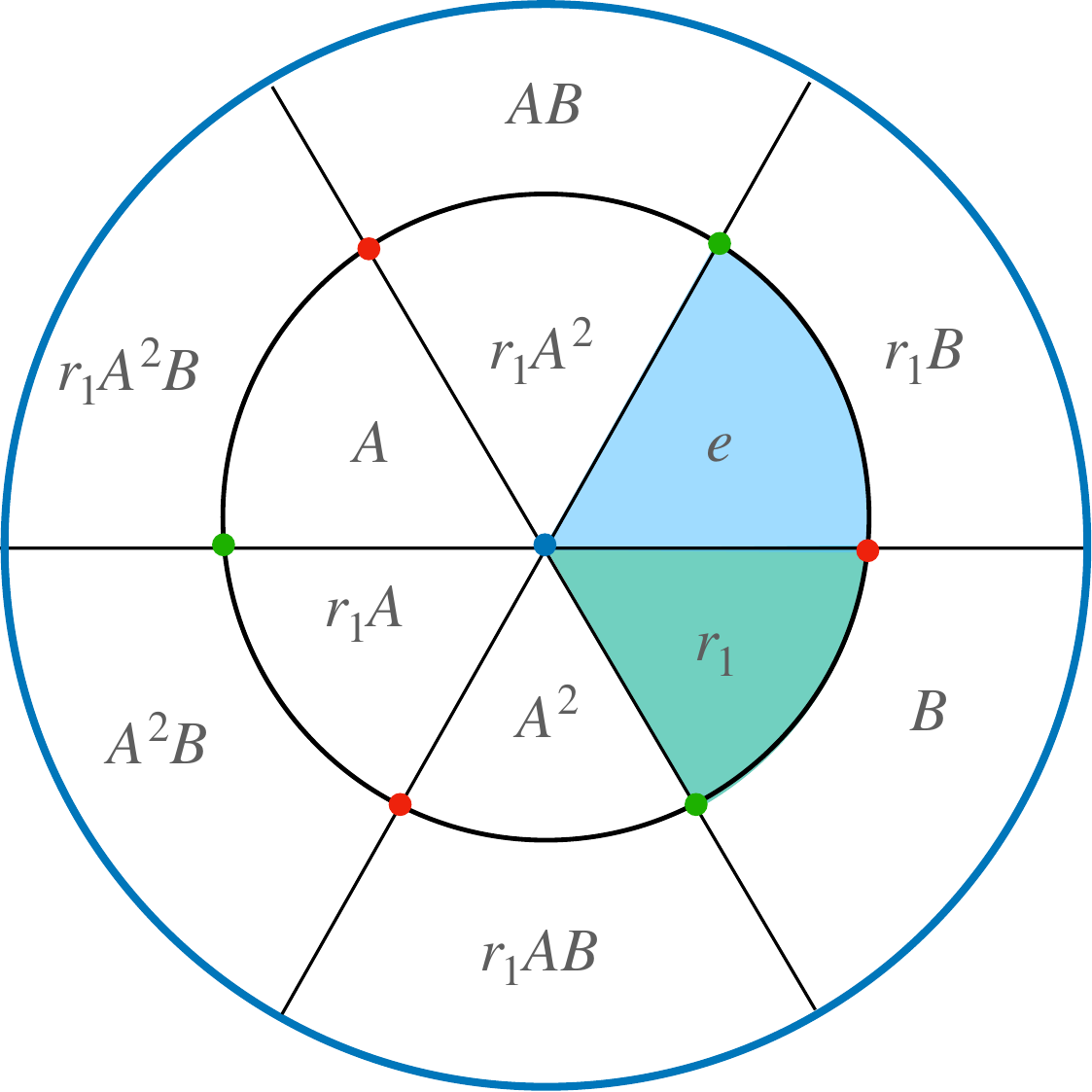}\quad
         \includegraphics[scale=0.25]{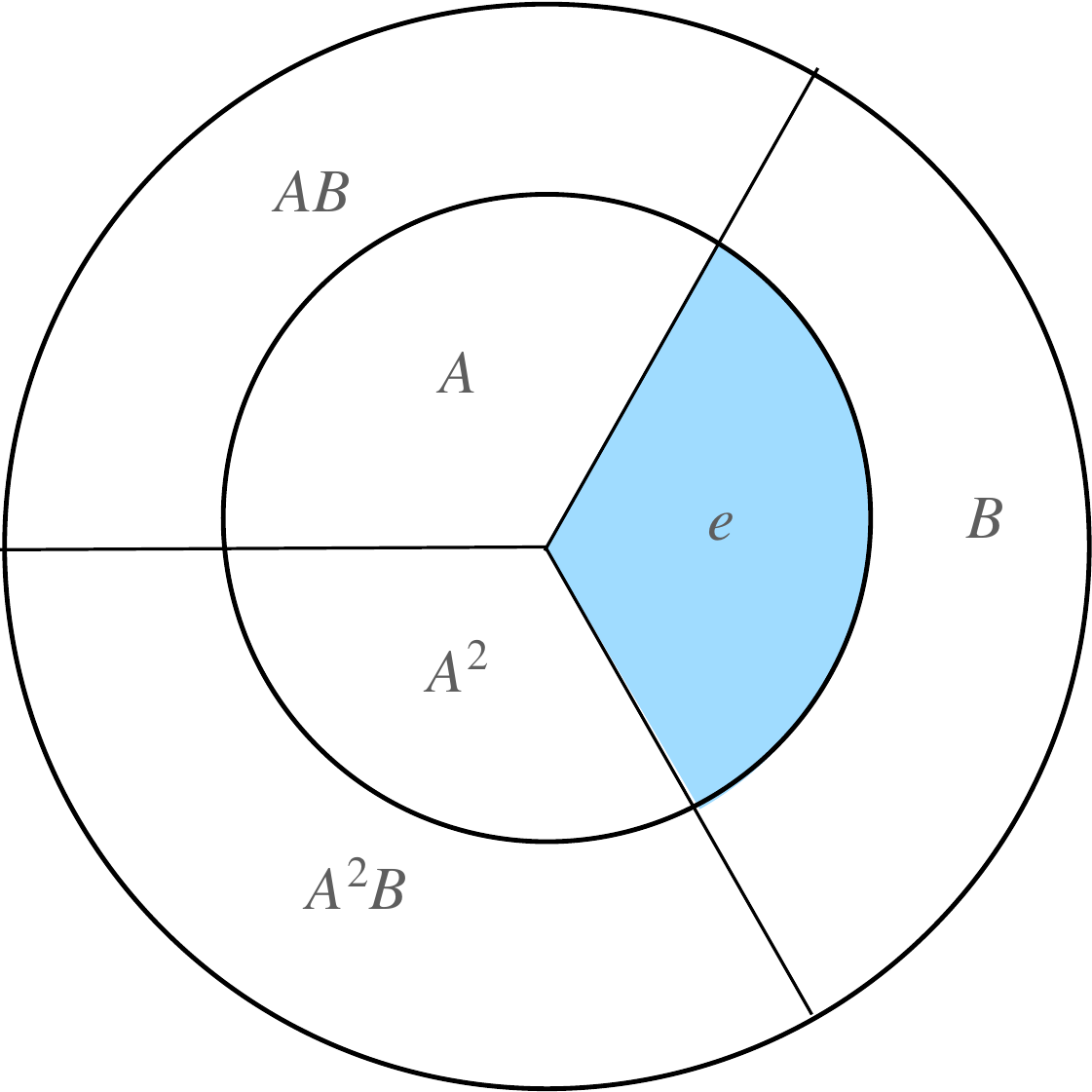}
    \end{center}
    \caption{Left: The extended Kleinian group $\hat\CK$ isomorphic to $\mathbb{D}_6\times\mathbb{Z}_2$. By acting on the fundamental region with the reflections $r_1,r_2,r_3$ the Riemann sphere is tessellated by (2,2,3) triangles. The outer circle is the point at infinity. Each of these triangles is assigned a unique group element which maps the fundamental region to it. Note that the labels we have chosen are always written in terms of products of $r_1,r_2r_1,r_3r_1$ this should be compared with \ref{eq:coxrsig}. The vertices of the triangles where the regions meet correspond to the elliptical fixed points of the elements of  $\hat\CK$. The different colors signify those fixed points which can be mapped into one another. In particular each color is the set of preimages of a twist operator under the action of the covering map. Middle: Defining $A=r_3r_1$ and $B=r_2r_1$  $\hat\CK$ can be written in terms of the conformal isometries $A,B$ and the reflection $r_1$. Note that $r_1$ only appears as the final (left most) map in the string. Right: The union of the regions $e$ and $r_1$ forms the fundamental region of the Kleinian group $\CK=\langle A^3=B^2=(AB)^2=e\rangle$ isomorphic to $\mathbb{D}_6$. The group consist only of conformal isometries and should be compared with \ref{eq:coxalternating}.}
    \label{fig:dCT}
\end{figure}

\subsection{Amalgamation}\label{amalgam-sec}
Thanks to the combination theorems of Klein and its generalization due to Maskit \cite{Maskit1987-ee}, two Kleinian groups can be ``combined'' to generate another Kleinian group\footnote{The groups we interested in are a particular subset of Kleinian groups called \emph{function groups} which support a non-constant automorphic function. All function groups can be generated by a finite number of applications of the combination theorems to the class of \emph{basic groups} \cite{Maskit1987-ee}. For completeness these include elementary groups (which include the finite groups), quasi-fuchsian groups and totally degenerate groups. The important point here is that the groups we are interested in can \emph{always} be constructed using the combination theorems applied to finite groups.}. For our purposes, we will be mainly using the generalization due to Maskit to generate virtually free Kleinian groups by combining multiple finite Kleinian groups.  Let us first state the combination theorem due to Klein which relatively easy to understand.
\begin{theorem}{(Klein's combination theorem)}
Let $\CK_1$ and $\CK_2$ be finitely generated Kleinian groups, and let $D_1$ and $D_2$ be their fundamental domains on $S^2$, respectively. Assume that the interior of $D_1$ contains the boundary and exterior of $D_2$ and vice versa. Then $\CK=\CK_1 * \CK_2$ i.e. the group generated by $\CK_1$ and $\CK_2$, is a Kleinian group with the fundamental domain $D = D_1 \cap D2$.
\end{theorem}
The Schottky quotient can be understood from the point of view this theorem. If we have only a single pair of circles, then the Kleinian group associated with it, $\CK_1$ is generated by a single element freely. The fundamental domain for it, $D_1$ is the exterior of the two circles. Consider another pair of circles such that their interiors are disjoint from the interiors of the circles in the first pair. Define $\CK_2$ and $D_2$ correspondingly. It is clear the conditions for the Klein's theorem are satisfied and $\CK_1,\CK_2$ together generate a free group on $2$ generators with the fundamental domain being $D_1\cap D_2$. This can be repeated $g$ times to get the genus $g$ surface. For our purpose, it is crucial to work with the extension of the theorem due to Maskit about the combination of Kleinian groups to produce virtually free Kleinian groups.

\begin{theorem}{(Maskit's combination theorem)}
    Let $\CK_1$ and $\CK_2$ be Kleinian groups and $J$ a cyclic subgroup of $\CK_1,\CK_2$ such that $J\neq \CK_1,\CK_2$. Let $D_1$ and $D_2$ be the fundamental regions of $\CK_1$ and $\CK_2$. If we can choose a $\gamma$, a simple closed curve dividing $\hat{\mathbb{C}}$ into two topological discs $B_1$ and $B_2$ such that: 1) $B_1$ and $B_2$ are invariant under $\CJ$, 2) $\gamma \cap D_1 = \gamma \cap D_2$, 3) $\CK_1-\CJ$ maps $B_1$ into $B_2$ and 4) $\CK_2-\CJ$ maps $B_2$ into $B_1$. Then $\CK=\CK_1*_\CJ \CK_2$ is a Kleinian group with fundamental region $D=(D_1\cap B_2)\cup (D_2\cap B_1)$.
\end{theorem}
Here the symbol $\CK_1*_\CJ \CK_2$ is referred to as the amalgamation of the groups $\CK_1$ and $\CK_2$ along the their common subgroup. It is defined as the quotient of the free product $\CK_1* \CK_2$ by the common subgroup $J$. If we take $\CK_1$ and $\CK_2$ to be finite Coxeter groups such that they both have a common abelian subgroup then the Maskit's theorem gives us $\CK=\CK_1*_\CJ \CK_2$ which is virtually free. This is because of the following result which relates virtually free groups and the amalgamation of Coxeter groups:
\begin{theorem}[\cite{coxetersurfacegroups}]\label{thm:vfcox}
Let $G$ be a Coxeter group. Then following are equivalent:
\begin{enumerate}
    \item $G$ is virtually free.
    \item $G$ is the amalgamation of any finite number of finite Coxeter groups over finite subgroups.
    \item $G$ does not contain a surface group (the fundamental group of genus $g\geq1$ surface).
\end{enumerate}
\end{theorem}
This follows as finite groups are virtually free and the amalgamation of any virtually free group over a finite subgroup is also virtually free \cite{ScottWall1979}.

It will be convenient for us to work with the fundamental domains of extended Kleinian groups $\hat \CK_1$ and $\hat \CK_2$ and obtain the fundamental domain of extended version of the amalgamation $\CK=\CK_1*_\CJ \CK_2$.
The fundamental domains of extended Kleinian groups are easy to characterize because they are bounded by loci that are fixed under reflections.

If a Kleinian group $\CK'$ that is obtained from $\CK$ by conjugation of a conformal isometry $A$ i.e. $\CK'=A\cdot \CK\cdot A^{-1}$, then $D'=A\cdot D$ where $D$ and $D'$ are fundamental domains of $\CK$ and $\CK'$ respectively. Maskit's recombination theorem allows for conjugation of one of the Kleinian groups, say $\CK_1$, relative to the other by a conformal isometry. As the fundamental domain $D$ of the amalgam $\CK$ is obtained by overlapping $D_1$ and $D_2$, conjugation of $\CK_1$ relative to $\CK_2$ produces a new overlap $D'$ that is not conformally related to the old one $D$. In this way, we can produce a family of inequivalent amalgams. After orbifolding, they produce a geometrically inequivalent family of bulk geometries. Their boundaries are conformally inequivalent. In other words, conjugation of $\CK_1$ relative to $\CK_2$ amounts to changing conformal moduli of the boundary. When the boundary is a sphere with twist operator insertions, the relative conjugation changes their conformal cross-ratios. 

In the following section we will give some concrete examples of amalgamation of finite Kleinian groups discussed in section \ref{fin-klein} to produce virtually free Kleinian groups\footnote{Our geometric constructions follow closely \cite{Maskit1987-ee, hid_explict_vschot, hid_2223}.}. We will also see how to identify free normal subgroups that are finite index. Before we move on, we give the formula for characteristic of $\CK$ that is constructed as an amalgam of two finite Kleinian groups $\CK_1, \CK_2$ along the common subgroup $\CJ$.
\begin{align}\label{two-amalgam}
    \chi(\CK)=\frac{1}{|\CK_1|}+\frac{1}{|\CK_2|}-\frac{1}{|\CJ|}.
\end{align}
Here $|G|$ stands for the order of $G$. In section \ref{general-tree}, we will give a general formula for the characteristic resulting from multiple amalgamations. 

\subsubsection{Amalgamation of two dihedral groups}\label{dndn-amalgam}
\begin{figure}[h]
    \begin{center}
        \includegraphics[scale=0.35]{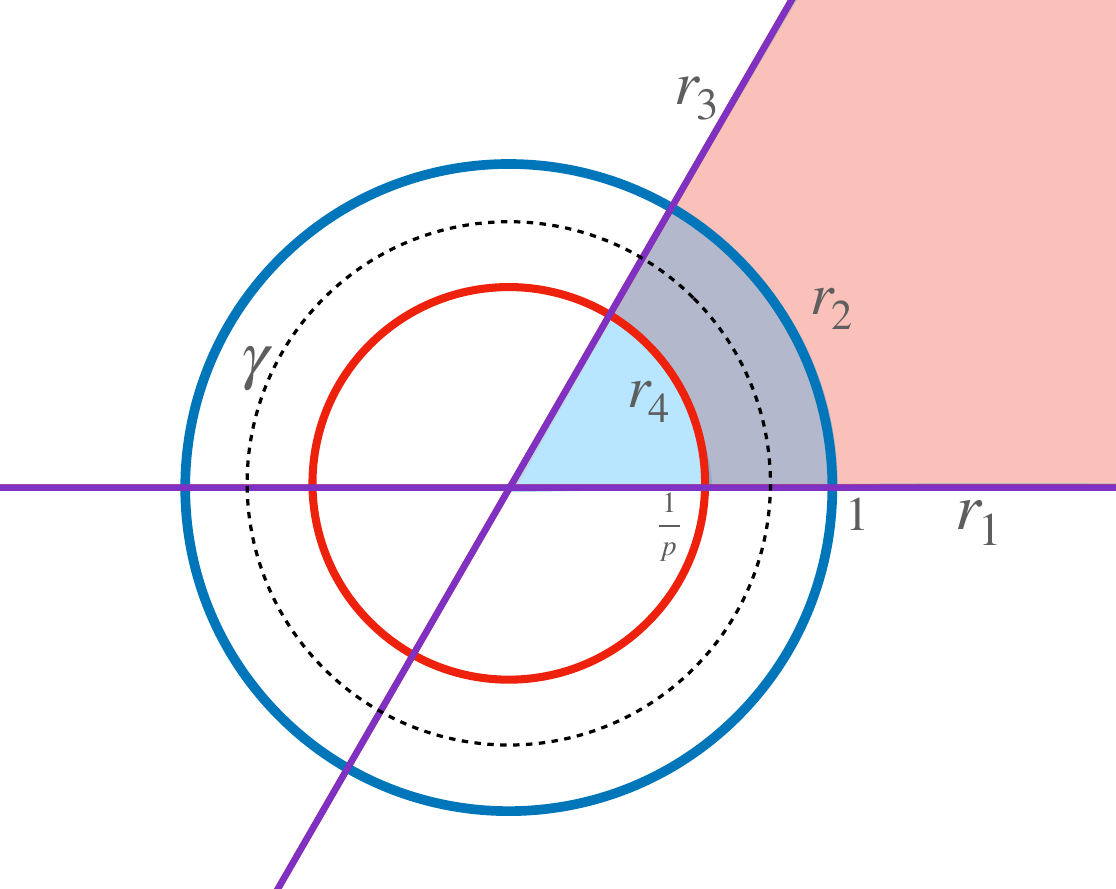}
        \includegraphics[scale=0.45]{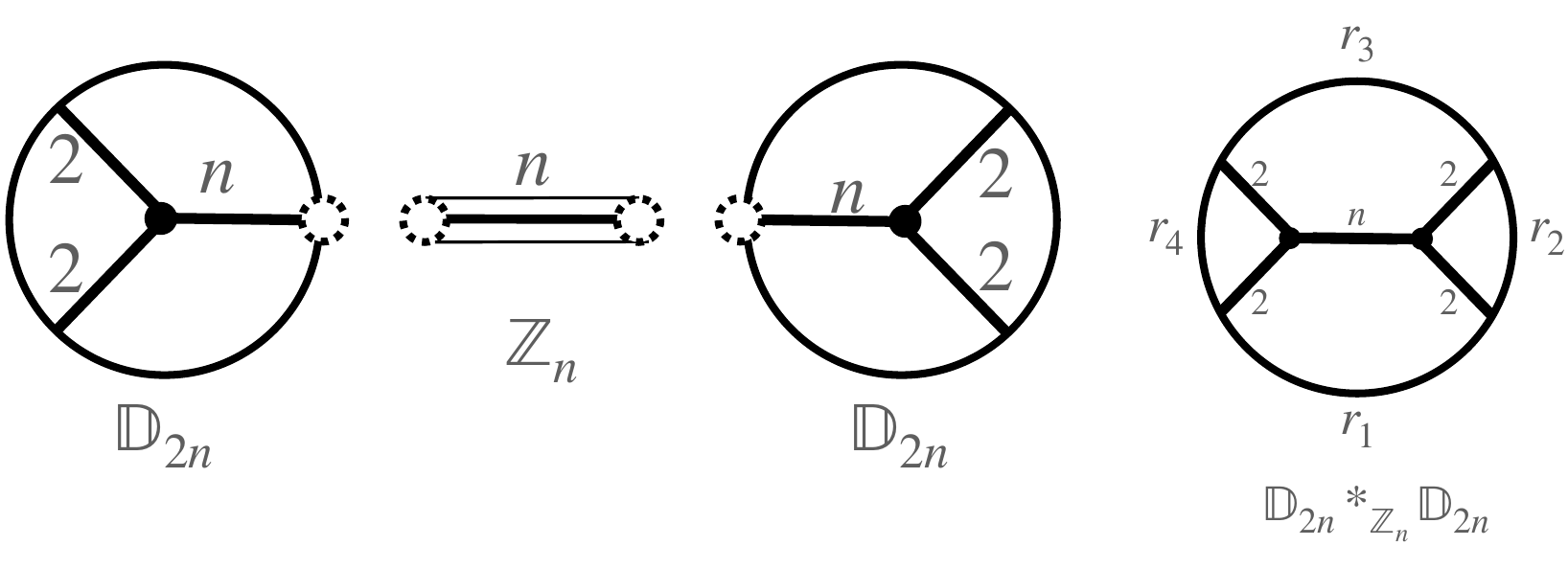}
    \end{center}
    \caption{
    In the first figure, we have indicated the fundamental regions for the action of two dihedral groups. For each side, we have indicated the reflection that it is fixed under. We have also denoted the curve $\gamma$ appearing in Maskit's theorem. In the second figure we have shown how the amalgamation procedure translates for the orbifold geometry $H^3/\CK$. We show how to ``glue'' the singular loci of the two dihedral actions to produce the singular locus of the Kleinian group $\mathbb{D}_{2n} *_{{\mathbb Z}_n} \mathbb{D}_{2n}$. The reflections associated to each party have been marked in the bottom right figure. They satisfy the relations given in \eqref{ext-klein-rel1}.}
    \label{ddamalgam-fig}
\end{figure}

In order to amalgamate two dihedral groups, it is useful to construct fundamental regions for both that have a non-trivial overlap. We keep one of the fundamental regions same as before but for the other, we change $r_2$ so that the fixed locus is not the unit circle, but rather a circle with radius $1/p$ centered at the origin with $p>1$. The fundamental regions of both groups are as shown in figure \ref{ddamalgam-fig}. The curve $\gamma$ is also shown in the figure. It is easy to see that it is invariant under ${\mathbb Z}_n$ and hence are the disks $B_1$ and $B_2$ bounded by $\gamma$. Condition $2$ of Maskit's theorem is obvious. We can see that conditions $3$ and $4$ also hold. This gives rise to the new Kleinian group isomorphic to $\mathbb{D}_{2n} *_{{\mathbb Z}_n} \mathbb{D}_{2n}$ whose fundamental regions is the quadrilateral shown in figure \ref{ddamalgam-fig}. Its quotient by ${\mathbb Z}_2$ i.e. the fundamental region of $(\partial H^3)/{\hat K}$ is also a quadrilateral with all angles being $\pi/2$. This shows that the amalgamation of two dihedral group produces an invariant with four party regions. 
The reflection generators corresponding to the four sides can be readily found. Two sides of the quadrilateral fundamental region are common with  the two sides of each of the triangular fundamental regions of $\hat \CK_1$ and $\hat \CK_2$ and each of the other two is shared with that of the triangular fundamental regions of $\hat \CK_1$ and $\hat \CK_2$ separately. If the reflections for $\hat \CK_1$ and $\hat \CK_2$ are $r^{(1)}_{i}$ and $r^{(2)}_{i}$ with $i=1,2,3$ respectively then it follows that the reflections for $\hat \CK_1 *_J \hat \CK_2$ are $r_1=r^{(1)}_1=r^{(2)}_1$, $r_3=r^{(1)}_3=r^{(2)}_3$, $r_2= r^{(1)}_2$ and $r_4=r^{(2)}_2$. These reflections obey the relations,
\begin{align}\label{ext-klein-rel1}
    (r_2 r_1)^{2}=1,\quad (r_3 r_1)^{n}=1, \quad (r_4 r_1)^2= 1, \quad (r_2 r_3)^{2}=1, \quad (r_3 r_4)^{2}=1,
\end{align}
all of which follow from the relations obeyed by the reflections of $\hat \CK_1$ and $\hat \CK_2$. The only product of generators that has no relation  is $r_2 r_4$. This is the group that is obtained by amalgamating two dihedral groups along the common cyclic subgroup ${\mathbb Z}_n$. Because the order of $r_2 r_4$ is infinite, the resulting Kleinian group $\hat \CK$ is infinite. See figure \ref{ddCT-fig}:
\begin{figure}[h]
    \begin{center}
        \includegraphics[scale=0.35]{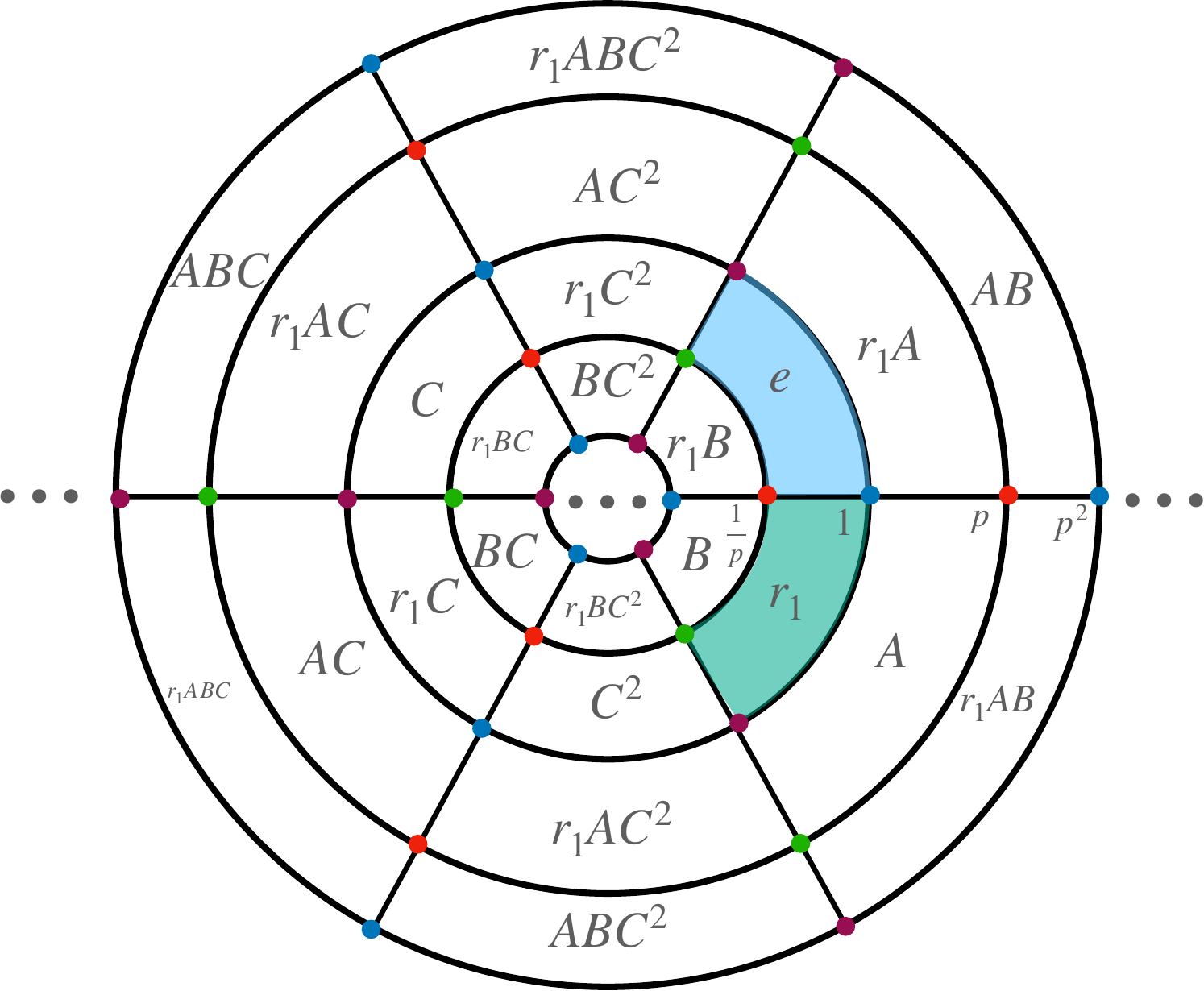}
    \end{center}
    \caption{The extended complex plane tessellated with (2,2,2,2) quadrangles according to the extended Kleinian group generated by $r_1,A=r_2r_1,B=r_4r_1,C=r_3r_1$ for the case $n=3$. Unlike the previous examples this group is infinite and there are an infinite number of rings of $2n$ copies of the extended fundamental region in to the center and out to infinity. This is because the group contains the hyperbolic element $AB$ whose hyperbolic fixed points are $0,\infty$ and which generates the maximal \emph{virtual Schottky group} $\CS=\langle AB\rangle$ of lowest index.}
    \label{ddCT-fig}
\end{figure}

Amalgamation of Kleinian groups can also be understood from the point of view of the singular loci in $H^3/\CK$. The singular loci in the case of the dihedral group are shown in the figure \ref{fin-cox-orb}. They consist of three segments labeled by $(2,2,n)$. The segments have  conical angles $(\pi, \pi, 2\pi/n)$. The singularity segment labeled by $n$ is fixed under the ${\mathbb Z}_n$ subgroup. As we are amalgamating $\mathbb{D}_{2n}$ and $\mathbb{D}_{2n}$ along the common ${\mathbb Z}_n$ subgroup, we take the two copies of tri-valent singular loci of $\mathbb{D}_{2n}$ and glue them along the segment that is fixed under ${\mathbb Z}_n$. This is described in the second sub-figure of figure \ref{ddamalgam-fig}. This gives rise to an ``s-channel'' like singularity structure where the external segments are labeled by $2$ and the internal one by $n$. 
This is expected because, the internal segment separates chambers $1$ and $3$ hence has a cone angle $2\pi/m_{13}$ and $m_{13}=n$. 

According to the formula \eqref{two-amalgam}, the characteristic of this Kleinian group is $0$. Using the equation \eqref{free-index}, this naively implies that the number of replicas is zero. The way out of this paradox is to note that the free normal subgroups of this group only have rank $1$ so the number of replicas is not fixed.

\subsubsection*{The Schottky subgroup}
\begin{figure}[t]
    \begin{center}
        \includegraphics[scale=0.18]{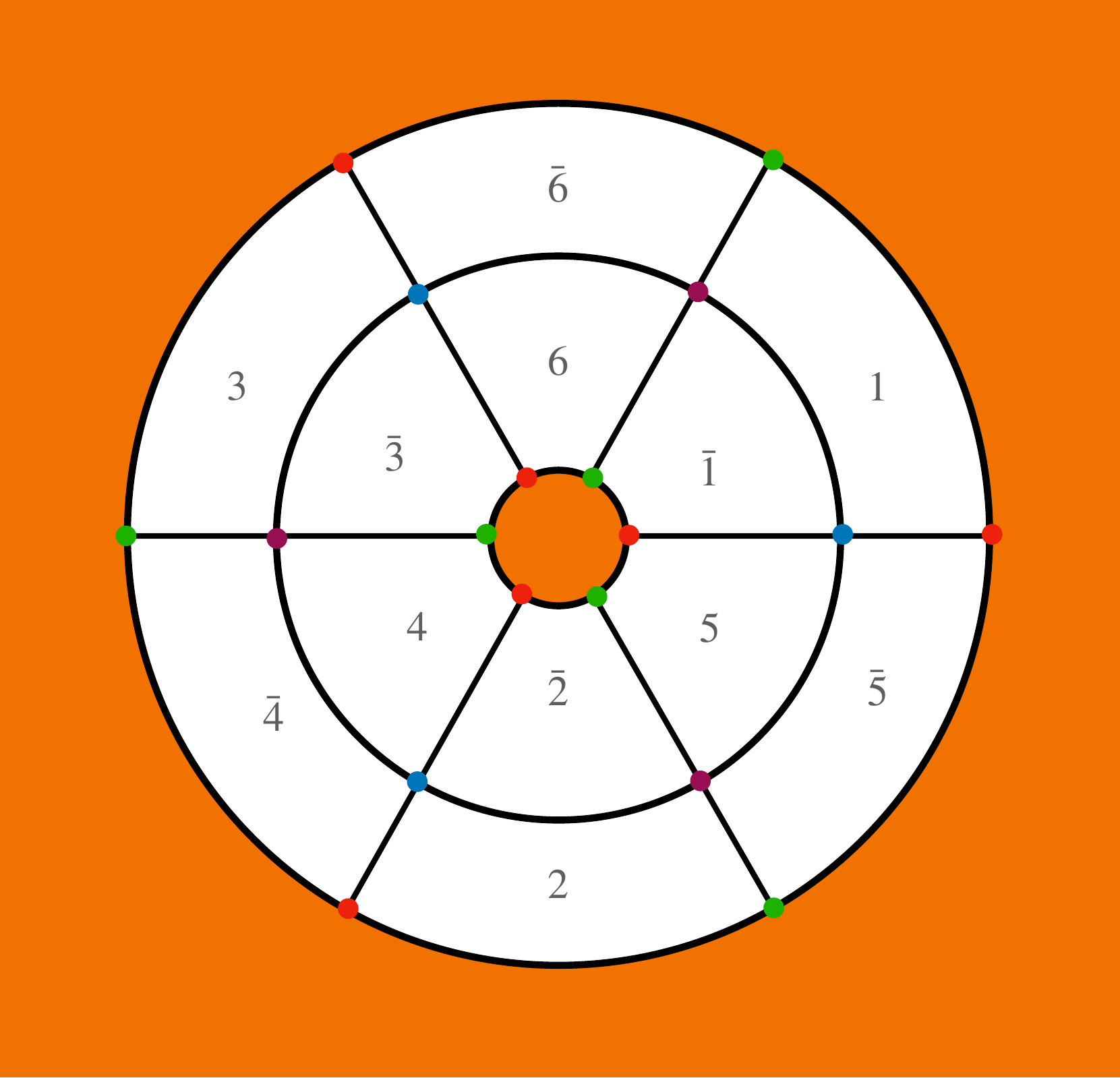}
        \includegraphics[scale=0.18]{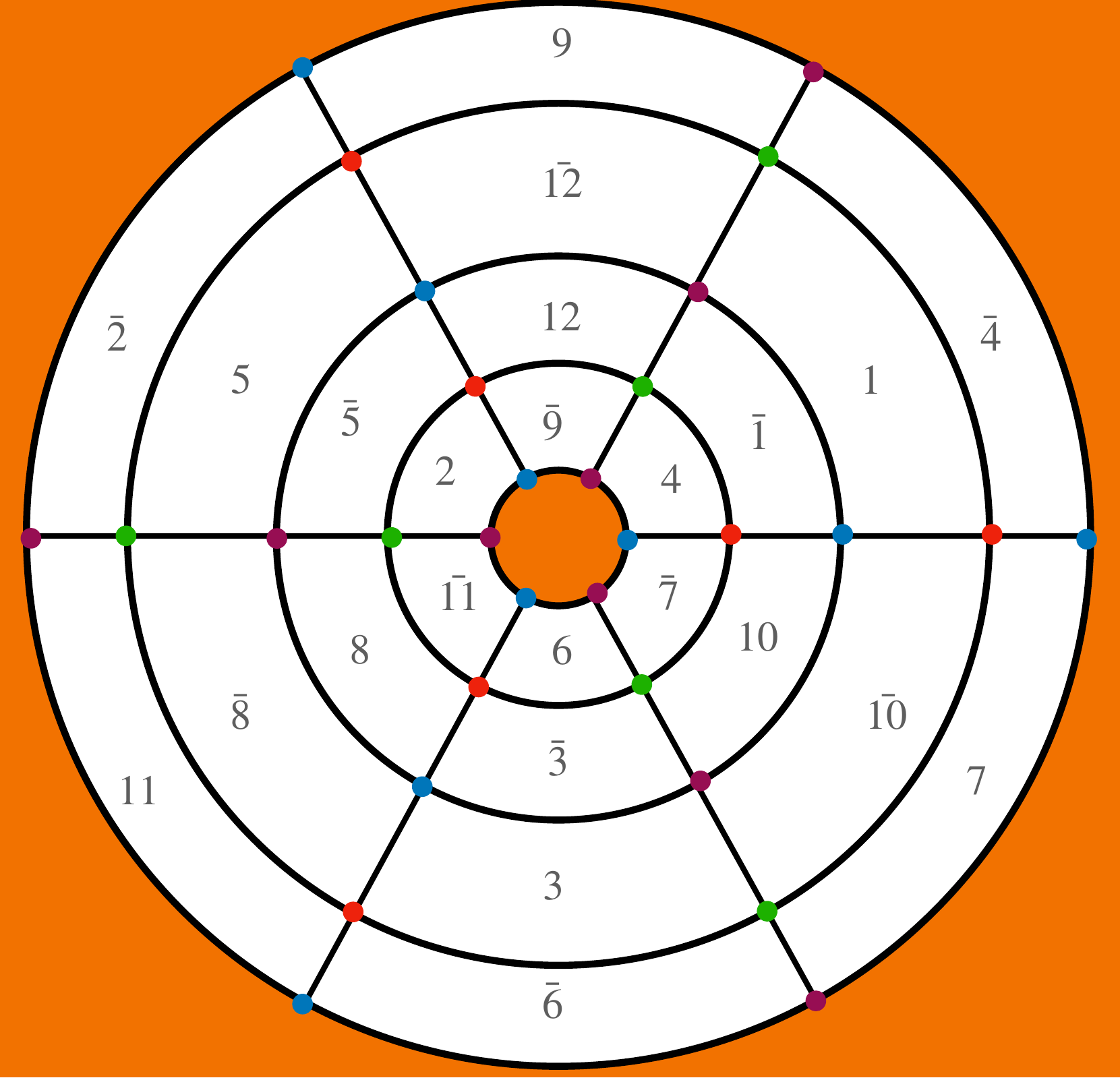}\\
        \includegraphics[scale=0.18]{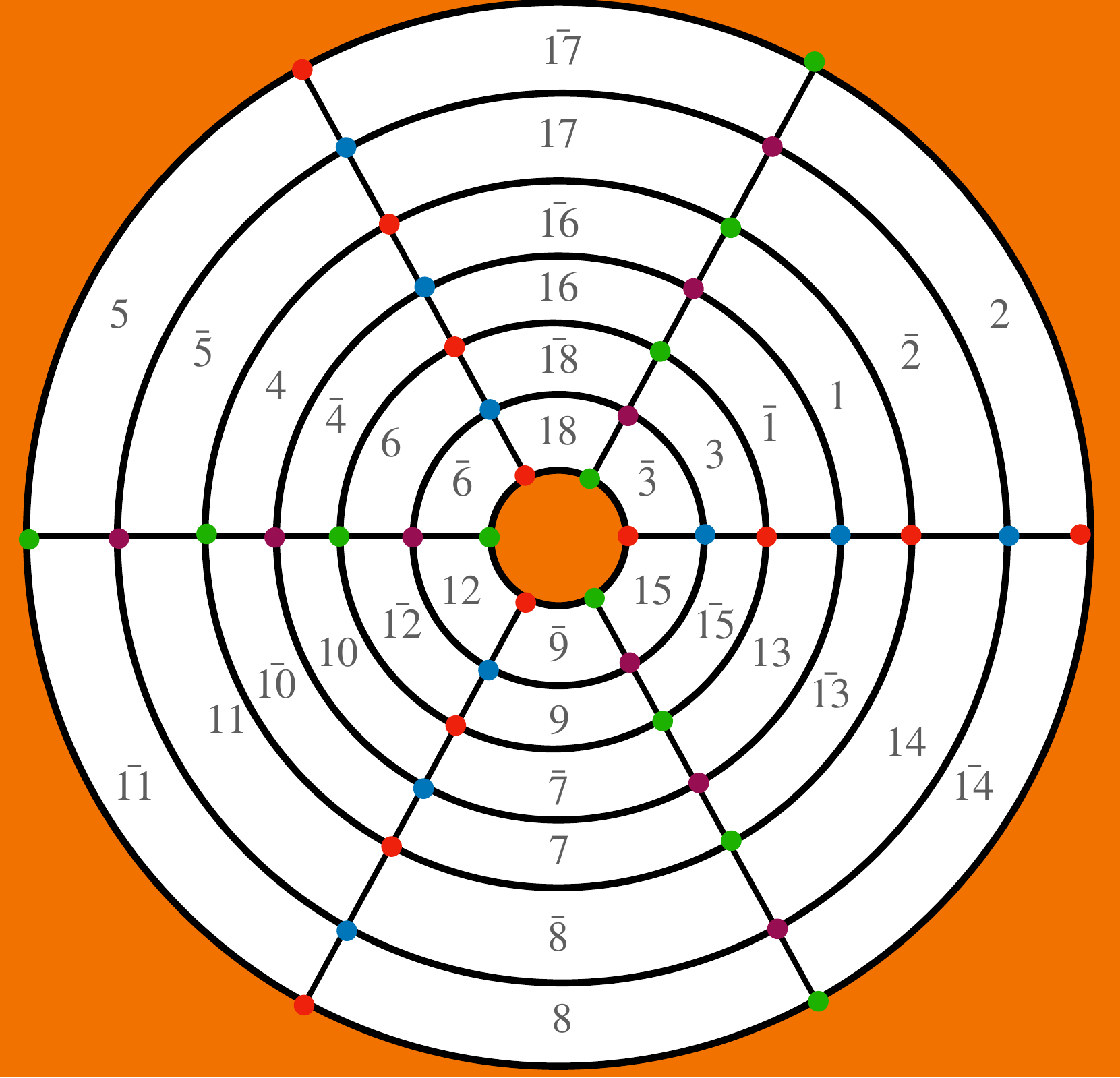}
        \includegraphics[scale=0.18]{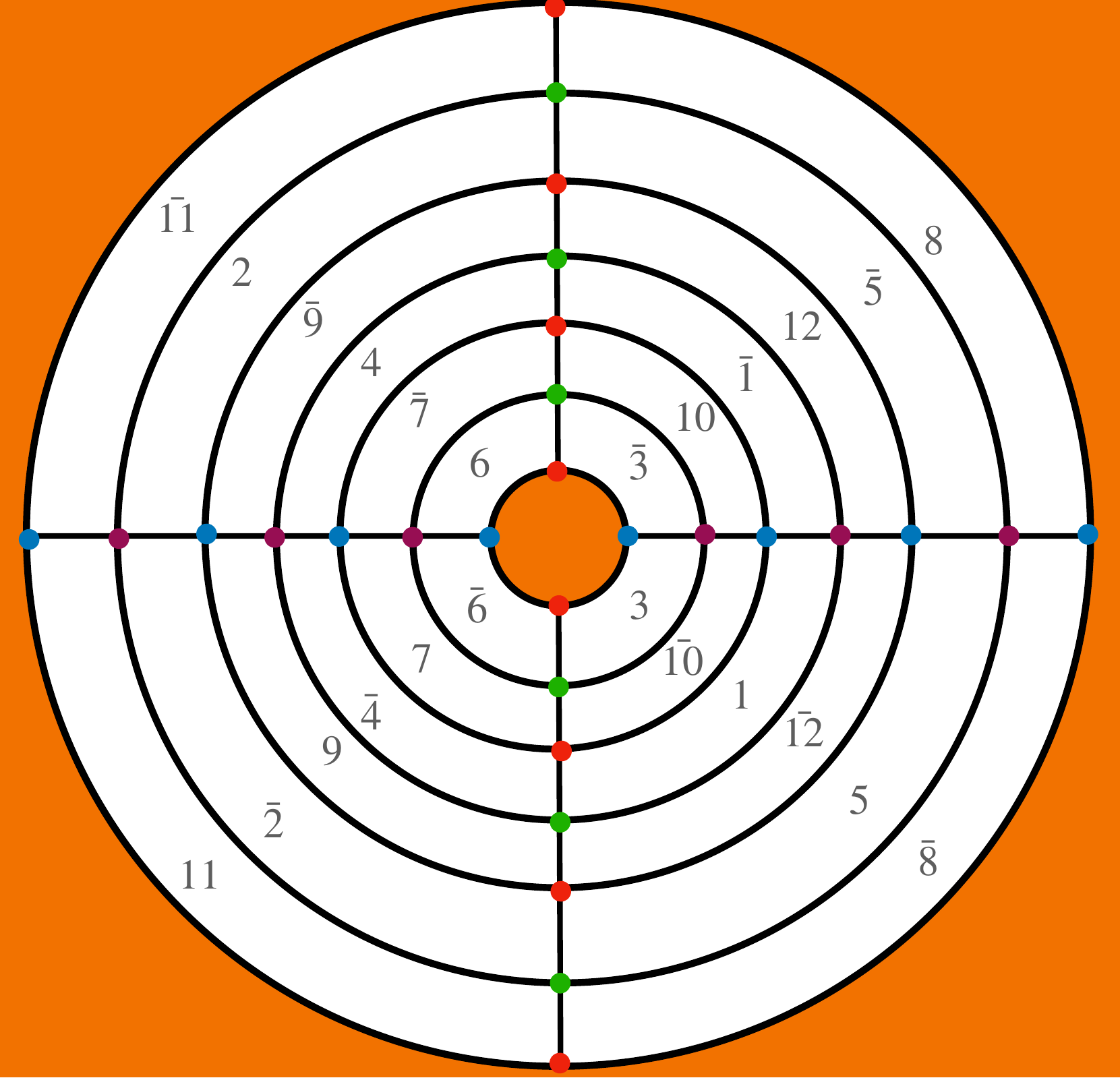}
    \end{center}
    \caption{\
    The two disconnected orange regions in these figures are the interiors of the two discs involved in the Schottky quotient. The remaining part is the fundamental domain of the Schottky quotient $H^3/{\mathbb Z}_m$. We have shown its tessellation  by the quadrilateral fundamental domains of $\mathbb{D}_{2n} *_{{\mathbb Z}_n} \mathbb{D}_{2n}$. The values of $(n,m)$ in these figures are $(3,1), (3,2), (3,3), (2, 3)$ respectively. As the extended replica symmetry group  are unchanged  under the exchange of $n$ and $m$ up to permutation of parties, the second and the fourth figure represent the same invariant up to permutation of parties. The dominate solution is controlled by the parameter $p$.}
    \label{schottky-quotient1}
\end{figure}

We can perform a quotient by a free normal subgroup $\CS$ by imposing a finite order, say $m$ on $r_2 r_4$. This corresponds to choosing the generator of $\CS$ to be $(r_2 r_4)^{m}$. The fundamental region for the action of this Schottky group is shown in figure \ref{schottky-quotient1}. As ${\hat \CK}/\CS$ acts on it, the fundamental region is tessellated by the quadrilateral fundamental region of $\hat \CK$. 
The rank of $\CS$ is $1$ in this case. Hence the genus of the quotient handlebody $H^3/\CS$ is $1$.
This relation yields a finite Coxeter group whose Dynkin diagram consists of two disconnected diagrams, one with two vertices connected by an edge with label $n$ and another with two vertices connected by an edge with label $m$. 

The genus of ${\CM}_{\CE}$ can also be computed from the cycle structure of the permutations  $\sigma_{12}, \sigma_{23}, \sigma_{34}$ and $\sigma_{41}$ associated with  the four twist operators using formula \eqref{riem-hur}. 
Recall that $\sigma_{\ta\tb}=r_\ta r_\tb$ \footnote{Please note the distinction between the reflections $r_a$ which generate the extended replica symmetry group and the generators $r_a$ used to construct the multi-invariants. See the discussion around Figure \ref{left-right}.}. Hence, in this case, $k_{\sigma_{\ta\tb}}=m_{\ta\tb}=2$ for all four $\sigma$'s.
\begin{align}
    \chi_{\CM_\CE}=n_r\Big(2-4\Big(1-\frac12\Big)\Big)=0
\end{align}  
The Euler characteristic $\chi_{\CM_\CE}$ does not depend on the number of replica. The number of replicas in the invariant changes if we change  the order of $r_2 r_4$, namely $m$. We saw earlier that the fact that the genus of the handlebody is genus $1$ depends only on the fact that the rank of the free normal subgroup is $1$ and not on the choice of the generator of this free subgroup.  
Another way to understand the genus is to note that the fundamental domain is a quadrilateral with all four angles $\pi/2$. This corresponds to a flat (euclidean) quadrilateral as the angles sum to exactly $2\pi$. The only compact Riemann surface that is flat is of genus $1$. Again, this conclusion rests only on the fundamental domain of $\hat \CK$ and not the choice of $\CS$. This happens only in the euclidean case. For hyperbolic boundaries  the genus of the quotient does depend on the number of replicas as we will see shortly.  

We can see the consequence of this quotient of $\CK$ by the Schottky group on the orbifold geometry. 
If we take the boundary regions $1,3$ to be smaller than $2,4$ then the dominant gravity solution changes. In terms of the Kleinian group this accomplished by adjusting the parameter $p$. Instead of filling the cycle that is invariant under $r_1 r_3$, it now fills in the cycle that is invariant under $r_2 r_4$. As a result, the singular locus also undergoes a phase transition from ``s-channel'' to ``t-channel''. As the internal segment is now between the chambers $2$ and $4$, its cone angle is $\pi/m_{24}$ and hence has the label $m_{24}=m$. All in all, we get a singular locus that is ``t-channel'' type. The external segments have the label $2$ as before, but the internal segment now has the label $m$.

\subsubsection{$\mathbb{D}_{4} *_{{\mathbb Z}_2} \mathbb{D}_{6}$}\label{ss:d4d6}
\begin{figure}[h]
    \begin{center}
        \includegraphics[scale=0.35]{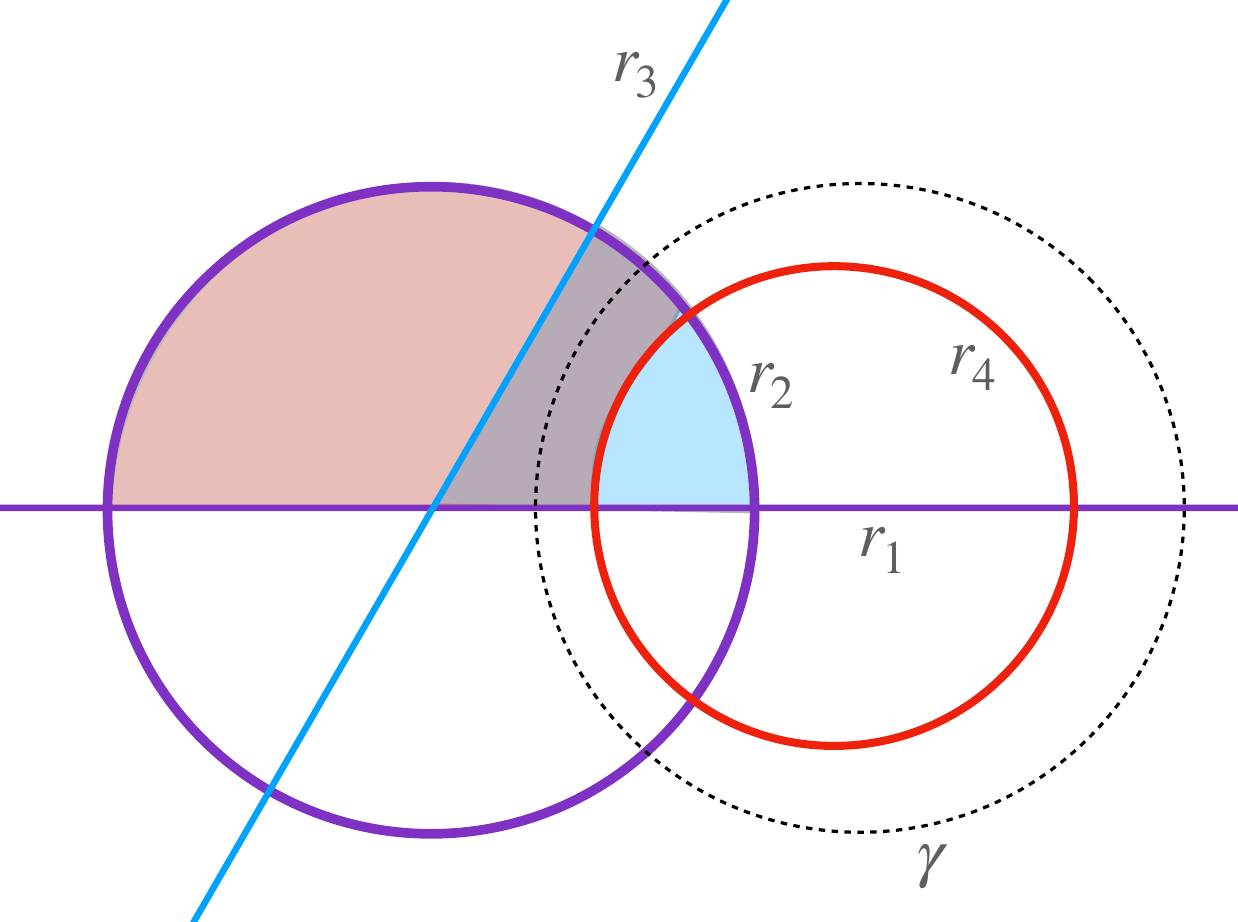}
        \includegraphics[scale=0.45]{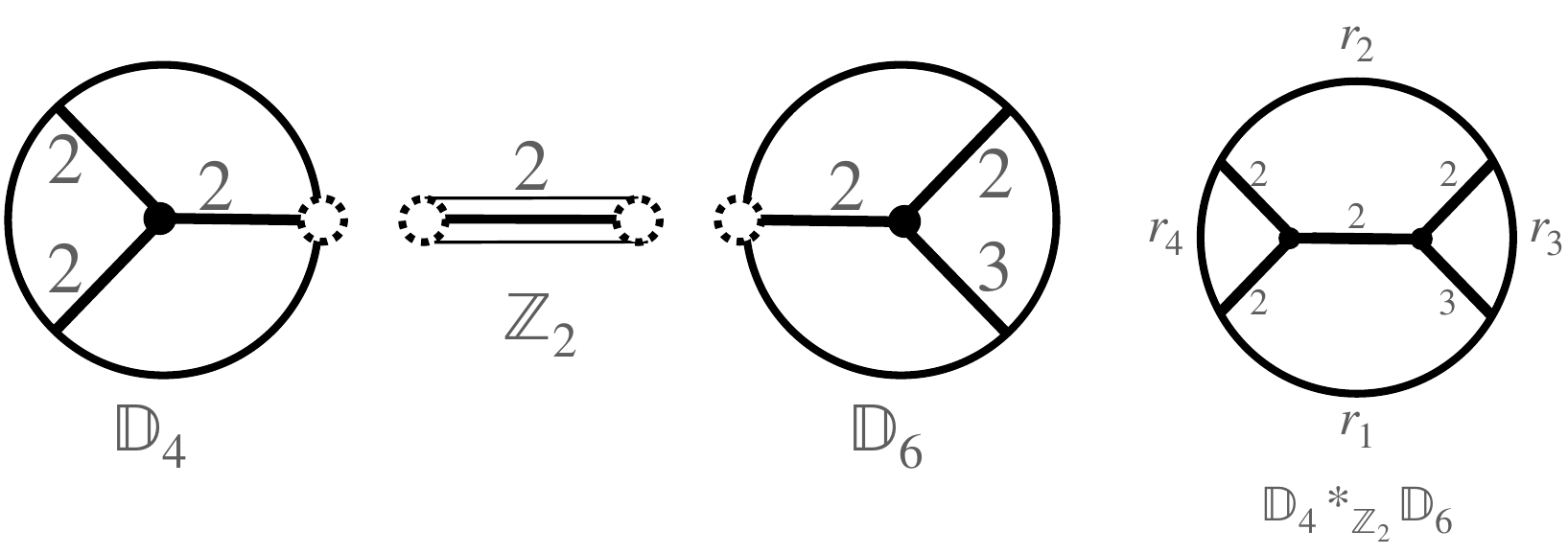}
    \end{center}
    \caption{
        In the first figure, we have indicated the fundamental regions for the action of two dihedral groups. For each side, we have indicated the reflection that it is fixed under. We have also denoted the curve $\gamma$ appearing in Maskit's theorem. In the bottom left figure we have shown how the amalgamation procedure translates for the orbifold geometry $H^3/\CK$. We show how to ``glue'' the singular loci of the two dihedral actions to produce the singular locus of the Kleinian group $\mathbb{D}_{4} *_{{\mathbb Z}_2} \mathbb{D}_{6}$ (bottom right figure). The reflections associated to each party have been marked in the bottom right figure. They satisfy the relations given in \eqref{hatk-gens-1}. }
    \label{d4d6amalgam-fig}
\end{figure}

\begin{figure}[h]
    \begin{center}
        \includegraphics[scale=0.5]{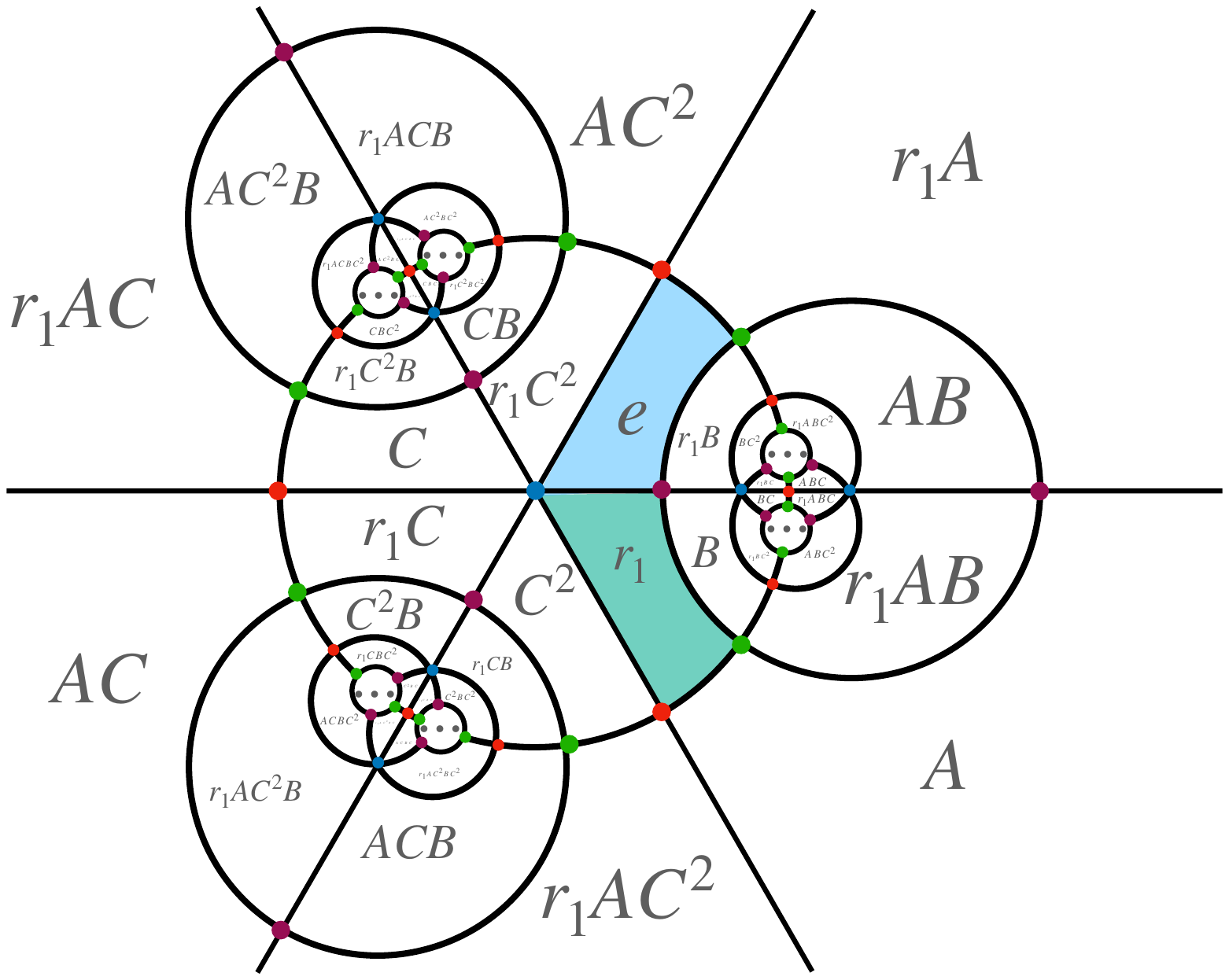}
    \end{center}
    \caption{The extended Kleinian group isomorphic to $\mathbb{D}_{4} *_{{\mathbb Z}_2} \mathbb{D}_{6}$. Here $A=r_2r_1,B=r_4r_1$ and $C=r_3r_1$. We have only shown the first few iterations of performing reflections on the fundamental region, but the pattern continues down infinitely as indicated by the ellipses in the six circles. The group contains an infinite number of hyperbolic fixed points all of which are on the unit circle contained within three circles: the circle of fixed points of $r_4$ and its two image under conjugation by $C$.).}
    \label{d4d6K}
\end{figure}

In this section, we will describe the amalgamation of dihedral group $\mathbb{D}_{4}$ and $\mathbb{D}_{6}$ but along the common ${\mathbb Z}_2$. The Coxeter tuples for  extended replica symmetry groups are $(2,2,2)$ and $(2,2,3)$. As before, the orbifold of $H^3$ by these groups have three conically singular segments of labels $(2,2,2)$ and $(2,2,3)$ respectively, meeting in the bulk in a trivalent junction. This is shown in the second figure in figure \ref{d4d6amalgam-fig}. The amalgamation along the common ${\mathbb Z}_2$ corresponds to gluing of these singular loci along the segment labeled $2$. The result is an ``s-channel'' like conical locus with the external labels $2,2,2$ and $3$ and the internal label $2$. This case is not covered in the previous examples because the external segments do not all have label $2$. 
The fundamental region of $\mathbb{D}_{4}$ and $\mathbb{D}_{6}$, along with the curve $\gamma$ is displayed in figure \ref{d4d6amalgam-fig}. It is easy to check that this choice satisfies all the conditions of Maskit's recombination theorem. The fundamental domain of the resulting Kleinian group which is isomorphic to $\mathbb{D}_{4} *_{{\mathbb Z}_2} \mathbb{D}_{6}$\footnote{This group is also known as the ``extended modular group" and it and its normal subgroups are particularly well studied see \cite{emodsub, emodsubii},  also connections with the automorphisms of Riemann surfaces \cite{May_1977}.} is again a quadrilateral as denoted in the figure \ref{d4d6amalgam-fig}. The angles of quadrilateral are $\pi/2,\pi/2, \pi/2, \pi/3$. As they add up to less than $2\pi$, the resulting surface $\partial (H^3)/{\hat \CK}$ is hyperbolic and hence has genus $\geq 2$. The figure also labels the four reflection generators of $\hat \CK$ given by
\be
r_1(z)=\bar{z},  \quad r_2(z)=\frac{1}{\bar{z}}, \quad r_3(z)=e^{\frac{2\pi i}{3}}\bar{z}, \quad r_4(z)=\frac{p\bar{z}-1}{\bar{z}-p}, \quad1\leq p \leq2
\ee
they obey
\begin{align}\label{hatk-gens-1}
    (r_2 r_1)^2=(r_3 r_1)^3 = (r_4 r_1)^2= (r_2 r_3)^2=(r_2 r_4)^2 =1.
\end{align}
The resulting extended Kleinian group is shown in figure \ref{d4d6K}.

The generator $(r_3 r_4)$ has infinite order and hence $\hat \CK$ is an infinite group. Using the formula \eqref{two-amalgam} for the characteristic of Kleinian groups resulting from amalgamation of finite groups, we see that the characteristic $\chi$ for this Kleinian group is $-1/12$. 

As in the previous example the parameter $p$ controls the shape of the fundamental region. The resulting Kleinian groups will be isomorphic, but not conjugate. For different values of $p$ the resulting handlebodies will have different inequivalent moduli which defines a one real dimension curve on the full $3g-g$ complex dimension moduli space. This is ultimately related to the cross ratio of the four twist operators.

\subsubsection*{The Schottky subgroup}
After quotienting by the Schottky group $\CS$, we get the finite group $\hat \CK/\CS$. In particular, while quotienting by $\CS$, we must impose a finite order, say $m$ on $r_3 r_4$. This condition is not sufficient to get a finite quotient but is definitely necessary. If we only impose this condition then the quotient group is described as being generated by $4$ generators $r_i$ obeying equation \eqref{hatk-gens-1} along with $(r_3 r_4)^m=1$. If $m=2,3,4,5$, this group is a finite Coxeter group. The Dynkin diagrams for these groups is given the figure \ref{dynk-d4d6} below. 
\begin{figure}[h]
    \begin{center}
        \includegraphics[scale=0.25]{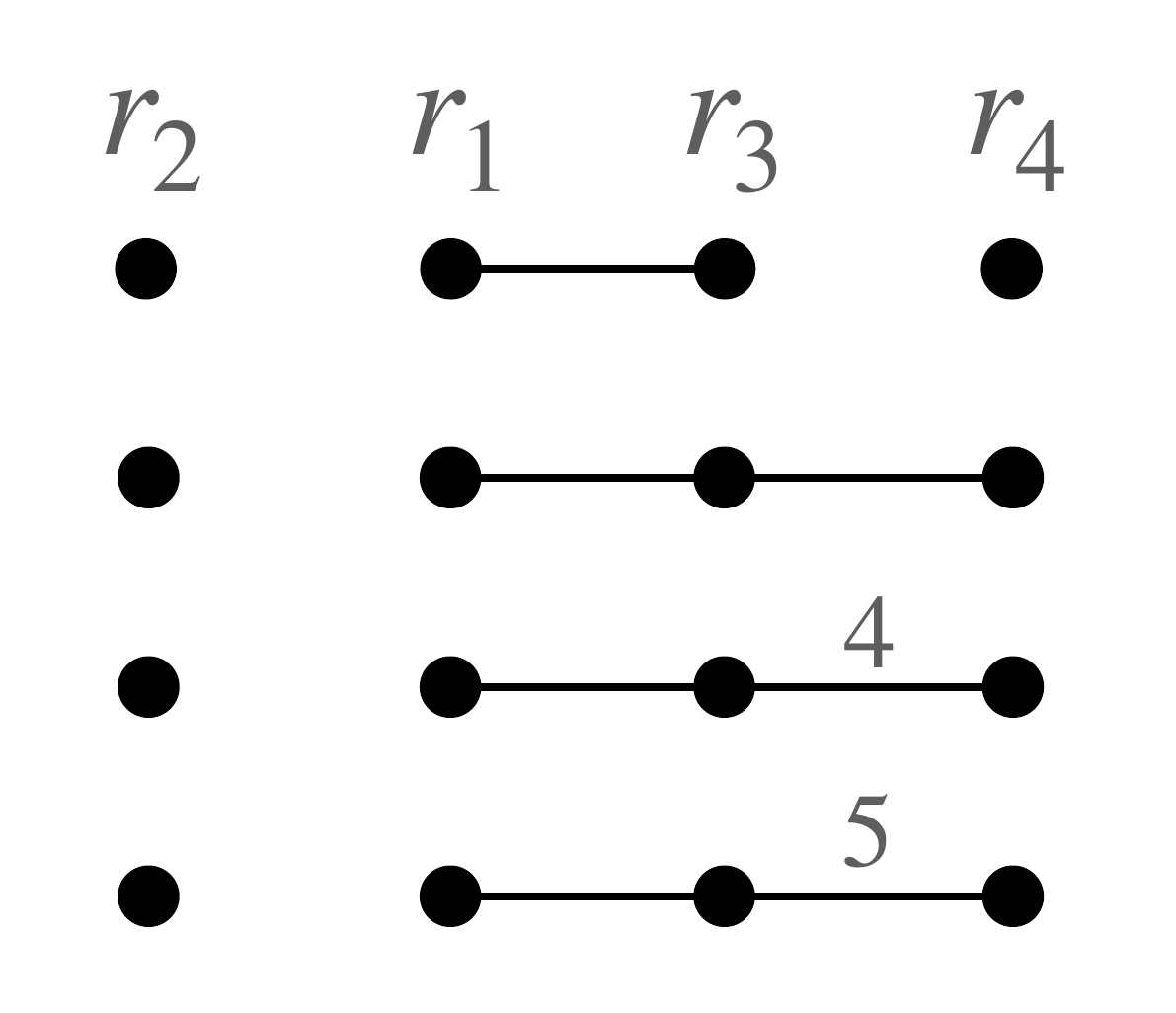}
    \end{center}
    \caption{
        Dynkin diagrams for the finite group resulting from the relation $(r_3 r_4)^m=1$ in addition to the ones given in equation \eqref{hatk-gens-1}, for $m=2,3,4$ and $5$ respectively.}
    \label{dynk-d4d6}
\end{figure}\noindent
The Coxeter group for $m\geq 6$ is infinite. This is because a connected Dynkin diagram with at least three vertices can not have an edge with label $6$. 

Interestingly, there is another way in which the case of $m\geq 6$ is different from the case of $m\leq 5$. To understand this, we take the regions corresponding to $r_1$ and $r_2$ to be smaller than those corresponding to $r_3$ and $r_4$ and transition to the ``t-channel''. The t-channel singular locus consists of external segments with labels $2,2,2, 3$ and an internal segment with the label $m$. For $m\geq 6$, the trivalent junctions in this diagram are not of allowed type and hence do not correspond to finite Coxeter groups. Hence the solution corresponding to this putative singular locus does not exist.  

The order of the orientation preserving subgroup of these Coxeter groups is $|\CR|= 12, 24, 48, 120$ respectively. Using equation \eqref{free-index} and the fact that the characteristic of the Kleinian group is $-1/12$, we see that the genus of the handlebody for $m=2, 3, 4, 5$ is $2,3, 5, 11$ respectively\footnote{We found \cite{lmfdb2223} useful as it lists automorphism group $G$ of Riemann surfaces for genus $2\leq g\leq 15$ and the boundary signature of the boundary orbifold after taking the quotient by $G$. For example the particular link is a search of all such surfaces with boundary signature (2,2,2,3) (the current case of interest). In practice given any symmetry group $G$ one can search for the possible boundary signatures and the genus of these Riemann surfaces. Using the boundary signature one can determine all possible replica symmetry preserving graphs of singularities and construct the corresponding Kleinian group $\CK$ using the combination theorems. Then given the genus $g$ one knows that the rank of the necessary virtual Schottky group is also $g$. Once the $g$ generators of $\CS$ are determined and it is verified that $\CS$ is a normal subgroup of $\CK$ one will have $\CK/\CS \sim G$ as desired. This gives a targeted way of constructing a large number of explicit examples using the same methods presented here.}. As explained earlier, this is also the rank of the free Schottky subgroup that we quotient the Kleinian group to get the finite Coxeter groups. 
Note that even though only one additional condition $(r_3 r_4)^m=1$ is imposed on the generators of $\hat \CK$, it does not correspond to quotient by a free group with only one generator. This is because, we need to generate the Schottky subgroup by $(r_3 r_4)^m$ and all its conjugates $g (r_3 r_4)^m g^{-1}, g\in \hat \CK$ because the Schottky subgroup is normal in $\hat \CK$\footnote{We thank Arvind Nair for discussion on this point.}. In practice, it is sufficient to construct the Schottky generators by taking $g$ to be generators of $\hat \CK$.
Now we will give the generators of the Schottky group in the cases $m=2,3$. 

\subsubsection*{$m=2$}
\begin{figure}[h]
    \begin{center}
        \includegraphics[scale=0.25]{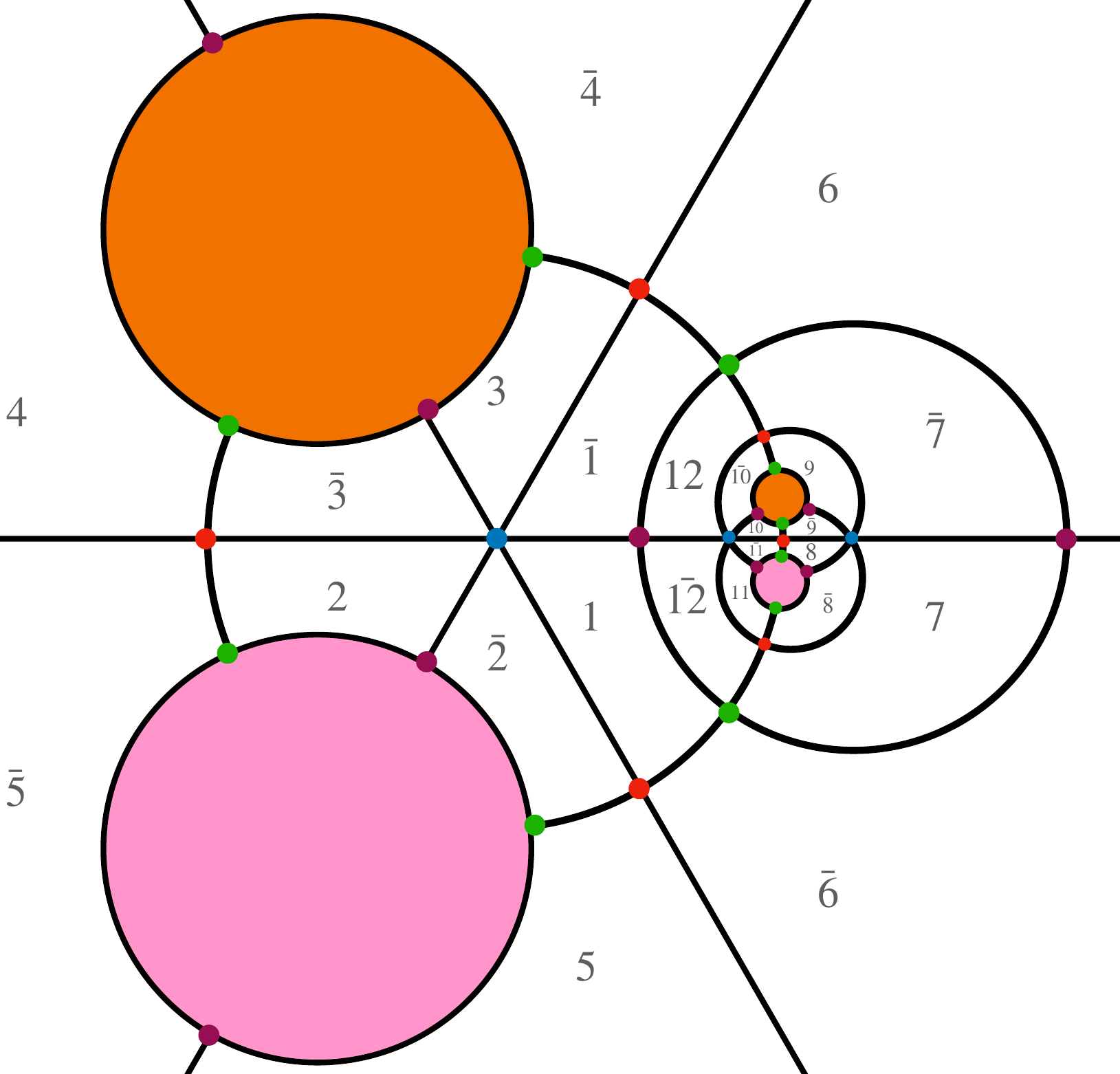}
    \end{center}
    \caption{
        The fundamental domain of this Schottky group tessellated by the images of fundamental domains of $\hat \CK$. A given Schottky generator maps a circle of some color to the other circle of the same color.}
    \label{d4d6-g2}
\end{figure}

One of the Schottky generators is obviously $L_1=(r_3 r_4)^2$ \footnote{A useful trick: The action of the Schottky generator should be to map the inside of one of the circles to the outside of the other. the idea is that if we were to ``move" outside of the fundamental region of the Schottky group we should instead identify the new region with something inside the fundamental region. Examining figure \ref{d4d6K}  if we move just inside the orange circle we will be in the region labeled $CB$ and then outside the other orange circle is $BC^2$. So we want to make the identification $CB=BC^2$. Thus, using the group algebra to move all of the transformation to one side the correct generator is given by $(CB)^2=(r_3r_4)^2$.}. If we conjugate it by $r_2$, we get the same element. This is because $r_2$ commutes with both $r_3$ and $r_4$ in $\hat \CK$. This is also seen from the fact that $r_2$ is not connected to either $r_3$ or $r_4$ in the Dynkin diagram. Conjugation by $r_1$ however yields a different element $L_2= r_1 (r_3 r_4)^2 r_1$. We have used the relation $r_1^2=1$ to replace $r_1^{-1}$ by $r_1$. There are no relations between $L_1$ and $L_2$. 
Conjugates by either $r_3$ and $r_4$ do not give any new elements. The Schottky group is then generated by $L_1$ and $L_2$. The fundamental domain of this Schottky group tessellated by the images of fundamental domains of $\hat \CK$ is shown in figure \ref{d4d6-g2}. 
It can be seen from the figure that one orange circle is mapped to the other orange circle  by  $L_1=(r_3 r_4)^2$.  The pink pair is obtained from the orange pair by reflecting them across $x$-axis i.e. by $r_1$. This is why the conformal transformation relating them is $r_1 L_1 r_1= L_2$.

\subsubsection*{$m=3$}
Again, one of the Schottky generators is given by $L_1=(r_3 r_4)^3$. The conjugation by $r_2$ yields the same element. The conjugation by $r_1$ gives a different element $L_2= r_1 (r_3 r_4)^3 r_1$. In this case, $r_3$ and $r_4$ don't commute. Their conjugation gives two new elements $L_3$ and $L_4$ but they are not independent. The only independent generators are $L_1, L_2$ and $L_3$. These are the three Schottky generators in this case. The fundamental domain of this Schottky group tessellated by the images of fundamental domains of $\hat \CK$ is shown in figure \ref{d4d6-g3}:
\begin{figure}[h]
    \begin{center}
       \includegraphics[scale=0.25]{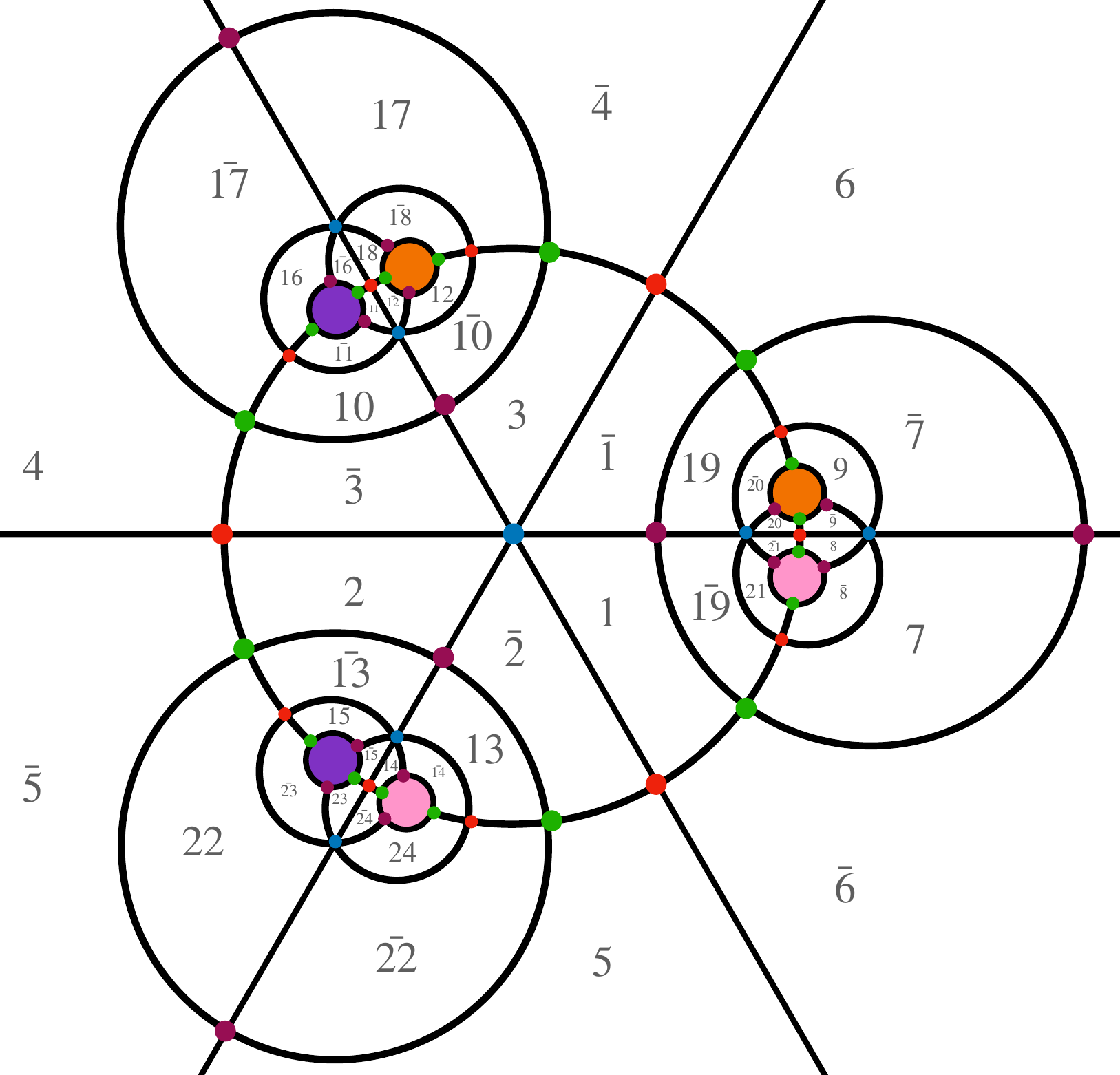}
    \end{center}
    \caption{
        The fundamental domain of this Schottky group tessellated by the images of fundamental domains of $\hat \CK$. A given Schottky generator maps a circle of some color to the other circle of the same color.}
    \label{d4d6-g3}
\end{figure} 
In this case also, it is seen from the figure that one orange circle is mapped to the other orange circle  by  $L_1=(r_3 r_4)^3$. 
The other Schottky generators $L_2$ and $L_3$ are understood as follows. 
The pink pair is the reflection of the orange pair by reflection $r_1$ as before i.e. $r_1L_1r_1 = L_2$ and the purple pair is the reflection of orange pair across the $x$-axis rotated by $2\pi/3$ i.e. by the reflection element $(r_3 r_1) r_1 (r_3 r_1)^{-1}=r_3r_1r_3=r_1r_3r_1$. Hence $(r_1r_3r_1)L_1(r_1r_3r_1)^{-1} = L_3$.

\subsubsection{Non-Coxeter replica symmetry groups}\label{non-cox}
\begin{figure}[h]
    \begin{center}
        \includegraphics[scale=0.25]{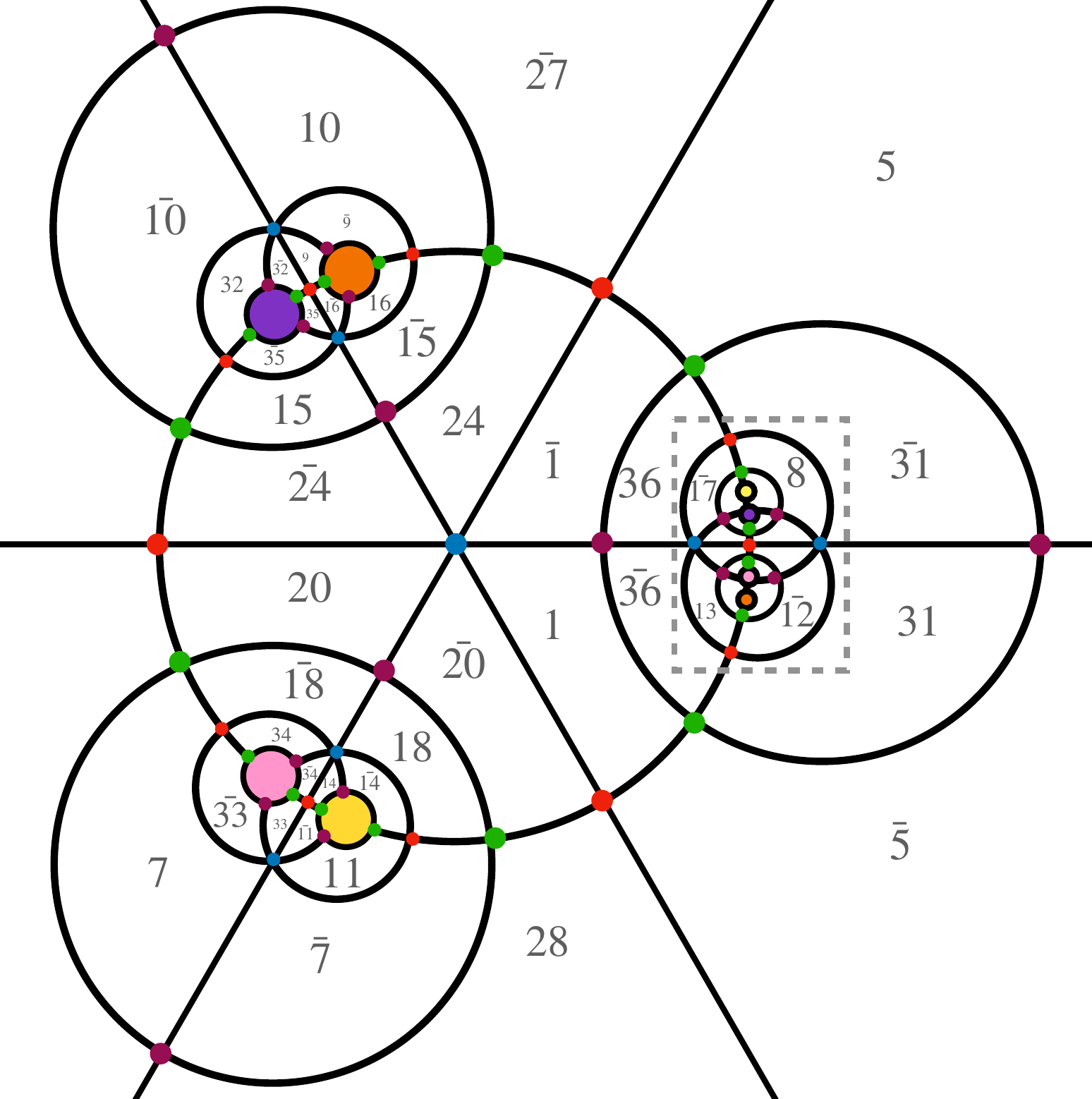}\qquad 
        \includegraphics[scale=0.25]{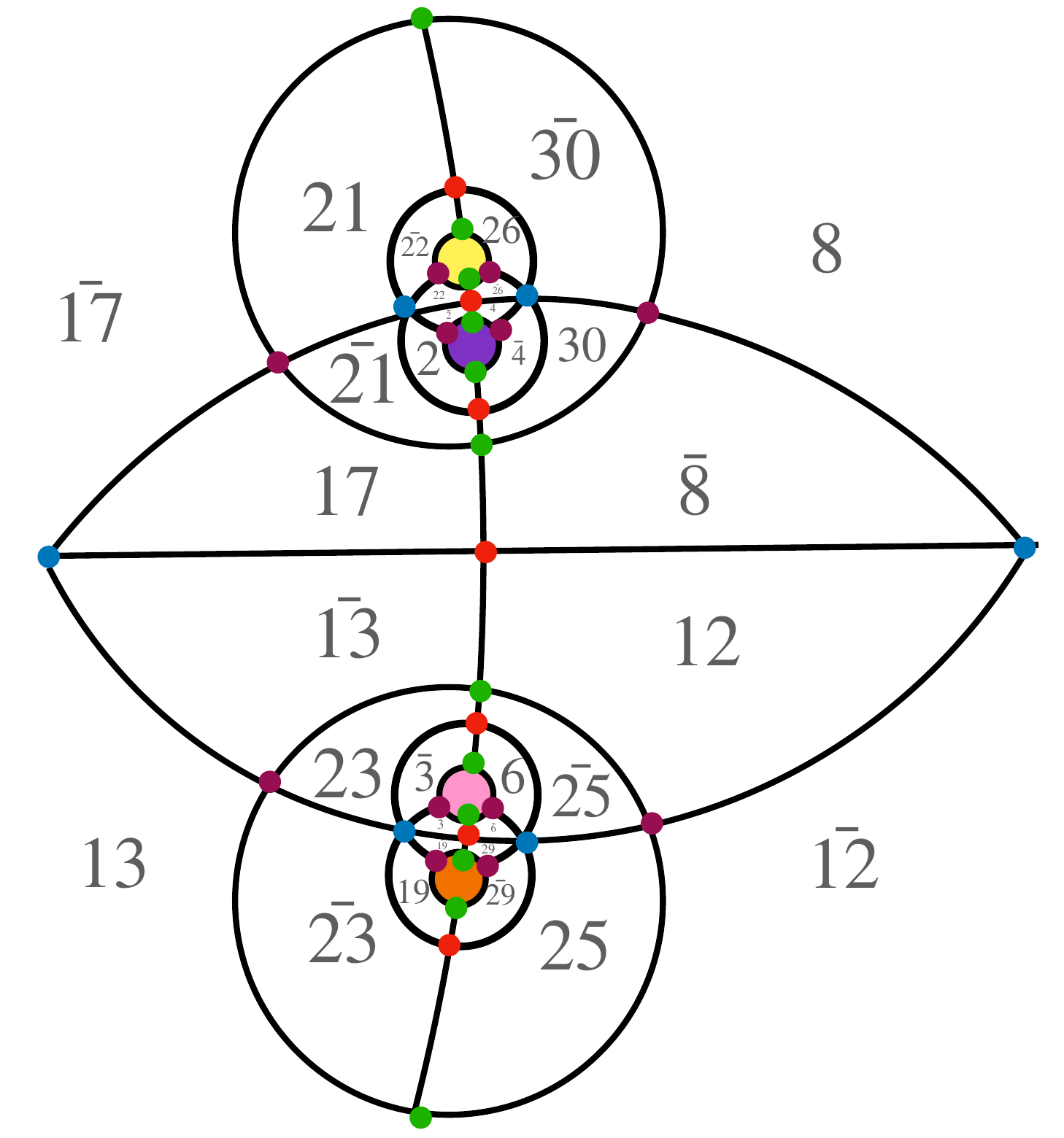}
    \end{center}
    \caption{
        The fundamental domain of this Schottky group tessellated by the images of fundamental domains of $\hat \CK$. A given Schottky generator maps a circle of some color to the other circle of the same color. In the second figure, we have zoomed into the part of the first figure indicated by the dotted rectangle.}
    \label{d4d6-g4}
\end{figure}

From the equation \eqref{free-index}, we see that for $m=4, 5$ the number of Schottky generators are $5$ and $11$ respectively. We will not construct them here. Instead, we will construct  Schottky subgroup with $4$ generators. The quotient group in this case is not a finite Coxeter group. See figure \ref{d4d6-g4} for the fundamental domain of this Schottky group and its tessellation by the fundamental domains of $\hat \CK$. The generators of the Schottky group are
\begin{align}\label{s3s3schottky}
    L_1= ((r_4 r_3)^2 r_1)^2,\quad L_2=(r_1(r_4 r_3)^2)^2,\quad L_3=(r_1 r_4 r_3)^3,\quad L_4= (r_4 r_3 r_1)^3 r_1. 
\end{align}
Using the fact that the characteristic of ${\CK}$ is $-1/12$ and using the formula \eqref{free-index}, we see that the order of the replica symmetry group is $12 (g-1)= 36$.\footnote{Note that the value of $m=6$ in this example, i.e. $(r_3r_4)^6=e$, which can be computed from the explicit representations of the group elements given in appendix \ref{app:reps} ($\sigma_{a^2c^2b}$). Therefore, to obtain a finite replica symmetry group one has to impose further constraints which are not of the Coxeter type, namely setting all the Schottky generators listed in \eqref{s3s3schottky} to $e$.} The finite group $\CK/\CS$ obtained is $\mathbb{S}_3\times \mathbb{S}_3$ and its order is indeed $36$.

\subsubsection{$\mathbb{D}_{6} *_{{\mathbb Z}_3} \mathbb{A}_{4}$}
\begin{figure}[h]
    \begin{center}
        \includegraphics[scale=0.50]{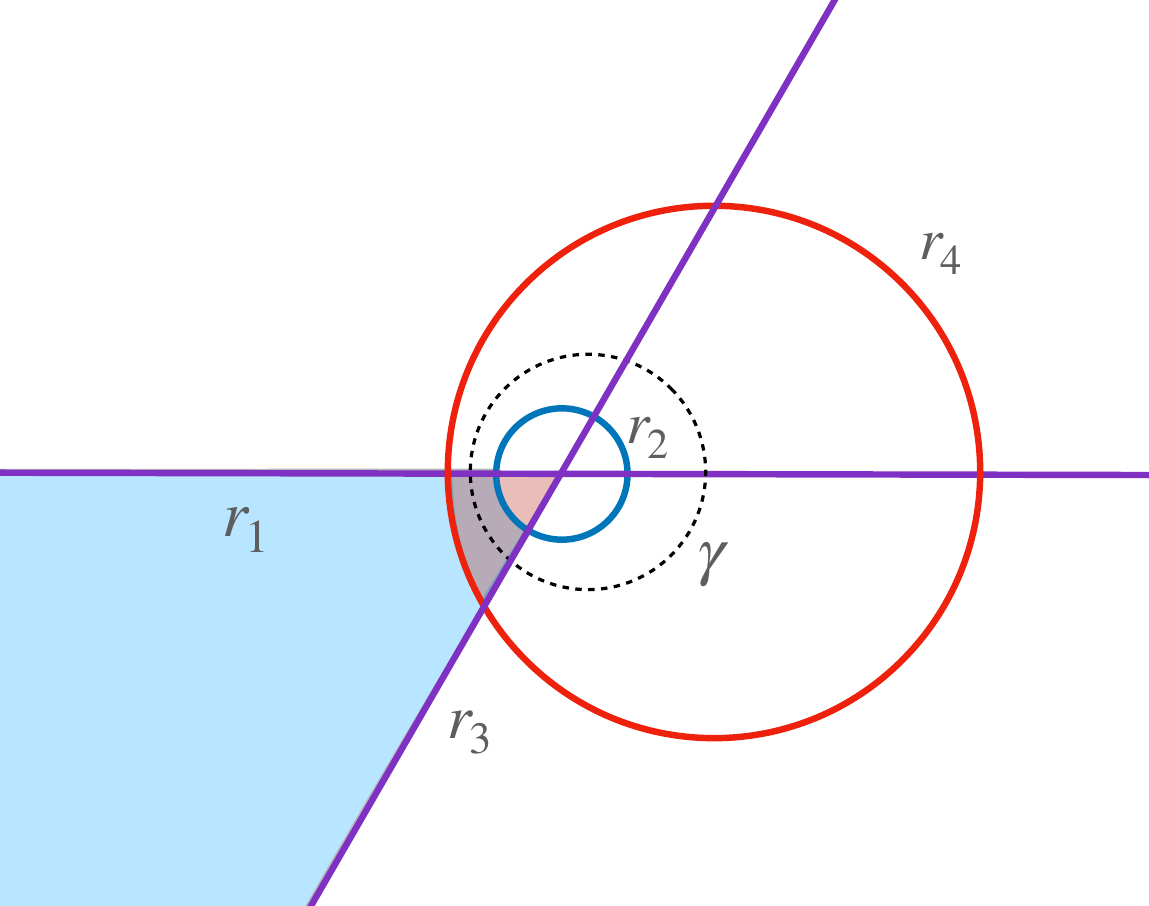}
        \includegraphics[scale=0.45]{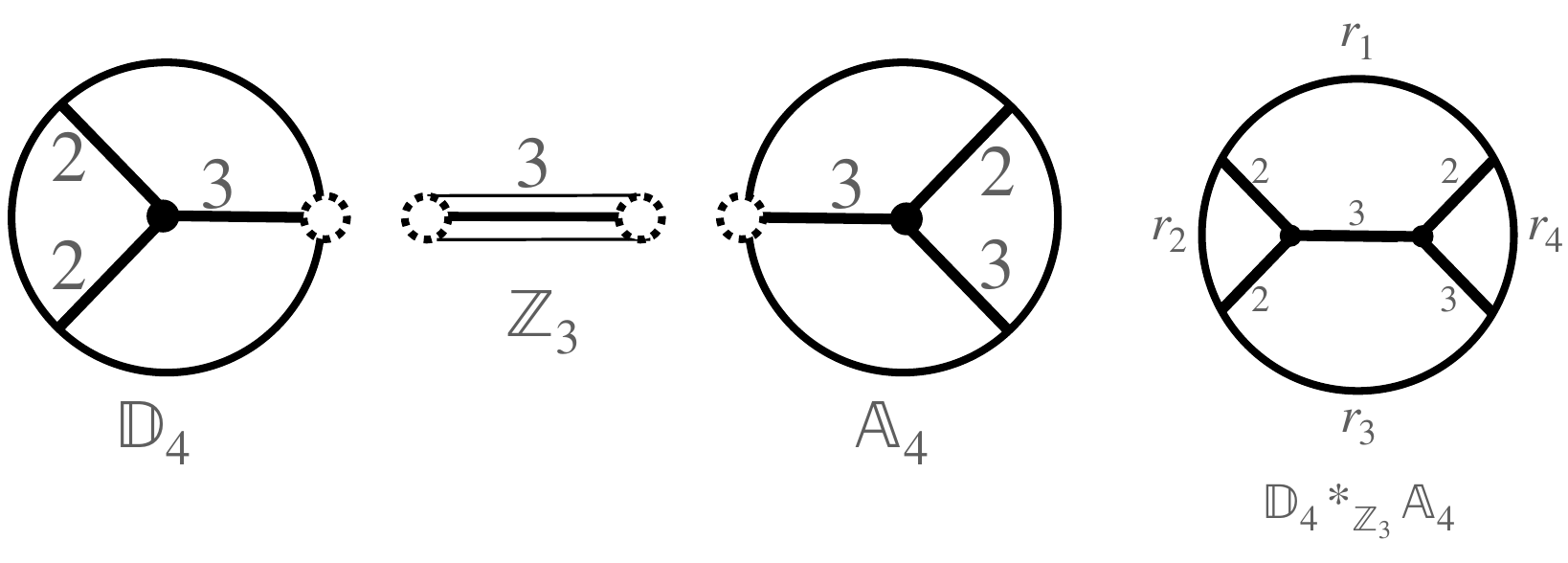}
    \end{center}
    \caption{
        The first figure shows the quadrilateral fundamental domain of $\mathbb{D}_{6} *_{{\mathbb Z}_3} \mathbb{A}_{4}$ bounded by geodesics that are fixed loci of reflections $r_1,\ldots, r_4$. 
        The second figure shows gluing of the singular loci of $\mathbb{D}_{6}$ and $\mathbb{A}_{4}$ to produce the singular locus of the Kleinian group $\mathbb{D}_{6} *_{{\mathbb Z}_3} \mathbb{A}_{4}$. The reflections corresponding to the four parties have been marked in the bottom right figure. They satisfy the relations listed in \eqref{ext-kle-d6a3}.}
    \label{d6a3-sol}
\end{figure}

In this section, we will describe the amalgamation of dihedral group $\mathbb{D}_{6}$ and $\mathbb{A}_{4}$ but along the common ${\mathbb Z}_3$. The Coxeter tuples for  extended replica symmetry groups are $(2,2,3)$ and $(2,3,3)$ which are amalgamated along the common conical locus with label $3$. See the second figure in figure \ref{d6a3-sol} for the corresponding orbifold solution. Using the formula \eqref{two-amalgam}, the characteristic for this group is $-1/12$. This is the same as the previous case of $\mathbb{D}_{4} *_{{\mathbb Z}_2} \mathbb{D}_{6}$.
In the first figure we have shown the quadrilateral fundamental domain of $\mathbb{D}_{6} *_{{\mathbb Z}_3} \mathbb{A}_{4}$ bounded by geodesics that are fixed loci of the reflections:
\be
r_1(z)=\bar{z}, \quad r_2(z)=\frac{p^2}{\bar{z}}, \quad r_3(z)=e^{\frac{2\pi i}{3}}\bar{z}, \quad r_{4}(z)=\frac{(1+\alpha)\bar{z}-2\alpha}{2\bar{z}-(1+\alpha)}, \text{ with } \alpha=\frac{1-\sqrt{3}}{1+\sqrt{3}}, \quad 0\leq p\leq -\alpha.
\ee
We also show the curve $\gamma$ involved in Maskit's combination of $\mathbb{D}_{6}$ and $\mathbb{A}_{4}$.  The angles of the quadrilateral are   $\pi/2, \pi/2, \pi/2, \pi/3$. They add up to less than $2\pi$ hence the boundary of the quotient surface $(\partial H^3)/\CS$ is hyperbolic. The resulting extended Kleinian group is given by
\be \label{ext-kle-d6a3}
\hat\CK=\langle r_1^2=r^2_2=r_3^2=r_4^2=e,\;(r_2r_1)^2=(r_3r_1)^3=(r_4r_1)^2=(r_2r_3)^2=(r_3r_4)^3=e\rangle.
\ee

\subsubsection*{The Schottky subgroup}
\begin{figure}[h]
    \begin{center}
        \includegraphics[scale=0.25]{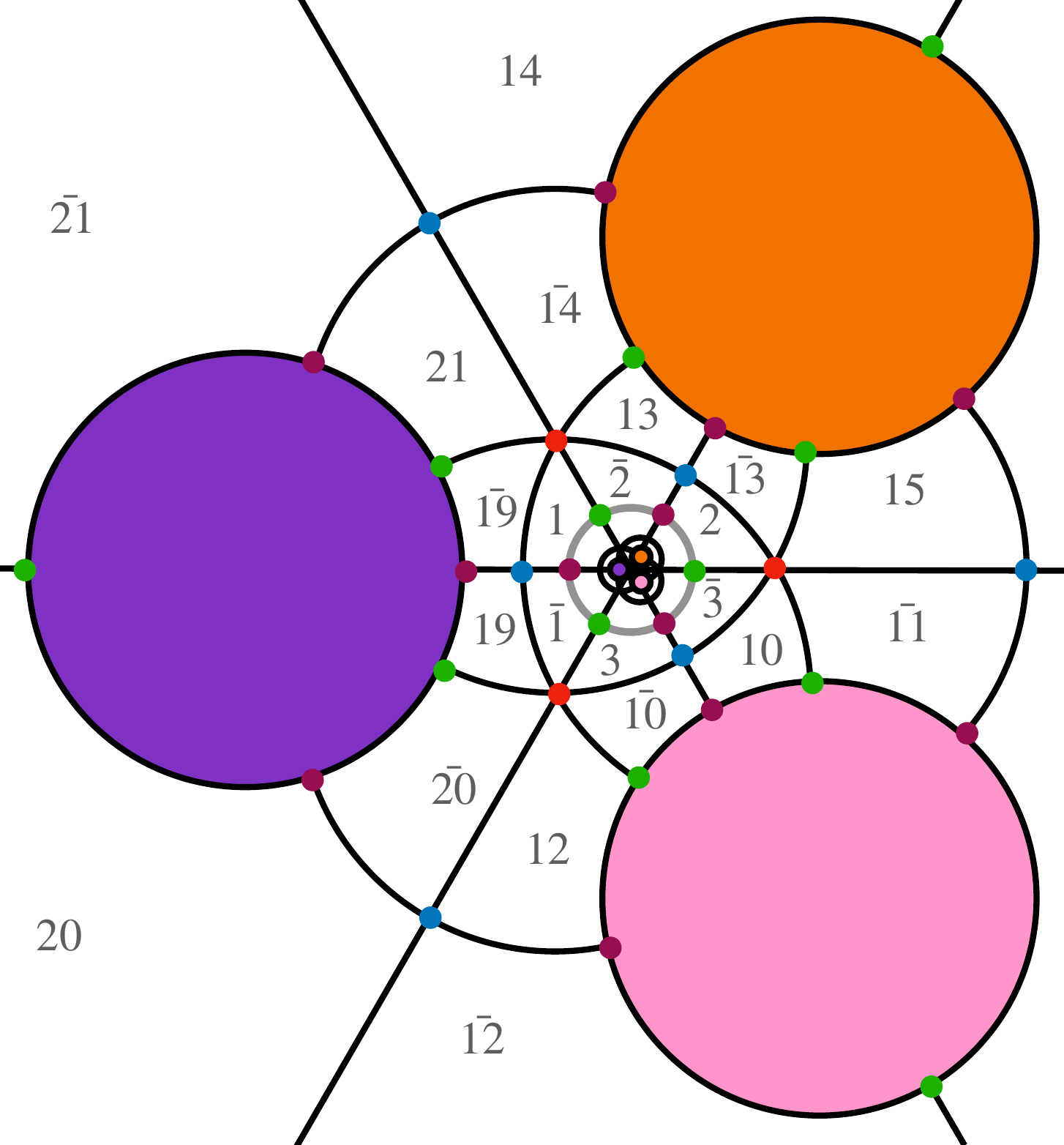}\qquad 
        \includegraphics[scale=0.25]{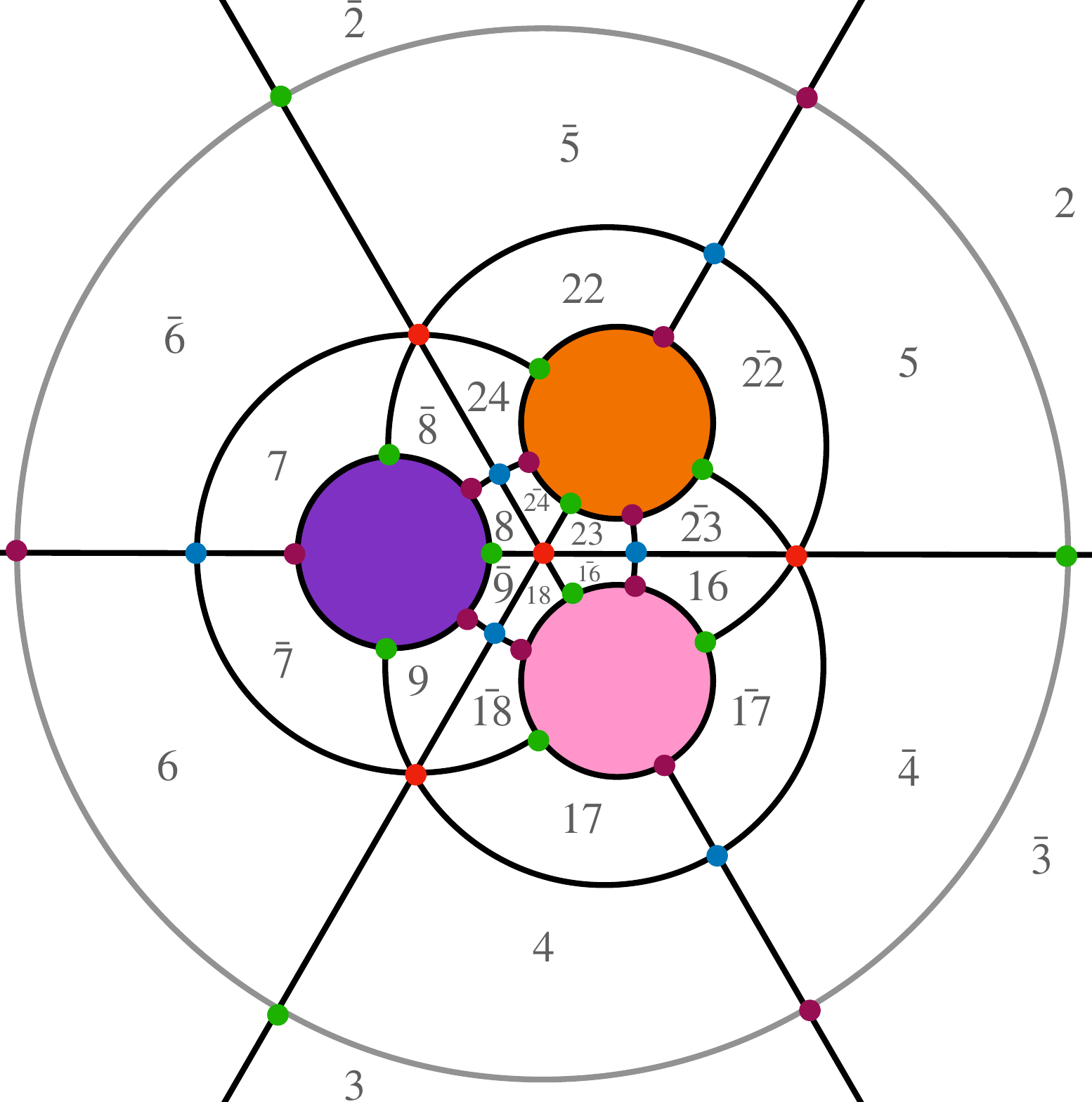}
    \end{center}
    \caption{
        The fundamental domain of this Schottky group tessellated by the images of fundamental domains of $\hat \CK$. A given Schottky generator maps a circle of some color to the other circle of the same color. In the second figure, we have zoomed into the part of the first figure marked  by the grey circle.}
    \label{d6a3-g3}
\end{figure}

In figure \ref{d6a3-g3}, we have chosen three Schottky generators. They are given by the conformal transformations that map a circle of some color to the other circle of the same color. In this way, we obtain the fundamental domain of this Schottky group tessellated by the images of fundamental domains of $\hat \CK$. The Schottky generator that maps the purple pair of circles is $L_1= (r_2 r_4)^2$. As the orange pair and pink pair is obtained from the purple pair by rotating by angle $2\pi/3$ around the origin, the other two Schottky generators are given by $L_2= (r_1 r_3) L_1 (r_1 r_3)^{-1}$ and $L_3=  (r_1 r_3)^2 L_1 (r_1 r_3)^{-2}$.
As the number of generators of the Schottky group are $3$, using equation \eqref{free-index}, we see that the order of the replica symmetry group is $12(g-1)=24$.

\section{Bulk replica symmetric invariants}\label{replica-solution}

\subsection{General amalgamation}\label{general-tree}
In the previous section we discussed in detail a number of examples of amalgamation of two spherical groups. In this section, we will discuss amalgamation of a general number of spherical groups. It is useful to think of this amalgamation in steps. 
The fundamental domain a single spherical group is triangular. The fundamental domain for the amalgamation of two such group is a quadrilateral that obtained by intersection of their individual triangular fundamental domains. If we amalgamate yet another spherical group with the resulting  Kleinian group along a common cyclic group, the fundamental domain is pentagon which is the intersection of the quadrilateral and the triangle and so on. Of course, for each amalgamation, we ensure that conditions of  Maskit's combination theorem are obeyed.

\begin{figure}[h]
    \begin{center}
        \includegraphics[scale=0.25]{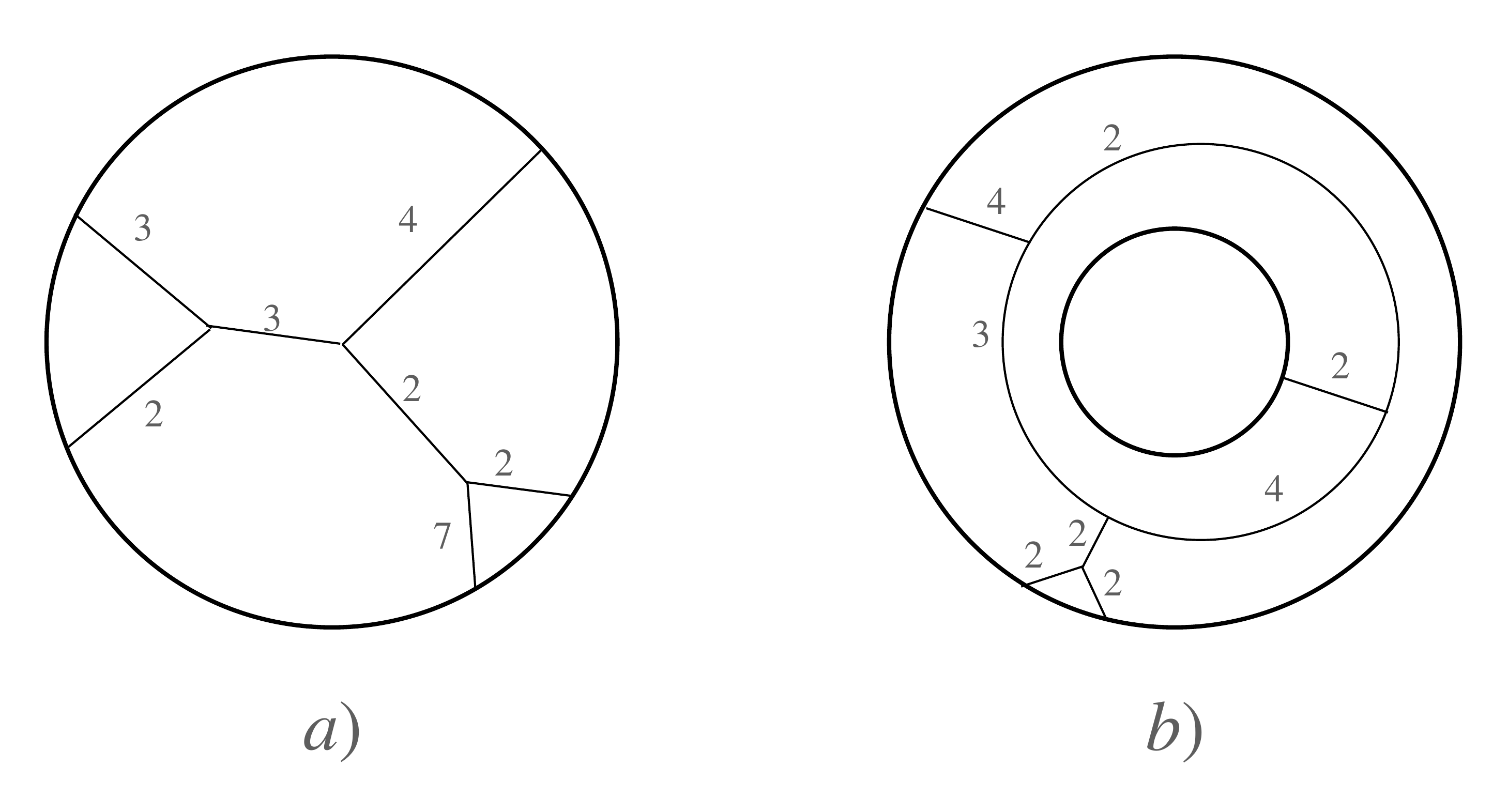}
    \end{center}
    \caption{An example of a generic singular locus where each vertex is one of the admissible ones. The list of admissible vertices is given in Figure \ref{fin-cox-orb}.}
    \label{gensingloc}\label{loop}
\end{figure}

It is straightforward to describe the singularity locus of the orbifold obtained by quotienting $H^3$ by the resulting Kleinian group. The singularity locus of a single spherical group consists of a trivalent graph of singular segments labeled by admissible tuples \eqref{spherical-tuples}. The label of each of the segment denotes the order of the cyclic group that it is fixed under. As a result, an edge with label $m$ has a conical angle $2\pi/m$ around it. 
The spherical groups are amalgamated along a common cyclic subgroup. As these subgroups are denoted by the edges, the singularity locus of the  amalgamated group is obtained simply by gluing the singularity loci of the two spherical groups along the edge that is fixed under the common cyclic subgroup. This is shown in the second sub-figures in figures \ref{ddamalgam-fig}, \ref{d4d6amalgam-fig} and \ref{d6a3-sol}. If we amalgamate another spherical group along a common cyclic group, its singular locus is given by gluing the individual singular loci along an edge with the common label. As a result, we get a tri-valent tree graph with three vertices and five external edges. As the spatial slice of the boundary is divided into five segments, this orbifold corresponds to an invariant with five party-regions. This procedure can be continued to produce orbifolds of more and more complicated Kleinian groups obtained by amalgamating multiple spherical groups. In general, the singular locus of the orbifold is a tree with tri-valent vertices with each edge labeled by the order of the cyclic group that it is fixed under. The labeling obeys the condition that the labels of the edges that meet at a vertex must be given by admissible tuples in equation \eqref{spherical-tuples}. See figure \ref{gensingloc} for an example of a general singular locus. It may also happen that the tri-valent tree formed by singular locus is disconnected. 

Note that from the singular locus, we can read off the complete information of the amalgamation. Each vertex gives us the individual spherical groups that are involved in the amalgamation and each edge gives the common cyclic subgroups that the ``adjacent'' spherical groups are amalgamated along. 

The amalgamation procedure that we have described so far produces only trees but it can be generalized to produce Kleinian groups whose singularity locus also has loops. An example of such a singular locus is shown in figure \ref{loop}. Even in this case, the resulting graph is always tri-valent with the vertices picked from admissible tuples. In more general context, this graph structure with groups associated to its vertices and edges is known as a ``graph of groups'' and the associated amalgamated group as its fundamental group.\footnote{The loops in the graph are not produced by amalgamation, but rather what is known as the HNN extension. There is another combination theorem which produces a Kleinian group isomorphic to a HNN extension but we will not make use of it this paper.} It is a theorem that all virtually free Kleinian groups are obtained as fundamental groups of graphs of group of above type.

The extended Kleinian group associated to graphs can also be described symmetrically, without reference to the amalgamation. We assign reflection elements to each party-region. These labels also label the chambers resulting from the embedding of the tree in a disk. Let us denote the label of the edge separating chambers $\ta$ and $\tb$ as $m_{\ta\tb}$ \footnote{Note that the labels $m_{\ta\tb}$ are not finite for all $\ta$ and $\tb$. The value of $m_{\ta\tb}$ is infinite i.e. there is no relation between $r_\ta$ and $r_\tb$, if the chambers corresponding to parties $\ta$ and $\tb$ do not share an edge.}. Then the only relations that are obeyed by these reflections are $(r_\ta r_\tb)^{m_{\ta \tb}}=1$. This gives an explicit presentation of the associated extended Kleinian group. This also shows that two chambers of the same label do not share any edge because this would assign the label $1$ to the edge that is common making it trivial as the angle around it would be $2\pi$. 

This discussion also explains certain equalities of the normalized multi-invariants enjoyed by the holographic states which are highlighted in section \ref{sym-handle}. If we take the reflection generators $r_\ta$ and $r_\tb$ associated to chambers that do not share a wall to be identical i.e. impose the relation $r_\ta r_\tb=1$ then the singular locus and hence the orbifold geometry remains unaffected. The identification of $r_\ta$ and $r_\tb$ enforces identification of the parties associated to those regions  giving us a $\tq-1$ partite state from a $\tq$-partite state. As a result we get an equality between $\tq$-partite normalized invariant and $\tq-1$ partite normalized invariant. This observation was also emphasized in \cite{Gadde:2022cqi, Gadde:2023zzj}.

In the rest of the paper, we will restrict ourselves to the case when the underlying space of the orbifold is a ball and the singular locus is a tree. These are the geometries relevant for describing multi-invariants of the vacuum state. The characteristic of the Kleinian group resulting from such amalgamation is given by the formula, 
\begin{align}\label{bulk-char}
    \chi(\CK)= \sum_v \frac{1}{|G_v|}-\sum_{ie} \frac{1}{|G_{ie}|}.
\end{align}
Here $G_v$ and $G_{ie}$ are the groups associated with the vertices and \emph{internal} edges of the graph respectively. This formula is used along with equation \eqref{free-index} to relate the order of the replica symmetry group $|\CR|$ and the genus $g$ of the handlebody on which it acts. 

Another formula that relates the genus $g$ to the order of the symmetry group $\CR$ is the Riemann-Hurwitz formula given in equation \eqref{riem-hur}. It reads,
\begin{align}\label{boundary-char}
    |G|=\frac{1}{\chi_\partial(\CK)} (1-g), \quad \chi_\partial(\CK)= 1-\frac12 \sum_{ee} \left(1-\frac{1}{|G_{ee}|}\right).
\end{align}
Here we have tailored the formula \eqref{riem-hur} to the current setting. We have set $\chi_\CM=2$ as we are interested in taking the initial manifold $\CM=S^2$. We have also used the fact that the order $k_\sigma$ of the twist operator is the label on the external edge of the singularity graph of the orbifold geometry that ends at the corresponding twist operator. This makes twist operators in one-to-one correspondence with the external edges and $k_\sigma$ is the order of the group associated with the corresponding external edge $G_{ee}$. 

Let us emphasize the following point: Given a $\tq$-point function of twist operators, a graph with the required boundary group order will exist only when either all orders occur in pairs or in the orders there are at least two disjoint occurrences of: $(2),(3,3),(3,4),(3,5)$ \cite{10.1112/plms/s3-59.2.373}. It is precisely in these cases that a bulk replica symmetry preserving handlebody solution will exist. For all choices of twist operators whose orders do not have this property the bulk replica symmetry must be broken.

Although, equation \eqref{boundary-char} is a formula relating the order of the symmetry group acting of the Riemann surface to its genus, just like the equation \eqref{free-index}, it is more widely applicable than equation \eqref{free-index}. This is because there could exist symmetric Riemann surfaces such that the symmetries can not be extended into the bulk. However, we do expect the equations  \eqref{boundary-char} and \eqref{free-index} to be consistent in the cases when the group action on Riemann surface does extend into the bulk. In such cases we get
\begin{align}\label{rh-connection}
    \chi(\CK)= \chi_\partial(\CK).
\end{align}
We check this in some examples below. 

For the Kleinian group $\mathbb{D}_{2n} *_{{\mathbb Z}_n} \mathbb{D}_{2n}$, we refer to  its singular locus in figure \ref{ddamalgam-fig} to compute both sides of equation \eqref{rh-connection} 
\begin{align}
    {\rm l.h.s.}= 2\frac{1}{2n}-\frac{1}{n}=0, \qquad {\rm r.h.s.}= 1-2\cdot 4\Big(1-\frac12\Big)=0.
\end{align}
For the Kleinian group $\mathbb{D}_{4} *_{{\mathbb Z}_2} \mathbb{D}_{6}$, we look at its singular locus in figure \ref{d4d6amalgam-fig}.
\begin{align}
    {\rm l.h.s.}= \frac14 +\frac16 -\frac12=-\frac{1}{12} \qquad {\rm r.h.s.}=  1-2\Big(3\Big(1-\frac12\Big)+\Big(1-\frac13\Big)\Big) =-\frac{1}{12}.
\end{align}
For Kleinian group $\mathbb{D}_{6} *_{{\mathbb Z}_3} \mathbb{A}_{4}$, the external singularities are the same as for the case of $\mathbb{D}_{4} *_{{\mathbb Z}_2} \mathbb{D}_{6}$ as seen from figure \ref{d6a3-sol}, so we only need to compute the l.h.s. of equation \eqref{rh-connection}. 
\begin{align}
    {\rm l.h.s.}= \frac16+ \frac{1}{12} -\frac13=-\frac{1}{12}.
\end{align}

In this section, we described the structure of the orbifold geometries $H^3/\CK$. Their underlying space is a three ball and they consist of singular locus that is a tri-valent tree. Each edge of the tree has an integer label which is the order of the group element that keeps the edge fixed. The integer labels of the three edges that meet at a vertex form an admissible tuple \eqref{spherical-tuples}. We expect that the general  orbifold solution can be deformed into a cone manifold, such that the cone angle around a singular edge is continuous value $2\pi \alpha, 0\leq \alpha \leq 1$ such that the parameters associated to the three edges that meet at vertex obey the inequality $\alpha+\beta+\gamma >1$. 

\subsubsection{Bound on the order of the symmetry group}

Note that using the formula \eqref{free-index}, we can put a bound on the order of the symmetry group  $G$ acting on a genus $g\geq2$ handlebody. This is done by minimizing  $|\chi(\CK)|$. It is shown in \cite{10.1112/plms/s3-59.2.373} that the minimum of $|\chi(\CK)|$ is attained by $\chi(\CK)=-1/12$, as in the examples above. The handlebodies resulting from this Kleinian group have the largest symmetry $\CR$, namely of order $12(g-1)$. That is for all handlebodies $g\geq2$:
\be
 |G|\leq12(g-1)
\ee
with equality achieved only for those finite groups which are quotients of the groups  $\mathbb{D}_{4} *_{{\mathbb Z}_2} \mathbb{D}_{6}$,  $\mathbb{D}_{6} *_{{\mathbb Z}_3} \mathbb{A}_{4}$ ,  $\mathbb{D}_{8} *_{{\mathbb Z}_4} \mathbb{S}_{4}$ , or $\mathbb{D}_{10} *_{{\mathbb Z}_5} \mathbb{A}_{5}$  by free normal subgroups. These are precisely the amalgams for which the boundary orbifold signature will be (2,2,2,3).

Some of the other results that are proved in \cite{10.1112/plms/s3-59.2.373}:
\begin{itemize}
    \item If $\CR$ contains no non-abelian dihedral subgroup then $|G|\leq 6(g-1)$
    \item If $|G|$ is odd, then $|G|\leq 3(g-1)$. 
\end{itemize}
All these results can be obtained by case-wise analysis of the singularity locus of virtually free Kleinian groups and using formula \eqref{bulk-char}.

One can contrast these results with the celebrated theorem of Hurwitz which states that the order of the symmetry group of genus $g$ Riemann surface is bounded by $84(g-1)$. The theorem rests on the Riemann-Hurwitz formula \eqref{riem-hur}. To maximize the order of the group acting on a Riemann surface, we minimize $\chi_\partial(\CK)$. However, in this case we do not require that the group action be extended into the bulk.  
Hurwitz showed, by case-wise analysis that symmetry groups resulting in the orbifold that is a sphere with three conical singularities with orders $(2, 3,7)$ minimize $|\chi_\partial (\CK)|$. This means that Riemann surfaces that cover this orbifold have symmetry groups with order $|G|=84(g-1)$. In all other cases,  $|G| < 84(g-1)$. 
The $(2,3,7)$ tuple of singularities does not appear as a result of orbifolding a handlebody as it is not one of the admissible tuples. 
However, it can certainly appear on the boundary as a result of orbifolding Riemann surface.

\subsection{Bulk replica symmetry for all configurations}
So far we have discussed handlebodies that have non-trivial symmetry group and the associated orbifolds. These orbifolds have conical singularities whose locus forms a trivalent graph. If we label the segments of the singular locus by the order of the cyclic group that stabilizes it, then the edge labels at any given vertex must be one of the admissible tuples \eqref{spherical-tuples}. 

\begin{figure}[t]
    \begin{center}
        \includegraphics[scale=0.25]{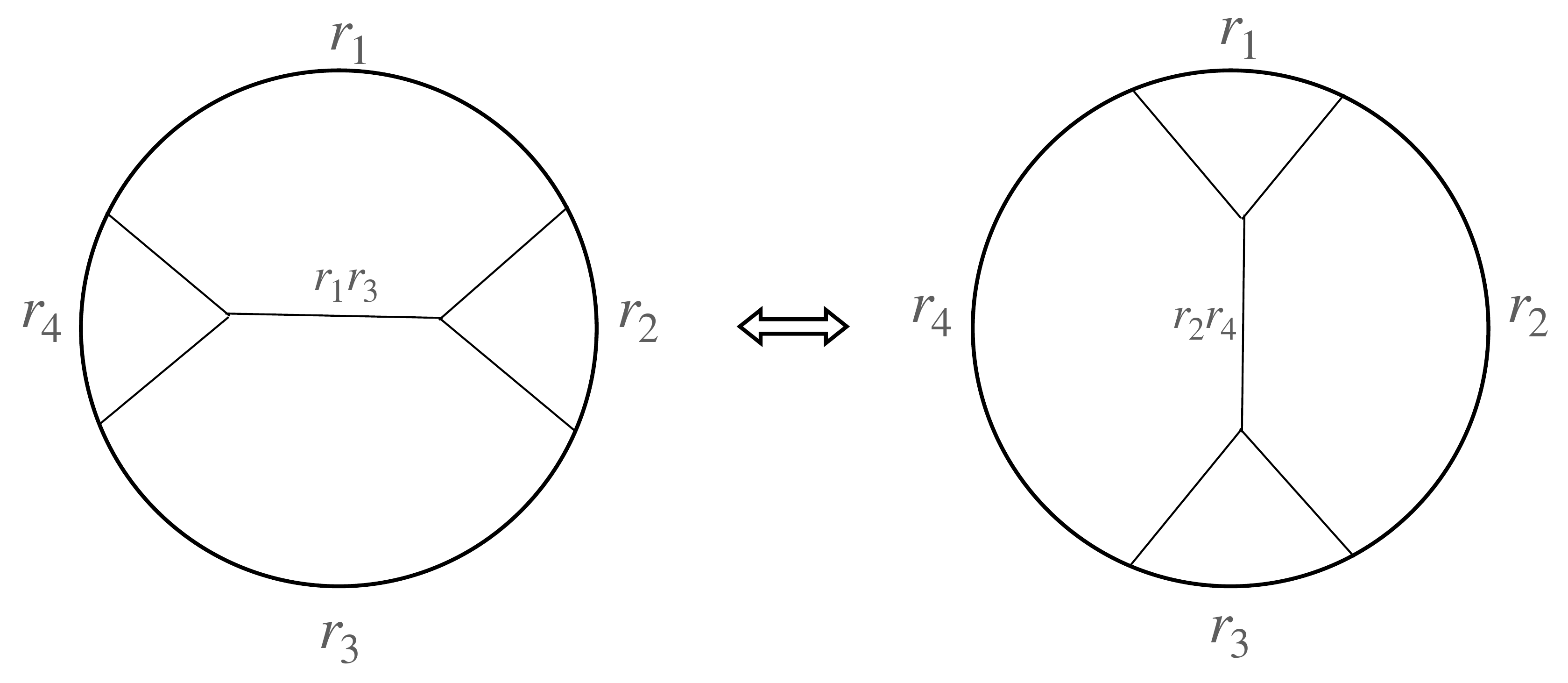}
    \end{center}
    \caption{
        As party regions $1,3$ become small compare to  party regions $2,4$, the singular locus that was ``s-channel'' transitions to ``t-channel''. We have labeled the internal singular segment by its stabilizing element. }
    \label{st-phase}
\end{figure}

Now we can vary the boundary points on which the external singular segments end. These are essentially positions of the twist operators. In varying these points, we are changing the definition of the party-regions on the boundary. Let us call the space of all party-regions the configuration space. As we move from one region of the configuration space to another, it may happen that the symmetric handlebody giving rise to some singular locus after orbifolding ceases to be the dominant solution. In such cases, the value of the multi-invariant undergoes a phase transition. We have already seen an example of such a phase transition in section \eqref{dndn-amalgam}. In that example, as party regions $1,3$ become small compare to  party regions $2,4$, the singular locus that was ``s-channel'' transitions to ``t-channel'' as shown in figure \ref{st-phase}.
As explained there, the order of the element $r_2 r_4$ is the label of the internal edge in the t-channel graph. If it is such that singularity graph in the t-channel involve tuples at vertices that are not admissible then such a putative solution does not exist. In that case, as the relative size of $1,3$ becomes small, the phase transition of ``s-channel'' bulk-orbifold would be to a phase where the replica symmetry is broken. This shows that the replica symmetry may not be preserved everywhere in the configuration space. The examples of this are the non-Coxeter invariants described in section \ref{non-cox}. In those cases, the order of element $r_2 r_4$ is high enough that it does not lead to an admissible tuple in the t-channel. 

Now we are in a position to ask the question about bulk replica symmetry preserving invariants more precisely. What are the multi-invariants for which the replica symmetry is preserved by the dominant bulk solution in all parts of the configuration space, or at least in all corners of the configuration space? We claim that the answer to above question is the family of Coxeter invariants defined in section \ref{coxeter}. 

\begin{figure}[t]
    \begin{center}
        \includegraphics[scale=0.25]{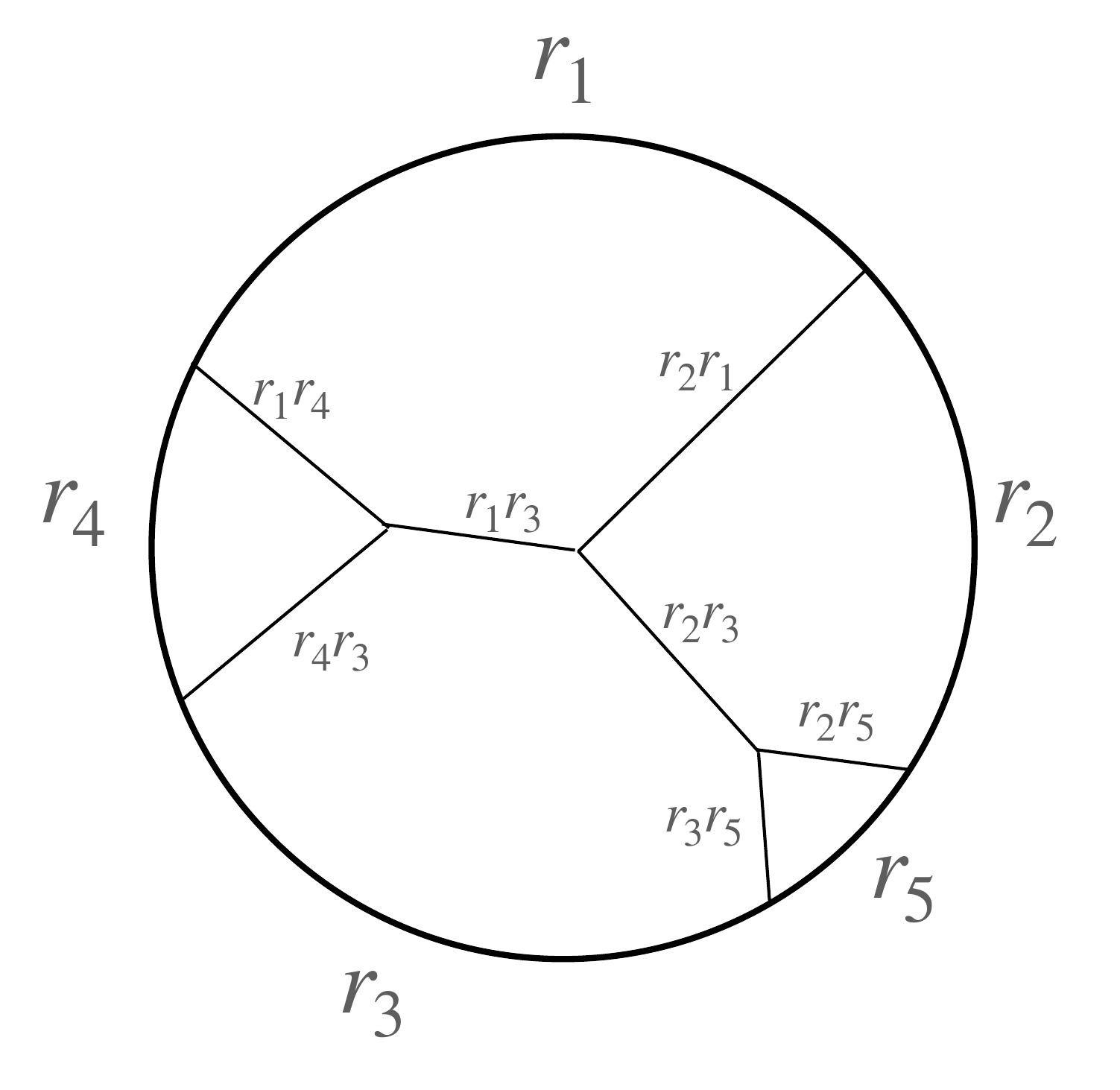}
    \end{center}
    \caption{A general singularity locus for the Coxeter invariant. 
        We have labeled all the edges by the element that stabilizes them.}
    \label{generaltree}
\end{figure}

Recall the definition of Coxeter invariants: A Coxeter invariant corresponds to choosing $\hat \CR$ to be a Coxeter group and the generators $r_\ta$ to be standard Coxeter reflections obeying $r_\ta^2=1$ and $(r_\ta t_\tb)^{m_{\ta\tb}}=1$ where $m_{\ta\tb}$ is encoded in the adjacency type matrix of the Dynkin diagram as explained in figure \ref{fin-cox}. These Coxeter groups and their direct products are the only finite Coxeter groups. The direct product of Coxeter groups correspond to a disconnected graph where each connected component is one of the Dynkin diagram shown in figure \ref{fin-cox}. 
Note that all the Dynkin diagrams and their disconnected sums, have the property that its subgraph on any three nodes is also a Dynkin diagram or a  disconnected sum of Dynkin diagrams. This shows that $(m_{\ta \tb}, m_{\tb \tc}, m_{\tc \ta})$ form an admissible tuple for any party regions  $\ta, \tb, \tc$. If the tri-valent vertex in the singularity graph separates the chambers corresponding to regions $\ta, \tb, \tc$ then the above property ensures that it is admissible. This means that the bulk replica symmetry is preserved everywhere in the configuration space or at least in the finite neighborhood of each corner of the configuration space. A general singularity locus for the Coxeter invariant is shown in figure \ref{generaltree}.

\subsection{Replica symmetry breaking}
In this section, we will discuss the case of symmetric invariants that do not preserve bulk replica symmetry. We will show that, in some situations, even in such cases, one can construct a bulk solution that is replica symmetric but it is not the dominant one. The dominant one on the other hand breaks the replica symmetry. 

\begin{figure}[t]
  \centering
   \includegraphics[scale=0.5]{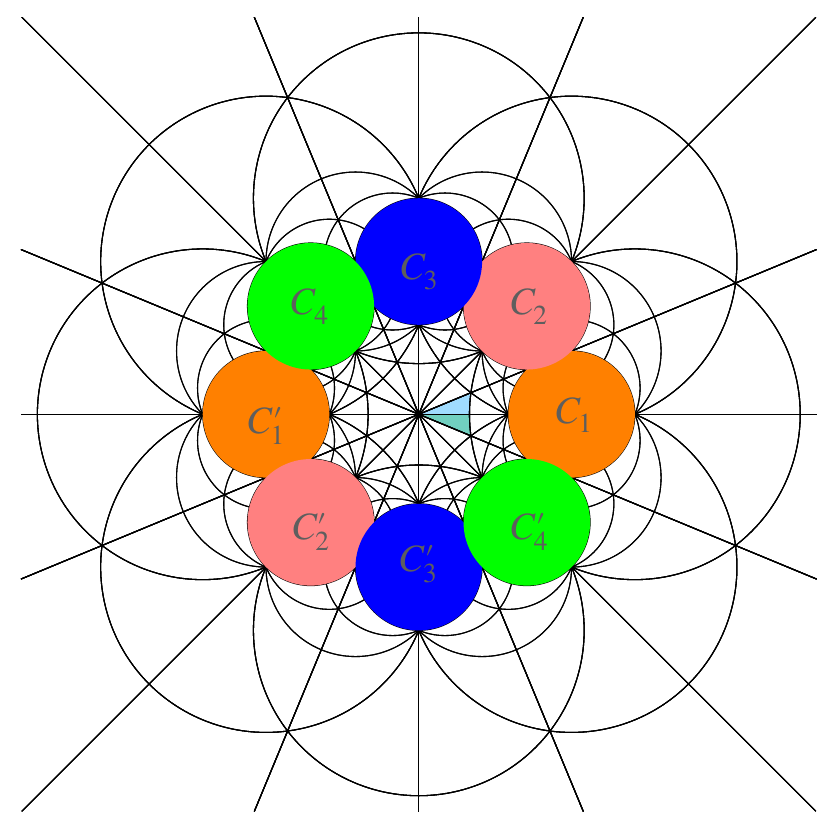}\qquad 
   \includegraphics[scale=0.5]{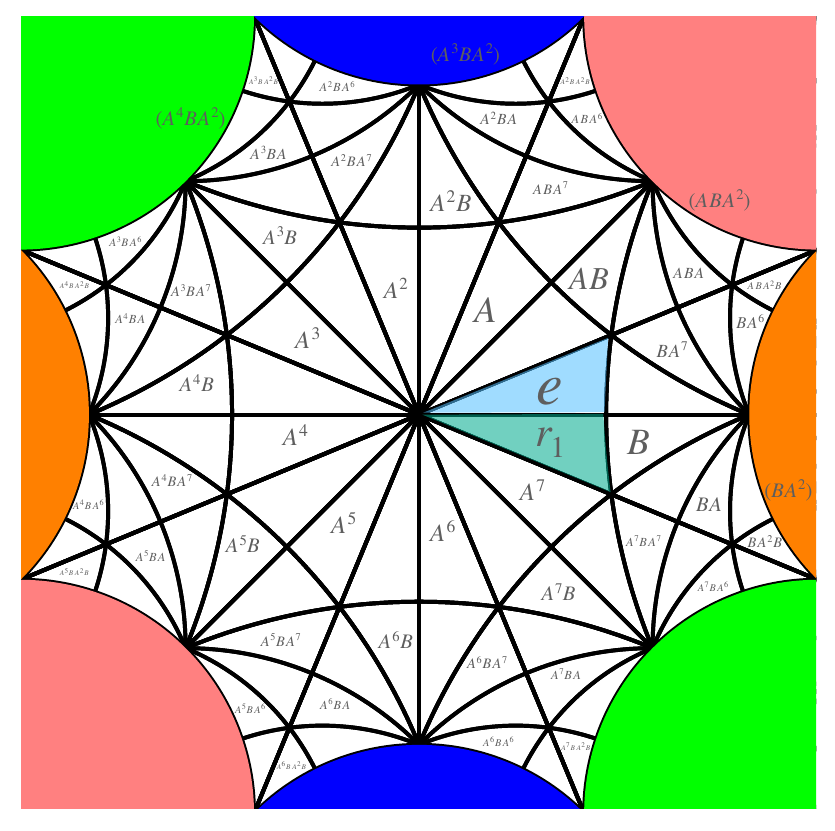}
  \caption{In the first sub-figure we have shown the action of $\Theta_2$ surface group described as transformations $L_i$ mapping pair of circles $C_i, C_i'$ for $i=1,\ldots, 4$. The second sub-figure shows the one of the two components of the fundamental domains of $\Theta_2$. It is a hyperbolic octagon. It is tessellated by hyperbolic triangles that are fundamental domains of the $(2,3,8)$ triangle group. Here $A=r_2r_1$ and $B=r_3r_1$ and $\CK=\langle A^8=B^2=e,(AB)^3=e\rangle$. the identification of the circles i.e. quotienting by the surface group $\Theta_2$ imposes the additional relation $(A^4B)^2=e$.} 
\label{fig:238}
\end{figure}

As before, we will embed the action of the group $\CR$ on the Riemann surface in a Kleinian group acting on $S^2$. However, this is not enough to produce a handlebody solution that preserves the replica symmetry. Recall that for this to be the case, we require the Kleinian group to be virtually free so that we can identify the free normal subgroup to the Schottky group, quotienting by which gives us a handlebody. Hence, the Kleinian group that allows embedding of $\CR$ in the replica symmetry breaking case can not be virtually free. A consequence of theorem \ref{thm:vfcox} is that if a Kleinian Coxeter group is not virtually free then it is virtually surface i.e. it has a surface group as a finite index normal subgroup. A surface group $\Theta_g$ is the fundamental group of a genus $g$ Riemann surface $\Sigma_g$. It consists of $g$ pairs of generators $X_i$ and $Y_i$ which enjoy a single relation 
\begin{align}
    \prod_i (X_i \cdot Y_i \cdot X_i^{-1} \cdot Y_i^{-1}) =e.
\end{align}
Geometrically, $X_i$ and $Y_i$ can be understood to loops anchored at some fixed point on the Riemann surface that wind around $i$-th $X$ cycle and $Y$ cycle respectively. As the fundamental group of $H^3$ is trivial, quotienting by the surface group $\Theta_g$ produces a three dimensional manifold that has the same fundamental group as the a genus $g$ Riemann surface. This manifold is nothing but a hyperbolic wormhole joining two copies of genus $g$ Riemann surface $\Sigma_g$. This can be understood as follows. The surface groups can be described, similar to Schottky groups, by specifying the transformations $L_i, i=1, \ldots , 2g$ that map a given circle $C_i$ to another $C_i'$. 
The total number of circles is $4g$. The surface groups have the property that their limit set is dense on some circle, say the unit circle. In order to remove the limit set from the quotient, the union of interiors of all the circles $C_i, C_i'$ must include the unit circle. As a result, all the circles $C_i, C_i'$ must form a necklace round the unit circle with neighboring circles slightly overlapping. An example of this necklace is given in the first sub-figure in figure \ref{fig:238}. 
Because of the overlap, the maps $L_i$ obey the condition
\begin{align}
    \prod_{i=1}^{4g} L(S_i)^{(-1)^i}=e,
\end{align}
The product is taken over all $4g$ circles $S_i$ in the order given by the necklace. The transformation  $L(S_i)$ maps $S_i$ to its partner. So $L(C_i)=L_i$ and $L(C_i')=L_i^{-1}$. The generators $L_i, i=1, \ldots, 2g$ provide a presentation of $\Theta_g$ that is alternative to the more conventional presentation with pairs of generators $(X_i, Y_i), i=1, \ldots, g$.
The exterior of all the circles, which is the fundamental domain of the surface group action is disconnected. It consists of a hyperbolic $4g$-gon  that is inside the necklace and another one that is outside the necklace. An example of ones of these two $4g$-gons, for $g=2$,  is given in the second sub-figure in figure \ref{fig:238}.  After quotienting each of these regions becomes a genus $g$ surface. The bulk solution is the hyperbolic wormhole geometry that interpolates between the two. One of its boundaries is the genus $g$ surface obtained from the $4g$-gon that is inside the unit circle and the other boundary is the genus $g$ surface obtained from the $4g$-gon that is outside. 
As the wormhole is topologically $\Sigma_g\times [0,1]$, its fundamental group is the same as that of $\Sigma_g$ i. e. $\Theta_g$ as expected.
A virtually surface Kleinian group contains normal surface subgroups $\Theta_g$ for infinitely many, but not all,  values of genus $g$. 
As in the case of handlebody quotients, the wormhole geometry has the remnant symmetry $\CK/\Theta_g=\CR$. This is analogue of the replica symmetry. Quotienting the wormhole further by $\CR$, we get a new wormhole interpolating between two spheres with the same set of singularities, with the singular points on both boundaries joined by a singular locus through the wormhole. 

Below we give an example of the Kleinian group that is $\CK=(2,3,8)$ triangle group. The smallest genus of the normal surface subgroup that it contains is $2$\footnote{For more properties of genus 2 surfaces and those triangle groups which contain genus 2 surface groups see \cite{Kuusalo1995}}.  Quotienting by this group, we get  a Euclidean wormhole geometry interpolating between two Riemann surfaces of genus $2$. The quotient group $\CK/\Theta_2$ is the symmetry group of the most symmetric genus $2$ surface known as the Bolza surface. Quotienting this wormhole geometry further by the replica symmetry group $\CK/\Theta_2$, we get a wormhole geometry between two spheres each with conical singularities of order $2, 3$ and $8$. The singularities of the same order on the two boundaries are joined by the singular locus through the wormhole. 
We take the fundamental triangular region of the (2,3,8) triangle group to be bounded by reflections,
\be
r_1=\bar{z}, \quad r_2=e^{\frac{2\pi i}{8}}\bar{z},\quad r_3=\frac{i \alpha \bar{z}-i\sqrt{ \alpha^2-1}}{i\sqrt{ \alpha^2-1}\bar{z}-i \alpha} \text{ with } \alpha=\sqrt{1+\frac{1}{\sqrt{2}}}.
\ee
It is easy to verify that the order of elements $r_1 r_3, r_2 r_3 $ and $r_1 r_2$ is $2, 3$ and $8$ respectively. As a result, the these reflection elements generate (2,3,8) triangle group as expected. As remarked earlier, this group contains surface groups for infinitely many values of genus. 
The  order of the quotient $\CK/\Theta_g$ for those values of $g$ is $48(g-1)$. We focus on the smallest value of $g$ i.e. $g=2$. 
We will describe the surface group using the generators $L_i, i=1, \ldots, 2g$ rather than the pair of generators $X_i, Y_i, i=1, \ldots, g$. 
\begin{align}
    L_1=(\sigma_{31}\sigma_{21}^4)^2, \qquad L_2= \sigma_{12} L_1\sigma_{12}^{-1}, \qquad L_3= \sigma_{12}^2 L_1\sigma_{12}^{-2}, \qquad L_4= \sigma_{12}^3 L_1\sigma_{12}^{-3}.
\end{align}
Here $\sigma_{ij}=r_i r_j$.
We can verify that they obey the sole relation $L_1L_2^{-1}L_3L_4^{-1} L_1^{-1}L_2 L^{-1}_3L_4=e$. 
The order of the replica symmetry group $\CK/\Theta_2$ is $48$. This is the symmetry group $GL(2,3)$ of the most symmetric genus $2$ surface called the Bolza surface. 

If we interpret this from the point of view of the invariant, it is constructed using $\CK/\Theta_2$ as the replica symmetry group. This gives a genus $2$ surface as the replicated surface. We can take twice the number of replicas so that we the replicated surface is a disconnected sum of two genus $2$ surfaces. Clearly, the dominant bulk solution filling in this pair is the disconnected one. It consists of handlebodies filling in the two genus $2$ surfaces separately. This solution breaks the replica symmetry. 
However, the connected solution that is the Euclidean wormhole \cite{2004JHEP...02..053M} does preserve the replica symmetry as demonstrated above.

All hyperbolic triangle groups - these are the groups specified by the tuple $(m_{12},m_{23},m_{31})$ that are not admissible - are virtually surface. 
Another notable example giving rise to a large replica symmetry is the group 
$(2,3,7)$. Its surface subgroup with smallest genus is ${\Theta}_3$. After quotienting $H^3$ by this $\Theta_3$, we again get a wormhole geometry but this time, connecting two genus $3$ boundaries. Each of the boundary components is a highly symmetric genus $3$ surface known as the Klein quartic.

\section{CFT computation}\label{cft-comp}

In this section we will discuss computation of multi-invariants from conformal field theory. The key formula that we will use is equation \eqref{uni-Z}. We reproduce this formula here for readers' convenience, 
\begin{align}\label{uni-Z-1}
    Z_{\CM_\CE} =  e^{-S_L^{(c)}[\phi]} \, Z_{\CM^{\rm uni}_\CE}. 
\end{align}
With this, $\CE= Z_{\CM_\CE}/(Z_\CM)^{n_r}$ where $n_r$ is the number replicas. 
As discussed in section \ref{cft-sec}, the quantity $Z_{\CM_\CE}$ can also be thought of as a correlation function of twist operators inserted at the boundaries of the party-regions. We will switch between these two view-points often. 

As explained in section \ref{cft-sec}, the formula \eqref{uni-Z-1} is most useful when the genus of $\CM_\CE$, denoted as $g(\CM_\CE)$ is $0$. In that case, we can use the universality of $Z_{S^2}$. In fact, we normalize the vacuum state such that $Z_{S^2}=1$. In this section, we will compute a number of symmetric multipartite invariants for which $g(\CM_\CE)=0$.
For higher genus surfaces, $Z_{\CM^{\rm uni}_\CE}$ is non-universal and depends on the details of the conformal field theory. For holographic CFTs we can compute $Z_{\CM^{\rm uni}_\CE}$ as the action of dominant handlebody solution filling in $\CM_\CE^{\rm uni}$. We will use this method to compute an invariant for which  $g(\CM_\CE)=1$. 

In the first part of the section, we will discuss computation of $e^{-S_L^{(c)}[\phi]}$ for surfaces that are specified by a covering map. A covering map is a complex map $\Gamma: \CM_\CE^{\rm uni} \to \CM$. 
This is particularly useful for us because the replicated surface $\CM_\CE$ for a multi-invariant $\CE$, naturally admits  a covering map to $\CM$ because the replicated surface $\CM_\CE$ is a branched cover of $\CM$. 
Given a covering map $\Gamma(z)$, the Weyl factor is given as $\phi(z)=\partial_z \Gamma(z)$. Calculation of the covering map for symmetric invariants for which $g(\CM_\CE)=0, 1$ is presented in section \ref{covering}. 

\subsection{Uniformization result}
In this section, we will give the formula for $e^{-S_L^{(c)}[\phi]}$  for a given genus $0$ covering map $\Gamma(z)$. 
Let us use the coordinates $z$ on $\CM_\CE^{\rm uni}$ and $x$ for $\CM$. 
Let the twist operators $\CO_{\sigma_a}$ be inserted on $\CM$ at positions $x_a, a=1, \ldots, q$. Let the twist operator  at $x_a$ correspond to a permutation element with $n/k_a$ cycles of length $k_a$. It has  $n/k_a\equiv m_a$ pre-images in  $\CM_\CE^{\rm uni}$, say at $z_{i, a}, i= 1, \ldots, m_a$. Strictly speaking the index $i$ should have a sub-index $a$ but we have dropped it to avoid clutter. It should be clear from the context which $a$ does the $i$ index belongs to. 
Let the expansion of the covering map at one of these pre-images be
\begin{align}\label{eq:cm_expansion}
    \Gamma(z)=x_a + \alpha_{i, a} (z-z_{i,a})^{k_{a}}+\ldots.
\end{align} 
The order of the twist operator at $x_a$ is $k_a$ and hence the order of ramification at this point is $k_a$. As a result, the expansion of the covering map near any of its pre-image starts from the $k$-th power of the local coordinate.  The Liouville action $e^{-S_L^{(c)}[\phi]}$ can now be computed for genus $0$ covering maps using the data $x_a, z_{i,a}, k_a$ and  $\alpha_{i, a}$. This calculation requires a careful regularization of the Liouville action $S_L$ and is done in detail in \cite{Lunin:2000yv, avery2010using}. We have reproduced important steps in this calculation in appendix \ref{l-uni-app} for readers' convenience.
Although it is not necessary, sometimes it is convenient to have a twist operator insertion at $x=\infty$. This case has to be treated a little differently from the case of twist operators inserted at finite points. Let  $\CO_{\sigma_q}$  be inserted at $x_q=\infty$. We also take one of its pre-images, say $z_{m_q, q}$,  to be $\infty$.
The expansion of the covering map at $z_{i,q}$ is, 
\begin{align}
    \Gamma(z)&= \alpha_{i, q} \,(z- z_{i,q})^{-k_q} +\ldots \qquad i=1, \ldots, m_q-1,\notag\\
    \Gamma(z)&= \alpha_{m_q, q} \, z^{k_q}.
\end{align}
As $x_q=\infty$ is treated differently, it is useful to introduce new label as follows: $\sigma_q\equiv \sigma_\infty$, $x_q=\infty \equiv x_\infty$, $k_q\equiv k_\infty$, $m_q\equiv m_\infty$, $\alpha_{i,q}\equiv  \beta_i, i= 1, \ldots, m_q-1$ and $\alpha_{m_q, q}\equiv  \beta_0$. 
With these definitions, the correlation function of normalized twist operators is
\begin{align}\label{l-formula}
    &\log \CE^{\rm reg} = \log \langle \CO^{\rm norm}_{\sigma_1}(x_1) \ldots \CO^{\rm norm}_{\sigma_{q-1}}(x_{q-1}) \CO^{\rm norm}_{\sigma_\infty}(\infty)  \rangle\notag\\
    &=-\frac{c}{12} \Big[ \sum_{a\neq q} \frac{k_a-1}{k_a}\log\Big(\prod_{i}^{m_a} |\alpha_{a,i}|\Big)+\frac{k_\infty+1}{k_\infty}\log\Big(\prod_{i}^{m_\infty-1} |\beta_{i}|\Big)- \frac{k_\infty-1}{k_\infty}\log\Big(|\beta_0|\Big) \notag\\
    &+\sum_a (n_r+m_a)\log (k_a) -(n_r-m_\infty+2)\log (k_\infty) 
    \Big].
\end{align}
Recall the definition of the normalized twist operators from section \ref{cft-sec}. Twist operators at finite points and at infinity are normalized as
\begin{align}
    \CO^{\rm norm}(x)=\CO^{\rm norm}/(\langle\CO(0)\CO(1)\rangle)^\frac12, \qquad \CO^{\rm norm}(\infty)= \lim_{x\to \infty} \CO^{\rm norm}(x) |x|^{2\Delta_\CO},
\end{align}
respectively. In the rest of the discussion, we will exclusively work with normalized twist operators and hence we will drop the superscript of  $\CO^{\rm norm}$.
Note that even if there is no twist operator inserted at $\infty$ i.e. even if $k_\infty=1$, we do get a non-zero contribution from all the finite  pre-images of $\infty$, namely from all the $\beta$'s except for $\beta_0$.

We are only interested in genus $0$ covering maps that result from symmetric invariants. We have seen in section \ref{fin-klein} that these invariants correspond to  finite Kleinian groups. Finite Kleinian groups  are precisely spherical groups. One of them is simply the cyclic group and corresponds to insertion of two twist operators. The rest are ${\mathbb Z}_2$ quotients of finite Coxeter groups with three nodes. They are given by the Coxeter tuples \eqref{spherical-tuples}.  They correspond to insertion of three twist operators and hence are computed by three point function of the appropriate twist operators. 

As the simplest example of the genus $0$ covering map, consider the bi-partite invariant with replica symmetry ${\mathbb Z}_n$. It is the familiar, ${\rm Tr}\rho^n$. It corresponds to two point function of two twist operators, $\CO_\sigma$ and its hermitian conjugate $\CO_{\sigma^{-1}}$, where $\sigma$ consists of a single cycle of length $n$. For convenience, we take $x_1=0$ and keep $x_2$  arbitrary. We do not have any twist operator inserted at $\infty$. Alternatively, to make the formula \eqref{l-formula} applicable, we take $k_\infty=1$. There are $n-1$ finite pre-images of 
\begin{align}
    \Gamma(z)=x_2\frac{z^n}{z^n-(z-1)^n}.
\end{align}
The point $x_1$ and $x_2$ each have a single pre-image, at $z=0$ and $z=1$ respectively. This means $k_1=k_2=n$ and $m_1=m_2=1$. 
The finite pre-images of $\infty$ are at $z_{\infty, j}=1/(1-\omega^j), j=1, \ldots, n-1, \omega= e^{2\pi i/n}$. Expanding at these points
\begin{align}
    \Gamma(z)|_{z=0}&= (-1)^{n+1} x_2 z^n,\qquad \Gamma(z)|_{z=1}=x_2 +x_2(z-1)^n,\notag\\
    \Gamma(z)|_{z=z_{\infty, j}}&=-\frac{x_2}{n}\frac{\omega^j}{(1-\omega^j)^2}\frac{1}{z-z_{\infty, j}}.
\end{align}
We read off,  $\alpha_{1,1}=(-1)^{n+1}x_2, \alpha_{1,2}=x_2$ and $\beta_j= (x_2/n)(\omega^j/(1-\omega^j)^2) $. 
Substituting in the formula \eqref{l-formula}, 
\begin{align}
    \log \CE^{\rm reg}= \log \langle \CO_\sigma(0) \CO_{\sigma^{-1}} (x_2)\rangle=-\frac{c}{6}\Big(n-\frac{1}{n}\Big)\log |x_2|. 
\end{align}
The coefficient in from of $\log|x_2|$ is precisely twice the conformal dimension of the twist operator with a single of cycle of length $n$. This is as expected of a correlation function where the twist operators are normalized canonically. 
The conformal dimension of a general twist operator can also be found using this method. Effectively a twist operator with multiple cycles is the normal ordered product of twist operators for each cycle. As a result the conformal dimension of a twist operator consisting of $p_k$ number of length $k$ cycles is,
\begin{align}
    \Delta_\sigma=\sum_k p(k) \frac{c}{12}\Big(k-\frac{1}{k}\Big).
\end{align}
Specializing to the case when the twist operator consists of cycles of fixed length $k_\sigma$ then $p(k_\sigma)=n_r/k_\sigma$. So we have,
\begin{align}
    \Delta_\sigma= \frac{n_r c}{12}\Big(1-\frac{1}{k_\sigma^2}\Big).
\end{align}

The rest of the spherical groups correspond to having three insertions of twist operators. Thanks to conformal invariance, the position dependence of this three point function of twist operators is completely fixed.
\begin{align}
    \langle \CO_{\sigma_1}(x_1) \CO_{\sigma_2}(x_2) \CO_{\sigma_3}(x_3)\rangle= C^{\rm OPE} x_{12}^{\Delta_3-\Delta_1-\Delta_2} x_{23}^{\Delta_1-\Delta_2-\Delta_3} x_{13}^{\Delta_2-\Delta_1-\Delta_3}
\end{align}
The coefficient $C^{\rm OPE}$ is undetermined and is known as the operator product expansion  coefficient. If we take $(x_1, x_2, x_3)$ to be at $(0,1,\infty)$. Then the position dependence part becomes $1$ and we get,
\begin{align}
    \langle \CO_{\sigma_1}(0) \CO_{\sigma_2}(1) \CO_{\sigma_3}(\infty)\rangle=C^{\rm OPE}.
\end{align}
For convenience, we will fix $(x_1, x_2, x_3)$ to be at $(0,1,\infty)$ and compute $\CE^{\rm reg}=e^{-S_L}=C^{\rm OPE}$. The position dependence can be restored using conformal symmetry. 
Now we move to the computation of the covering map for the covering maps for invariants corresponding to spherical groups. 
 
\subsection{Covering map}\label{covering}
For symmetric invariants the covering maps are defined using the replica group action. We can use the invariance of the covering maps under the group action to determine them completely. We will see how to do it in the case of the Dihedral group $\mathbb{D}_{2n}$. The extended replica symmetry in this case is the Coxeter group with the tuple $(2,2,n)$. This means the extended replica symmetry group is generated by three reflections $r_1, r_2, r_3$ and  the three twist operators corresponds to the group elements $\sigma_1= r_1r_2$, $\sigma_2 =r_2r_3$ and $\sigma_3= r_3 r_1$. They have orders $2, 2$ and $n$ respectively. The order of the replica group $\mathbb{D}_{2n}$ -  which is the total number of replicas - is $2n$. The twist operators written explicitly as permutation elements in $\mathbb{S}_{2n}$ are
\begin{align}
    \sigma_1 &= (1, 2n)(2, 2n-1)(3, 2n-2)\ldots(n, n+1)\notag\\
    \sigma_2 &= (1,n+1)(2, 2n)(3, 2n-1)\ldots (n, n+2) \notag\\
    \sigma_3 &= (1, 2, \ldots, n)(n+1, n+2, \ldots, 2n).
\end{align}
The number of cycles in $\sigma_1, \sigma_2$ and $\sigma_3$ are $n, n$ and $2$ as expected. This means that the number of pre-images of $x_1, x_2$ and $x_3$ is also $n, n$ and $2$ respectively. 
In section \ref{dihedral1}, we have explicitly constructed the reflection generators $r_1, r_2$ and $r_3$ on the covering sphere. Their fixed point loci are given by three circles (including straight lines which is a circle with infinite radius). These circles and their images under reflection across each tesselate the sphere into triangles. The points where the  circles and their images intersect are exactly the pre-images of the location of the twist operators. Hence they are also the fixed points of the elliptical symmetry generators. 
From subfigure 1 in figure  \ref{fin-cox-fig}, we can simply list the pre-images of each of the twist operators. Letting $s_a$ be the set of pre-images of $x_a$ we have,
\begin{align}
    s_1=\{1, \omega^2, \omega^4, \ldots, \omega^{2(n-1)}\},\qquad s_2=\{\omega, \omega^3, \ldots, \omega^{2n-1}\},\qquad s_3=\{0,\infty\},
\end{align}
where $\omega=e^{i\pi /n}$.  

A property of covering maps is that they are \emph{automorphic} with respect to the Kleinian group $\CK$ which defines the fundamental region. That is:
\be
\Gamma\left(f(z)\right)=\Gamma(z), \quad \forall f(z)=\frac{az+b}{cz+d} \in \CK
\ee
This leads to a number of consequences \cite{ford_automorphic_1929}:
\begin{itemize}
    \item Within the fundamental region the covering map must take each value of $\hat{\mathbb{C}}$ exactly once.
    \item A point is an elliptical fixed point of $\CK$ of order $k_a$ iff the expansion of any automorphic function around that point is of the form \eqref{eq:cm_expansion}. Such points occur only at the vertices of the extended fundamental region and the images of these vertices under $\hat \CK$. The angle of the extended fundamental region at that vertex is $\frac{\pi}{k_a}$.
    \item All covering maps of genus $g=0$ are rational functions. Such a function is unique up to conformal isometries.
\end{itemize}

Make use of this the covering map can be constructed by constructing polynomials $p_a(z)$ such that the set of its roots is the set $s_a$, each root occurring  with multiplicity $k_a$. 
\begin{align}
    p_1(z)=(z^n-1)^2,\qquad p_2(z)=(z^n+1)^2, \qquad p_3(z)= z^n. 
\end{align}
We have note included the point $z=\infty$ in $s_3$ into the root set of $p_3(z)$. As we will see now, it is not necessary. 
We then impose the equations
\begin{align}
    \Gamma(z)=A\frac{p_1(z)}{p_3(z)}, \qquad 1-\Gamma(z)= B\frac{p_2(z)}{p_3(z)}.
\end{align}
on the covering map $\Gamma(z)$. 
This ensures that the points in the set $s_1, s_2$ and $s_3$ are mapped to $0, 1$ and $\infty$ as desired. The multiplicity of each root ensures the correct order of ramification around each point. From this equation, 
\begin{align}
    A p_1(z)+B p_2(z) = p_3(z).
\end{align}
This equation needs to obey for all values of $z$. This may seem like very stringent condition, however the solution to it always exists. This is because of the mathematical theorem which states that a covering map is of genus $0$ if and only if it is a rational function.
Verifying the equation at $z=0, 1$ we get $A= -1/4$ and $B=1/4$. So we have the covering map,
\begin{align}\label{eq:dihcovering}
    \Gamma_{(2,2,n)}(z)=-\frac{(z^n-1)^2}{4z^n}. 
\end{align} 
Expanding the map at the pre-images of $0, 1$ and $\infty$, we can compute the coefficients $\alpha$'s and $\beta$'s. Substituting in equation \eqref{l-formula},
\begin{align}
    \log C^{\rm OPE}_{(2,2,n)}\rangle= -\frac{c}{3}\left(n-\frac{1}{n}\right) \log 2.
\end{align}

The OPE coefficient $C^{\rm OPE}$ for twist operators corresponding to other spherical groups is evaluated in the same way. The covering maps\footnote{For the full derivation of the covering maps using geometric properties of the platonic solids see \cite{Sansone1969-fm}.} and OPE coefficients in each of these cases are
\begin{table}[H]
\begin{center}
    \begin{tabular}{ |c||l|l|l| } 
     \hline
     ${\hat \CR}$ & $\Gamma(z)$ & $\frac{1}{c}$($\Delta_1, \Delta_2, \Delta_3$) & $\frac{1}{c}\log C^{\rm OPE}$ \\
     \hline \hline
     $(2,2,n)$ & $-\frac{(z^n-1)^2}{4z^n}$ &($\frac{n  }{8},\frac{n  }{8},\frac{n^2-1}{6n}$) & $-\frac{1}{3}(n-\frac{1}{n}) \log 2$  \\ 
     \hline
     $(2,3,3)$ & $\frac{(z^4+2i\sqrt{3}z^2+1)^3}{12i\sqrt{3} z^2(z^4-1)^2}$ &($\frac34, \frac89, \frac89$)&$-\frac{1}{2}\log 2 - \frac{9}{8} \log 3$  \\ 
     \hline
     $(2,3,4)$ & $\frac{1}{108}\frac{(z^8+14 z^4+1)^3}{z^4(z^4-1)^4}$ & ($\frac32, \frac{16}{9}, \frac{15}{8}$)&$-\frac{119}{36}\log 2 - \frac{9}{8}\log 3$  \\ 
     \hline
     $(2,3,5)$ & $-\frac{1}{1728}\frac{(z^{20}-228 z^{15}+494z^{10}+228 z^{5}+1 )^3}{z^5(z^{10}+11z^{5}-1)^5}$ &($\frac{15}{4}, \frac{40}{9}, \frac{24}{5}$) &$-\frac{1668 }{360}\log 2-\frac{1134}{360}\log 3-\frac{625}{360}\log 5$  \\ 
     \hline
    \end{tabular}
    \end{center}
    \caption{Covering maps, dimensions of the twist operators and the OPE coefficient $C^{\rm OPE}$ for all the spherical groups}
    \label{spherical-table}
\end{table}

\subsubsection{Schwarz triangle function}
Another way to find the covering map in the above cases is to use the so-called Schwarz triangle function. It is a function that conformally maps the upper half plane to a triangle in the upper half plane having lines or circular arcs for edges. The vertices of the triangle form angles $\pi \alpha, \pi\beta, \pi\gamma$ respectively. Denoting such a map as $S_{\alpha, \beta, \gamma}(x)$, the covering map is its inverse,
\begin{align}\label{from-stf}
    \Gamma_{(m_{12}, m_{23}, m_{31})}(z)=S^{-1}_{1/m_{12}, 1/m_{23}, 1/{m_{31}}}.
\end{align} 
This map is given explicitly as
\begin{align}
    &S_{\alpha, \beta, \gamma}(x)=z^\alpha \frac{\,_2F_1(a',b',c';x)}{\,_2F_1(a,b,c;x)}\notag\\ 
    &a= (1-\alpha-\beta-\gamma)/2, \quad b=(1-\alpha+\beta-\gamma)/2, \quad c= 1-\alpha\notag\\
    &a'= (1+\alpha-\beta-\gamma)/2, \quad b'=(1+\alpha+\beta-\gamma)/2, \quad c'= 1+\alpha.
\end{align}
The covering map that we get from equation \eqref{from-stf} agree with the ones given in table up to conformal transformation. 

\subsubsection{Genus $1$}\label{ssec:g1}
It turns out that the formula \eqref{l-formula} can be adapted to the case of covering maps that are genus one. 
\begin{align}\label{l-formula-1}
    &\log \CE^{\rm reg} = \log \langle \CO^{\rm norm}_{\sigma_1}(x_1) \ldots \CO^{\rm norm}_{\sigma_{q-1}}(x_{q-1}) \CO^{\rm norm}_{\sigma_\infty}(\infty)  \rangle\notag\\
    &=-\frac{c}{12} \Big[ \sum_{a\neq q} \frac{k_a-1}{k_a}\log\Big(\prod_{i}^{m_a} |\alpha_{a,i}|\Big)+\frac{k_\infty+1}{k_\infty}\log\Big(\prod_{i}^{m_\infty-1} |\beta_{i}|\Big)- \frac{k_\infty-1}{k_\infty}\log\Big(|\beta_0|\Big) \notag\\
    &+\sum_a (n_r+m_a)\log (k_a) -(n_r-m_\infty)\log (k_\infty) 
    \Big]+\log Z_{\rm torus}.
\end{align}
The only difference is the lack of $2$ multiplying $\log(k_\infty)$ in equation \eqref{l-formula} and of course the additional factor of $\log Z_{\rm torus}$.

In studying symmetric invariants, the genus $1$ covering maps appear when we consider four party regions and the neighboring regions being separated by twist operators with order $2$ i.e. with $k_\sigma=2$. The extended Kleinian group in this case corresponds to the Coxeter group whose Dynkin diagram is disconnected with four nodes, with pairs of nodes connected by edges with integer labels $m$ and $n$. The order of the replica group is $n_r=2mn$. In summary we have the correlation function four twist operators, each with $k_\sigma=2$ and $m_\sigma=n_r/k_\sigma= mn$. We will first consider the case with $m=1$\footnote{This coincides with the computable cross norm negativity (CCNR) see for example \cite{PhysRevD.108.054508,2023PhRvL.130m1601Y, 2022arXiv221211978M}. }.

We insert three of the twist operators  at $0,1, \infty$ and the fourth one at $0\leq \eta\leq 1$.  The covering map from the torus to the sphere is such that the rectangular fundamental domain of the extended Kleinian group $\hat \CK$ is mapped to the upper half plane. Analogous to the Schwarz triangle function, the inverse map that maps the upper half plane to a given rectangle is written using the Schwarz-Christoffel mapping,
\begin{align}
    f(x)=-\int_{x}^\infty dz\frac{1}{2\sqrt{(z-e_1)(z-e_2)(z-e_3)}}, \qquad e_1+e_2+e_3=0.
\end{align} 
which can be analytical continued to the entire complex plane. 
The inverse of this function is the so called Weierstrass elliptic function $\wp(z)$. It obeys the differential equation,
\begin{align}
    \wp'(z)^2=4\wp^3(z)-g_2\wp(z)-g_3,\quad g_2\equiv -4(e_1e_2+e_2e_3+e_3e_1), g_3\equiv 4 e_1e_2e_3.
\end{align}
The $\wp(z)$ function is doubly periodic, as expected of a covering map of the torus, with periods $(\omega_1, \omega_2)=(1,\rho)$ where $\rho=ik$ is a purely imaginary number. It also obeys,
\begin{align}
    \wp(0)=\infty, \quad \wp\left(\frac12\right)=e_1, \quad \wp\left(\frac{\rho}{2}\right)=e_2,\quad \wp\left(\frac12(1+\rho)\right)=e_3.
\end{align}
We would like to transform the covering map so that the pre-images of the twist operator locations $(0,1,\eta, \infty)$ are given by $(\frac{\rho}{2}, \frac12, \frac12(1+\rho), \infty)$ respectively. The other $n-1$ pre-images are given by translating the above points by $j\rho, j=1,\ldots, n-1$. 
The following covering map $\Gamma(z)$ does the job
\begin{align}
    \Gamma(z)=\frac{\wp(z)-e_2}{e_1-e_3}.
\end{align}
The cross-ratio $\eta$ and the lattice period parameter $\rho$ are related to each other by the maps,
\begin{align}
    \eta\equiv \frac{x_{12}x_{34}}{x_{13}x_{24}}= \frac{e_3-e_2}{e_1-e_2}=\Big(\frac{\theta_2(\rho)}{\theta_3(\rho)}\Big)^4,\quad \rho=i\frac{K(1-\eta)}{K(\eta)}.
\end{align}
Here $\theta$'s are the standard Jacobi theta functions and $K$ is the complete elliptic integral of the first kind. 
The expansions of the covering map to leading order at the pre-images of the twist operators is
\begin{align}
    \begin{aligned}
    &\frac{1}{e_1-e_2}\frac{1}{z^2}, \quad &z&=0\\
   &\frac{1}{e_1-e_2}\left(3e_i^2-\frac{1}{4}g_2\right)(z-z_i)^2 ,\quad &z_i&=\frac{1}{2}\{1,\rho,1+\rho\}
    \end{aligned}
\end{align}
Substituting in equation \eqref{l-formula-1},
\begin{align}
    \log&(\langle\CO_{\sigma_1}(0)\CO_{\sigma_2}(\eta)\CO_{\sigma_3}(1)\CO_{\sigma_4}(\infty)\rangle)\\
&=-\frac{c}{12}\Biggr\{
\frac{n}{2}\log |\alpha_0||\alpha_\eta||\alpha_1|
+\frac{3n}{2}\log |\beta_1|
+8n\log 2\Biggr\}+\log(Z_{\text{torus}})\\
&=-\frac{cn}{12}\log(\eta(1-\eta))-\frac{2cn}{3}\log2+\log(Z_{\text{torus}})
\end{align}
Here we have used the fact
\begin{align}
    \prod_{i=1}^3|3e_i^2-\frac14 g_2|=\frac{\Delta}{16}, \qquad \quad \Delta=g_2^3-27g_3^2=16(e_1-e_2)^6|\eta|^2|1-\eta|^2.
\end{align}
The quantity $\Delta$ is known as the modular discriminant. 

\subsubsection*{$Z_{\rm torus}$ from Holography}
As remarked in section \ref{cft-sec}, the term $Z_{\rm torus}$ is not universal i.e. depends on the details of the theory. It requires a separate computation. In this section, we will compute it from Holography assuming that the 2d CFT is holographic. 
The action of the dominant bulk solution that fills in a torus with modular parameter $\tau$ is given by \cite{1998JHEP...12..005M, 2020JHEP...02..170D}
\begin{align}
    S_{\rm grav}(\tau)=\min_{a,b,c,d\in\mathbb{Z},
    \;ad-bc=1}\left[\frac{i\pi c}{12}\left(\frac{a\tau+b}{c\tau+d}-\frac{a\overline{\tau}+b}{c\overline{\tau}+d}\right)\right], \quad Z_{\text{torus}}(\tau)=e^{-S_{\rm{grav}}(\tau)}.
\end{align}
There are infinitely many bulk solutions that fill a torus. This corresponds to the choice of the cycle that becomes contractible in the bulk. We have to compute action for each of these choices and minimize the gravitational action over them. This is what the above formula does. 
Here the lattice parameter $\rho$ and the modular parameter of the torus replica manifold are related by $\tau=n\rho=ink$. 
\be
\begin{split}
S_{\text{grav}}(\tau)=\min_{a,b,c,d\in\mathbb{Z},
\;ad-bc=1}\left[\frac{i\pi c}{12}\frac{2n\rho}{d^2-c^2n^2\rho^2}\right]
=-\max_{a,b,c,d\in\mathbb{Z},
\;ad-bc=1}\left[\frac{\pi c}{6}\frac{2nk}{d^2+c^2n^2k^2}\right]
\end{split}
\ee
The contributing phases correspond to $c=0,d=1$ and $c=1,d=0$ with the phase transition occurring when $nk=1$. This occurs when $\tau=i$ or $\rho=\frac{i}{n}$ which happens when the cross-ratio reaches the value
\be
\eta_*=\left(\frac{\theta_2(\frac{i}{n})}{\theta_3(\frac{i}{n})}\right)^4
\ee
which is increasing with $n$. 
\be
\log(Z_{\text{torus}})=
\begin{cases}
        \frac{\pi nc}{6}\frac{K(1-\eta)}{K(\eta)} & \eta\leq \eta_*\\
        \frac{\pi c}{6n}\frac{K(\eta)}{K(1-\eta)}  & \eta\geq \eta_*
    \end{cases}
\ee
As the torus that we are working with is rectangular, the phase transition occurs if its length becomes larger than its width and vice versa. In order to produce the orbifold solution for both these phases, we have to use different Schottky groups. As remarked in section \ref{sym-handle}, the choice of the filling is determined by the Schottky representation. In particular, the circles $C_i$ and $C_i'$ defining the Schottky presentation become contractible in the bulk. On both sides, choosing the right Schottky presentation, we get the following orbifold geometries. They are precisely in correspondence with the filled cycle and hence the choice of the Schottky elements.  
The system prefers to stay in the connected phase for longer as $n$ increases.
\begin{figure}[h]
  \centering
   \includegraphics[width=.8\textwidth]{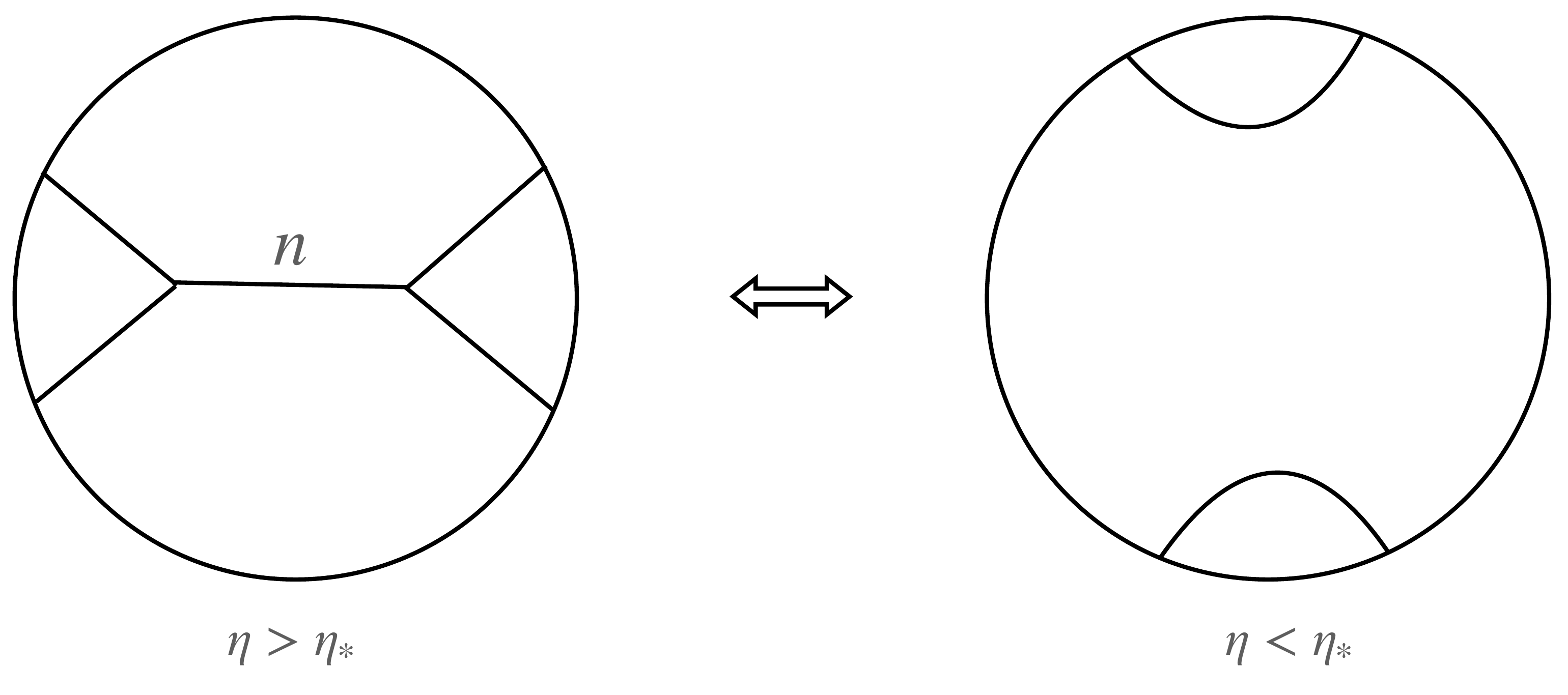}
  \caption{Expected phases for bulk orbifold} 
\label{fig:4pt_sphere_phases}
\end{figure}

\subsubsection*{General $m$}
The four point function of the twist operator is readily extended to general values of $m$. In this case, the modular parameter of the torus is $\tau=\frac{m}{n}\rho$. We get,
\begin{align}
    \log&(\langle\sigma(0)_{g_0}\sigma(\eta)_{g_\eta}\sigma(1)_{g_1}\sigma(\infty)_{g_\infty}\rangle)\\
&=-\frac{cnm}{12}\log(\eta(1-\eta))-\frac{2cnm}{3}\log2+\log(Z_{\text{torus}}).
\end{align}
where the torus partition function is given by
\be
\log(Z_{\text{torus}})=
\begin{cases}
        \frac{\pi cm}{6n}\frac{K(1-\eta)}{K(\eta)} & \eta\leq \eta_*\\
        \frac{\pi cn}{6m}\frac{K(\eta)}{K(1-\eta)}  & \eta\geq \eta_*
    \end{cases}, \quad \eta_*=\left(\frac{\theta_2(\frac{in}{m})}{\theta_3(\frac{in}{m})}\right)^4
\ee

\begin{figure}[h]
    \centering
     \includegraphics[width=.8\textwidth]{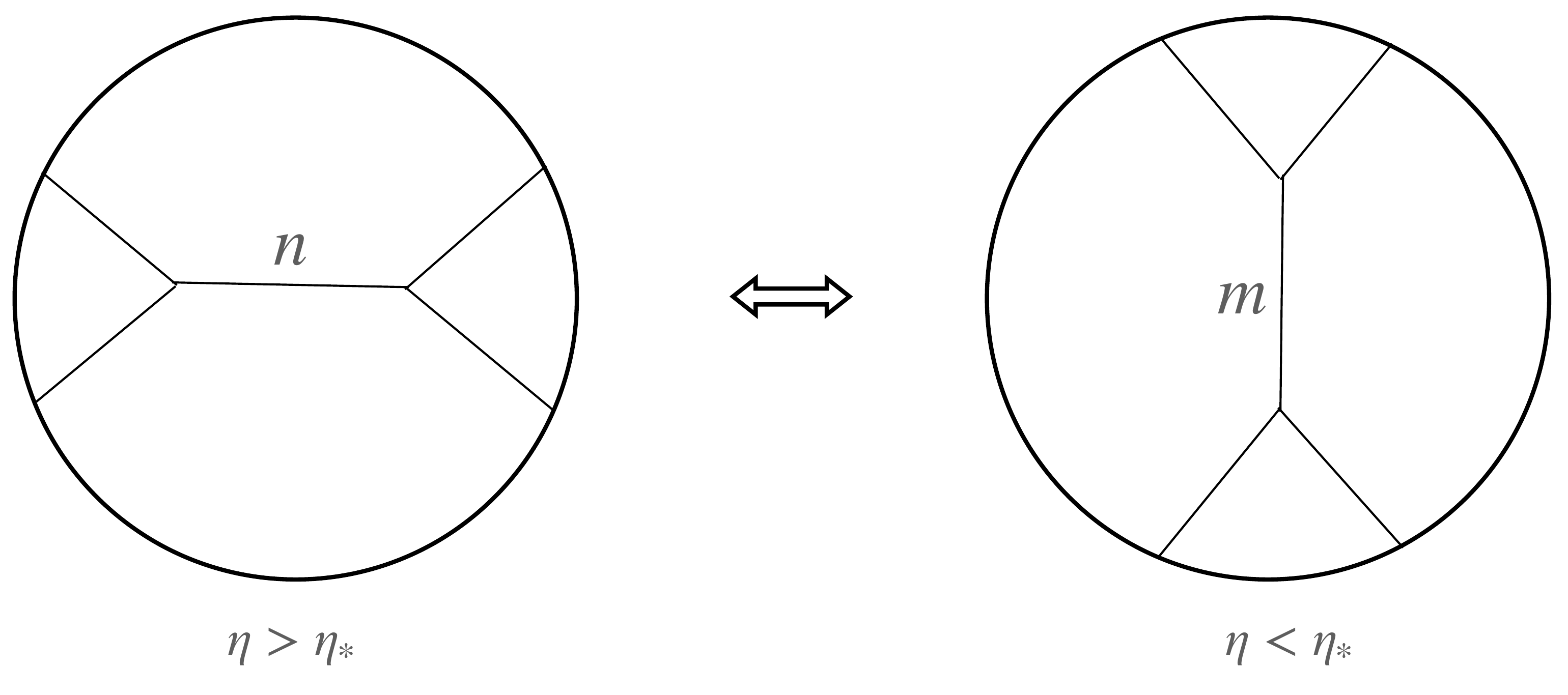}
    \caption{Expected phases for bulk orbifold} 
  \label{fig:nmtorus4locus}
  \end{figure}

\begin{figure}[h]
\begin{tabular}{ccc}
\centering
\includegraphics[width=.29\textwidth]{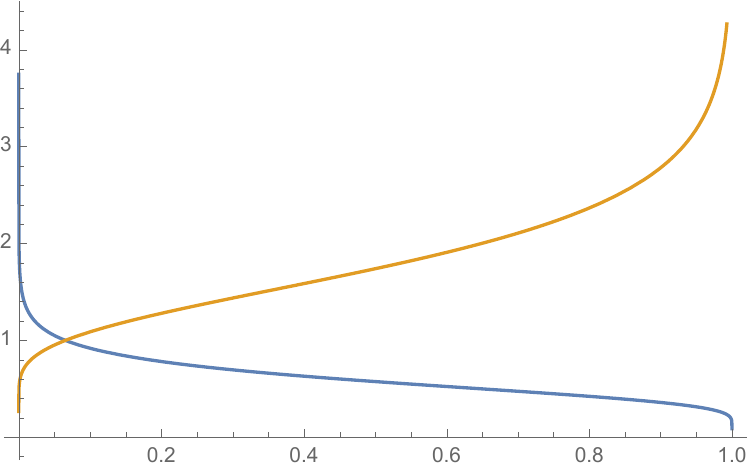}&\quad
\includegraphics[width=.29\textwidth]{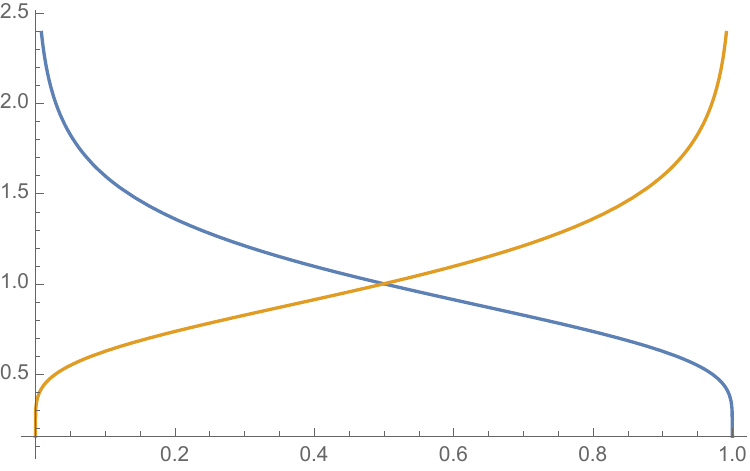}&\quad
\includegraphics[width=.29\textwidth]{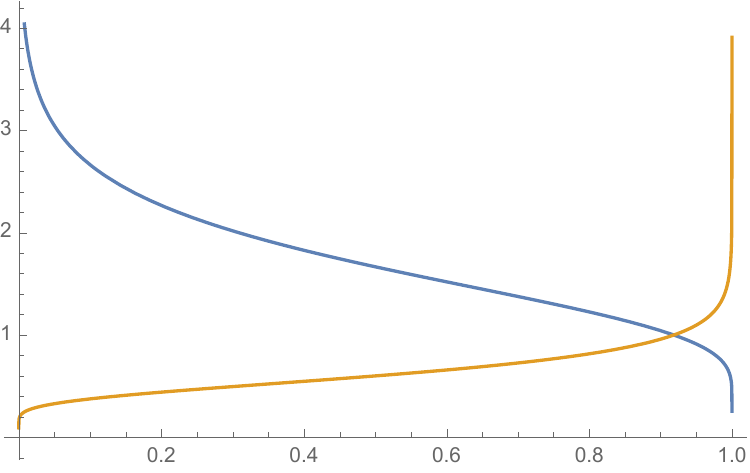}
\end{tabular}
\caption{\label{fig:mntorusphases} $\log(Z_{\text{torus}})$ as a function of cross ratio $\eta$. The value corresponding to the internal singular locus of order $n$ is shown in orange and $m$ in blue. L: $\frac{n}{m}\geq 1$ M: $\frac{n}{m}= 1$ R: $\frac{n}{m}\leq 1$. For each choice of cross ratio the correct phase will be the one of larger value. When $n$ and $m$ are not the same the phase corresponding to $\max(n,m)$ will be maximal over a large  range of the cross-ratio.}
\end{figure}

\subsection{Check: Bulk computation for spherical groups }\label{bulk-orb-compute}
In this section, we check the invariants $\CE^{\rm reg}$ for spherical groups computed in table \ref{spherical-table} against gravity calculation. The gravitation calculation is that of the action of the orbifolds with the singular locus that is a tri-valent junction shown in figure \ref{fin-cox-orb}. The gravitational action for the ``cone-manifolds''  consisting a tri-valent junction of conical singularities  with arbitrary angles is evaluated in \cite{2016arXiv160401774C}. We borrow their result here. 

Let us assume that the $2d$ CFT we are working with is dual to three dimensional Einstein gravity. In this theory, a mass $M$ sources a conical singularity around its world-line in $3d$ gravity. The mass $M$, conformal dimension $\Delta$ of the dual CFT operator and the conical angular deficit $4\pi \eta$ are related to each other as
\begin{align}
    \Delta= \frac{c}{12}M, \qquad \eta=\frac12(1-\sqrt{1-M})\qquad \Rightarrow \qquad \Delta=\frac{c}{3}\eta(1-\eta). 
\end{align}
In \cite{2016arXiv160401774C}, the gravitation action is evaluated for general values of the cone angle i.e. for general values of $\Delta$. Eventually, we will be interested in specializing to the $\Delta$ values appearing in Table \ref{spherical-table}, in order to check the CFT results. Note that for $M>1$, the conical definite turns complex. This is precisely the black-hole threshold. Expectedly, it corresponds to $\Delta=c/12$. We will be working with $M<1$ so that the mass sources a conical singularity and does not form a black hole. 

In \cite{2016arXiv160401774C}, the authors use the rewriting of $3d$ gravity in terms of $2d$ Liouville theory. For this, it is useful to foliate the metric by equidistant slices from the tri-valent junction of singularity.
The metric is parametrized as
\begin{align}
    ds^2=\frac{4}{1-r^2}(dr^2+r^2 e^{\phi(x,\bar x)} dx d\bar x). 
\end{align}
The singularity is imposed by the following boundary conditions on $\phi$.
\be
\phi(z,\bar{z})\rightarrow
\begin{cases}
-2\log|x|^2 & x \rightarrow \infty\\
-2\eta_i\log|x-x_i| & x\rightarrow x_i
\end{cases}
\ee
Einstein equations  reduce to the Liouville equations, away from the boundary
\be
\partial\bar{\partial}\phi=-\frac{1}{2}e^{\phi}.
\ee
This equation is solved and the action is obtained in \cite{2016arXiv160401774C}. We will reproduce their answer below
\begin{align}
    -S_{\rm grav}(\tilde{\mathcal{B}}_{\mathcal{E}})&=\frac{c}{6}(F(2\eta_1)-F(\eta_2+\eta_3-\eta_1)+(1-2\eta_1)\log(1-2\eta_1)+\text{perms}\notag\\
    &+F(0)-F(\sum_i\eta_i)-2(1-\sum_i\eta_i)\log(1-\sum_i\eta_i))\notag\\
    {\rm where}\qquad F(y)&=-G(y)-G(1-y), \quad G(y)=\int^{y}_{\frac{1}{2}}\log\Gamma(z)dz
\end{align}
This formula appears daunting due to the appearance of the integral of the $\Gamma$ function. However, we have checked that it correctly reproduces $C^{\rm OPE}$ given in Table \ref{spherical-table} after specializing to appropriate conformal dimensions and using
\begin{align}
    \log \langle\CO_{\sigma_1}(x_1)\CO_{\sigma_2}(x_2)\CO_{\sigma_3}(x_3)\rangle = -S_{\rm grav}({\mathcal{B}}_{\mathcal{E}}) = -n_r\, S_{\rm grav}(\tilde{\mathcal{B}}_{\mathcal{E}}).
\end{align}

\section{Outlook}\label{outlook}
In this paper we have constructed a family of multi-invariants for $2d$ conformal field theories which have the property that their dual $3d$ geometry preserves replica symmetry. We have done so by first analyzing the bulk geometries that have non-trivial isometries. The theory of Kleinian group played an important role in this analysis.  Importantly, the methods presented, following from the works of \cite{10.1112/plms/s3-59.2.373, reni-zim, reni-zim2}, make clear that from only the monodromies of twist operators it is possible to fully diagnosis whether a measure will have a dual replica symmetry preserving bulk solution. We present some possible future directions:

\paragraph{Further connections with quantum information theory}
In quantum information theory there is a large zoo of entanglement measures that have been defined and proposed. In many cases they involve optimizations over infinite spaces of auxiliary states (e.g. purifications) which makes them, even though theoretically well motivated, in practice nigh impossible to calculate even in simple systems. In stark contrast, since multi-invariants are fully determined just by the state of the system and the operation of partial trace, they are straightforward to calculate and thus should have broad applications beyond holography to other quantum systems. Even so one would desire a better understanding of their properties and connections to quantum information operations (distillation, recovery protocols, etc.) In particular it would be interesting to attempt something similar to what was done for the reflected entropy and Markov gap in \cite{Hayden:2021gno}.

\paragraph{Equalities}
It also becomes clear from our approach to bulk replica symmetry (see section \ref{sym-handle}) that there are infinitely many infinitely large families of (normalized) multi-invariants such that each family evaluates identically on the holographic state. These equalities can be used as a tool to diagnose the holographic nature of any quantum state, since a general quantum state might violate most of them.  On a similar note, there has been a significant interest in understanding the space of holographic states by studying the holographic entropic cone \cite{Bao:2015bfa, Hubeny:2018trv, 2019PhRvD.100b6004H, 2019JHEP...10..118H, Czech:2022fzb, Bao:2024azn, 2022CmPhy...5..244C, 2024JHEP...08..238H, 2024arXiv240113029C}. It would be interesting if possible to devise a complete set of equalities and inequalities that can be used as a diagnostic tool to identify quantum states which exhibit holographic nature. Interestingly, these equalities seem to only require the saddle point nature of gravitational path integral. In other words, the dual $2d$ CFT needs to have a large central charge, but not necessarily a large gap in the spectrum. It would be interesting to check the robustness of this results by considering a string theory in $AdS_3$. The D1-D5 brane system that gives rise to the duality between symmetric product orbifold and $AdS_3\times S_3\times {\mathbb T}^4$ might be a good testing ground. Also, it would be interesting to compute the replica preserving invariants in free large $N$  gauge theories or free $O(N)$ vector models at large $N$ to see if the equalities between normalized invariants remain unaffected.

\paragraph{Higher genus calculations}
The explicit calculation of multi-invariants requires knowledge of the partition function of the covering space and the covering map. In the case of genus 0 and 1 both of these quantities are explicitly known. In fact in this paper we have already completed calculations for measure of all possible boundary signature corresponding to a genus 0 or 1 covering space. As such to explore other measures it vital to begin to attempt to generalize to the case of higher genus. As a first step the class of measure presented in subsection \ref{ss:d4d6} with boundary signature (2,2,2,3) are maximally symmetric with handlebodies saturating the automorphism bound $12(g-1)$. The lowest genus $g=2$ case in particular seems like the natural invariant to next pursue for explicit calculation. We leave this to on-going work.

\paragraph{Phase transitions and replica symmetry breaking}
Considering a generic replica symmetry preserving multi-invariant it becomes clear that by considering the incidental limit of twist operators often measures must transition to a replica symmetry breaking phase. This will occur whenever the fusion of the twist operators will cause the resulting boundary orbifold to no longer permit a graph of orbifold singularities built from the spherical types. It would be interesting to explore this phenomenon and more broadly general phase transitions between handlebody solution.

\paragraph{Higher dimensions}
In this paper, we entirely focused on multi-partite entanglement in two dimensional CFTs. What about higher dimensional CFTs? 

Even in higher dimensions, the manifold $\CM_\CE$ is constructed following the same cutting and pasting prescription along the party regions as specified by the permutation tuple $(g_1,\ldots, g_\tq)$ just as in two dimensions. As in two-dimensions, the partition function can also be understood as a correlation function of twist operators supported on the boundary of regions which are co-dimension $2$. 

Can we perform a similar analysis in that case, based on higher dimensional Kleinian group. The answer to this question is yes. All the mathematical analysis proceeds parallelly to that of the $2d$ CFT. The only issue is in connecting the mathematical analysis to physical questions. Let us elaborate. 
One of the important steps in the analysis of multi-invariants in $2d$ CFTs is to uniformize the replicated $\CM_{\CE}$. The uniformized space $\CM_\CE^{\rm uni}$ is then obtained as a Schottky quotient along with the dominant handlebody that fills it. 

For higher dimensional case, it is a priori not clear to us in which cases can the replicated manifold $\CM_{\CE}$  be ``uniformized'', if at all. If it can be uniformized then we $\CM_\CE^{\rm uni}$ i.e. if the space $\CM_{\CE}$ with conical excesses at co-dimension two loci is conformally equivalent to a smooth space $\CM_\CE^{\rm uni}$ then we can engineer the bulk replica symmetry preserving invariants in the same way as in two dimensions. 

The Schottky construction generalizes to higher dimensions straightforwardly. We pick $g$ pairs of $S^{d-1}$'s on $S^{d}$ such that their interiors are disjoint. For a pair $a$, consider the conformal transformation $h_a$ that maps the interior of one to the exterior of the other. As before, these conformal transformations generate a free group with $g$ generators. The fundamental domain of this group is $S^{d}$ with the interiors of each of the $2g$  $S^{d-1}$ removed. Quotienting by this group produces the analog of genus $g$ surface in higher dimension. More precisely, the topology of the resulting manifold is the direct sum of $g$ copies of $S^{d-1}\times S^1$. For $d=2$, this gives the genus $g$ surface as expected. 
Extending this action uniquely to the hyperbolic ball that is bounded by $S^d$ and quotienting by it, produces the analog of genus $g$ handlebody in higher dimension. This is the direct sum of $g$ copies of $H^{d}\times S^1$. We will call this higher dimensional handlebody.   For $d=2$, this is indeed the topology of a genus $g$ handlebody. It plays the role of the dominant gravitational solution filling in $\CM_\CE^{\rm uni}$.

Just like in two dimensions we would obtain the handlebodies'' with non-trivial symmetry by considering virtually free higher Kleinian group $\CK$ i.e. subgroups of $SO(d+1,1)$ which contain a free group as its normal subgroup. This subgroup would play the role of the Schottky group $\CS$. Quotienting $H^{d+1}$ by this Schottky group produces the symmetric handlebody. The  quotient group $\CK/\CS$ is the symmetry of this handlebody. The question is if we can reverse engineer a multi-invariant  as in the two dimensional case which produces the replicated manifold $\CM_\CE$ that is Weyl equivalent to the boundary of this symmetry handlebody. The uniformization question in higher dimensions is often tied to the so-called Yamabe problem. The Yamabe problem asks if one can find a Weyl factor $e^\phi$ for any Riemannian manifold which  turns it into a manifold with constant scalar curvature. It would be worth investigating the uniformization  question in the above context as it will give us insight into multi-partite entanglement of the vacuum state of higher dimensional holographic theories. 

\section*{Acknowledgements}
We are grateful to Ning Bao, Matt Headrick, Shiraz Minwalla, Takato Mori, Rob Myers, Arvind Nair, Onkar Parrikar, Pratik Rath, Jonathan Sorce, Tadashi Takayanagi, Zixia Wei and Beni Yoshida for discussion.  AG and VK are supported by the Infosys Endowment for the study of the Quantum Structure of Spacetime. AG and VK would like to acknowledge the support of the Department of Atomic Energy, Government of India, under Project Identification No. RTI 4002. The work of JH is supported by MEXT KAKENHI Grant-in-Aid for Transformative Research Areas (A) through the ``Extreme Universe'' collaboration: Grant Number 21H05187. VK is supported in part by the U.S. Department of Energy under grant DE-SC0007859. We would like to thank the APCTP, the Perimeter Institute, Brandeis University, Leinweber Center for Theoretical Physics and, ICTS-TIFR Bangalore where a part of the present work was completed. AG and VK would also like to acknowledge their debt to the people of India for their steady support to the study of the basic sciences.

\appendix
\section{Review of Liouville uniformization}\label{l-uni-app}

In this appendix, we will review the computation of the Liouville action associated to the replicated manifold $\CM_\CE$ obtained as a ramified cover of $\CM$ \cite{Lunin:2000yv, avery2010using}. As remarked in section \ref{cft-sec}, the partition function of the CFT on $\CM_\CE$ can be equivalently thought of as a correlation function of appropriate twist operators inserted at the ramification point. The space $\CM_\CE$ has points of conical excess. Uniformization theorem guarantees that $\CM_\CE$ is conformally equivalent to another space $\CM_{\CE}^{\rm uni}$ of uniform curvature i.e.
\begin{align}
    g_{\CM_\CE}=e^\phi \, g_{\CM_\CE^{\rm uni}}.
\end{align}
The constant uniform curvature on $\CM_{\CE}^{\rm uni}$ is $1, 0$ or $-1$ if $\CM_\CE$ has genus $0,1$ or $\geq 2$ respectively. The partition functions on $\CM_\CE$ and $\CM_\CE^{\rm uni}$ are related as
\begin{align}
    Z_{\CM_\CE}&=e^{-S_L(\phi)} \, Z_{\CM_\CE^{\rm uni}},\notag\\
    S_L&=\frac{c}{96\pi } \int dz^2\sqrt{-g}(g_{\mu\nu}\partial^\mu \phi\partial^\nu\phi)+2R\phi.
\end{align}
Here $S_L$ is the Liouville action for the Weyl factor $\phi$ evaluated on $\CM_\CE^{\rm uni}$. 
In this appendix we will review the calculation of $S_L(\phi)$ for $\CM_\CE$ that corresponds to the ramified cover of $\CM$ specified through twist operator insertions on $\CM$.  
For a general ramification $\CE$, it is difficult to compute even $\phi$, let alone $S_L(\phi)$. To compute $\phi$, one first computes the covering map $\Gamma:\CM_\CE^{\rm uni}\to \CM$. 
In section \ref{cft-comp}, we have seen how $\Gamma$ is computed for spherical group invariants. The Weyl factor is then $\phi= 2\log|\partial \Gamma|$. In the rest of the section we will assume that we know the covering map $\Gamma$.

We will denote $\CM$ as the base space  and $\CM_\CE^{\rm uni}$ as the covering space. We will take both the spaces to have genus $0$. 
We will not use the round metric for them but rather use the following fiducial metric. 
We use a \emph{flat} circular disc of radius $1/\delta$ centered at the origin of the base space  
and another flat circular disc of the same radius covering the rest of the sphere.  The two discs are glued at their boundaries which is a circle of radius of $1/\delta$. We will use the complex coordinate $x$ and $\tilde x$ for the two discs respectively.
\be
ds^2=\begin{cases}
			dxd\bar{x}, & |x|<\frac{1}{\delta}\\
            d\tilde{x}d\bar{\tilde{x}}, & |\tilde{x}|<\frac{1}{\delta}
		 \end{cases}, \quad \tilde{x}=\frac{1}{\delta^2}\frac{1}{x}
\ee
The entirety of the curvature of the sphere is concentrated at this gluing locus.
Similar fiducial metric with two flat disc patches with coordinates $z$ and $\tilde z$ is also taken for the covering space. In this case, we take the radius of the two discs to be $1/\delta'$. So we have $\tilde z=1/(\delta'^2 z)$. 
The point at infinity in the base and covering space is mapped to $\tilde x=0$ and $\tilde z=0$ respectively. 

The reason for using the fiducal metric is that it isolates the kinetic contributions to the Liouville action from the curvature contributions. Instead of integrating over the entirety of $\CM_\CE^{\rm uni}$ we can instead expand the covering map around any singular points and after integrating the kinetic term of the Liouville action by parts calculate the contour integral
\be
\frac{c}{96\pi}i\int_{\partial\CM_\CE^{\rm uni}}\phi\partial\phi dz
\ee
where $\partial\CM_\CE^{\rm uni}$ is given by preimage of \emph{all} boundaries in the base space in \emph{both} coordinate charts (including the boundary of both disks) as well as the boundary of the disks of the covering space. In addition it is necessary to keep track of the orientation of each boundary. The convention is to choose all contours such that their normals point inward. Any contours which originally have external pointing normals will thus have their orientation reversed introducing an extra minus sign into the calculation. Note that this orientation is with respect to $x,\tilde{x}$. In particular there will only ever be two external contours which occurs at the boundary of the two discs at $\tilde{x}=\frac{1}{\delta}$.

We take all but one twist operators to be inserted at finite points $x_a$, all of which are taken to be in the first disc, $|x|<1/\delta$. One is inserted at infinity, which is mapped to $\tilde x=0$. One of the pre-images of $\tilde x=0$ is at $\tilde z=0$. The rest of the pre-images are at finite points are taken to be in the first disc, $|z|<1/\delta'$. The pre-images of $x_a$ is denoted at $z_{i,a}, i=1, \ldots, m_a$. The expansion of the covering map there is
\begin{align}
    \Gamma(z)&=x_a + \alpha_{i, a} (z-z_{i,a})^{k_{a}}+\ldots.
\end{align}
The finite pre-images of $\tilde x=0$ are denoted as $z_{i,q}, i= 1, \ldots, m_q$. The expansion of the covering map there is 
\begin{align}
    \Gamma(z)&= \alpha_{i, q} \,(z- z_{i,q})^{-k_q} +\ldots.
\end{align}
The expansion of the covering map at $z=\infty$ is, 
\begin{align}
    \Gamma(z)&= \alpha_{m_q, q} \, z^{k_q}.
\end{align}
Because we are working with twist operators with cycles of constant length, we have $k_a m_a=n_r$, where $n_r$ is the number of replica. 
As we are treating the twist operators differently from the rest. We use the notation $k_q=k_\infty$, $m_q= m_\infty$ and $\alpha_{i,q}\equiv  \beta_i, i= 1, \ldots, m_q-1$ and $\alpha_{m_q, q}\equiv  \beta_0$..  
We drill holes of radius $\epsilon$ around each of the twist operators at finite points. We take the radius of the hole around the twist operator at infinity i.e.  at $\tilde x=0$ to be  $\tilde \epsilon$.
All these holes have pre-images on $\CM_\CE^{\rm uni}$. Their sizes depend on the covering map. 
We cover all the pre-image holes by small flat discs. 
\begin{figure}[h]
    \begin{center}
        \includegraphics[scale=0.20]{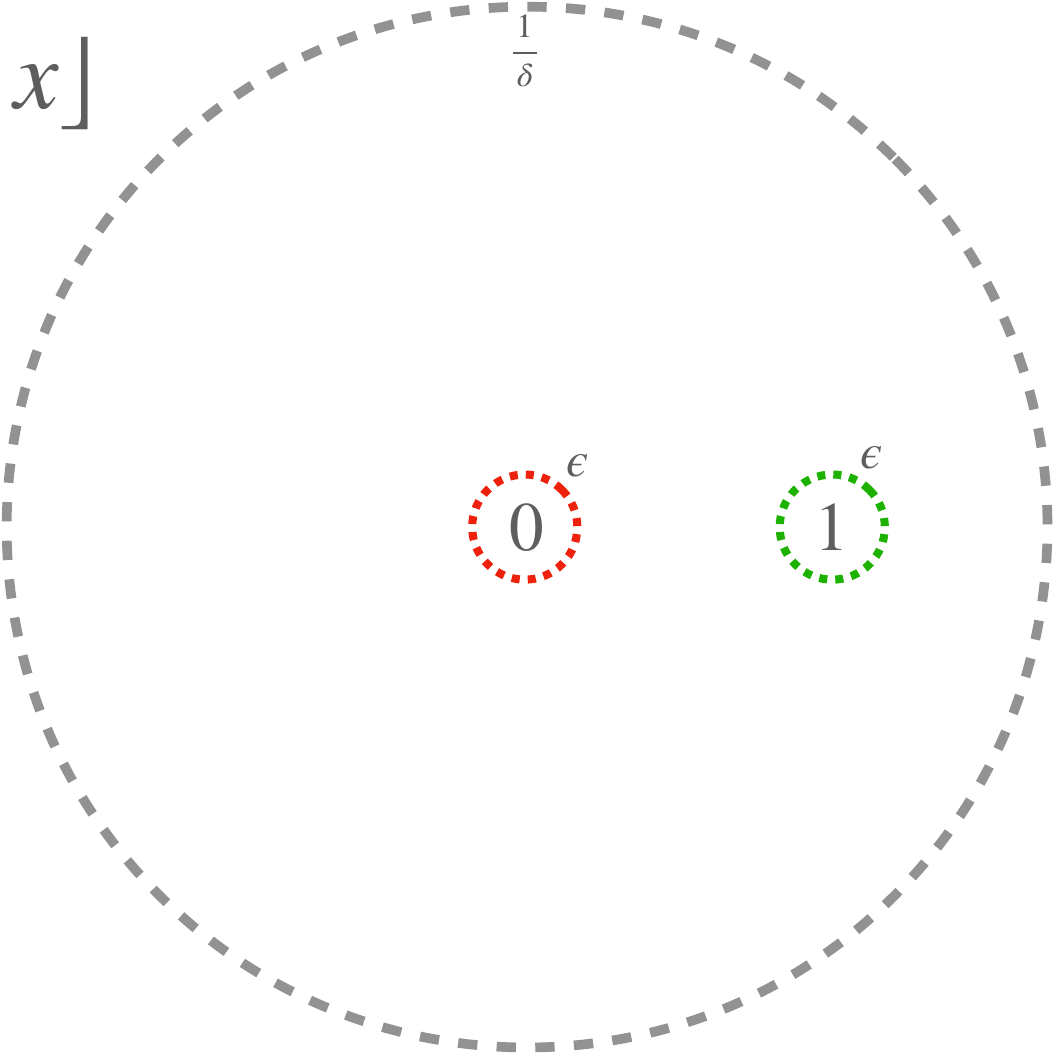}\,\,
        \includegraphics[scale=0.20]{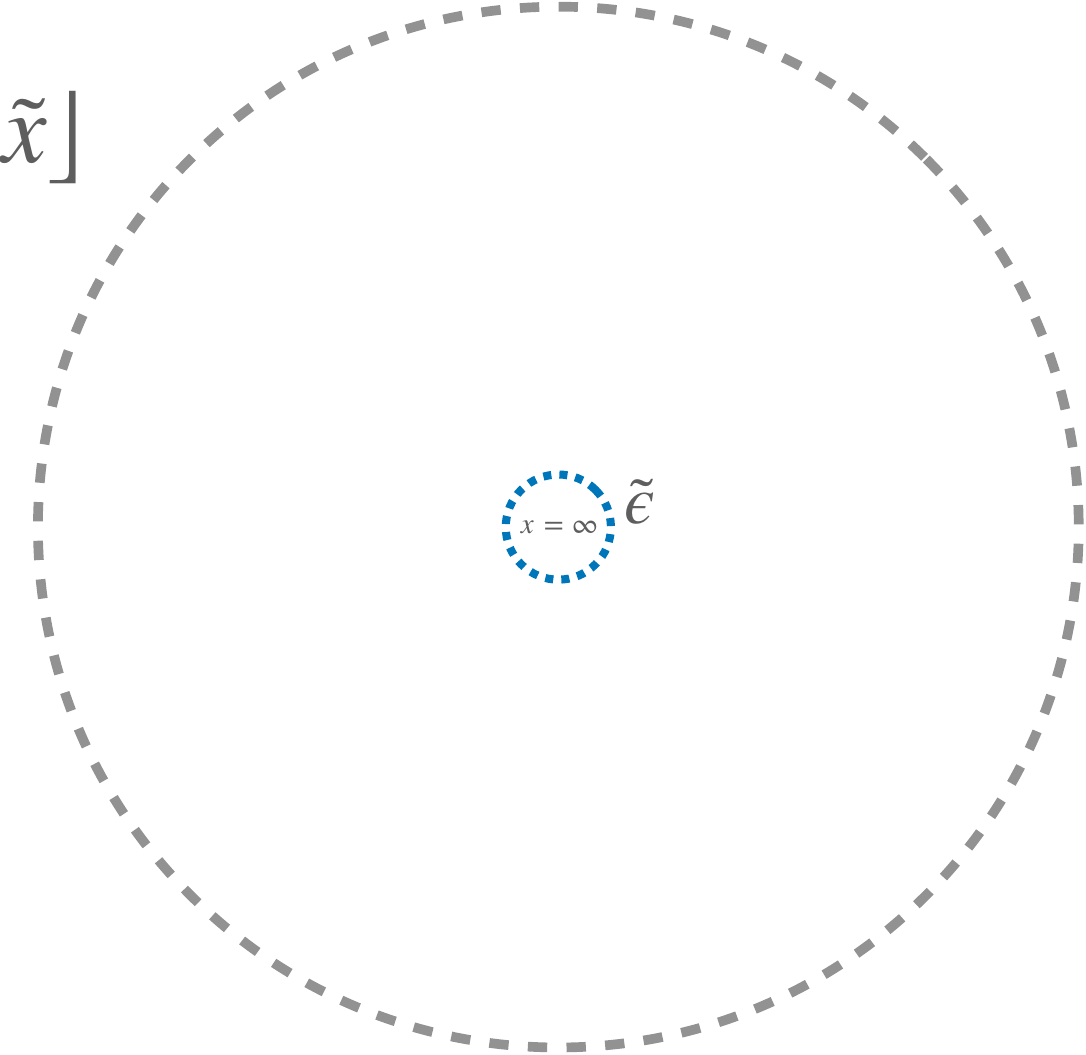}\,\,
        \includegraphics[scale=0.20]{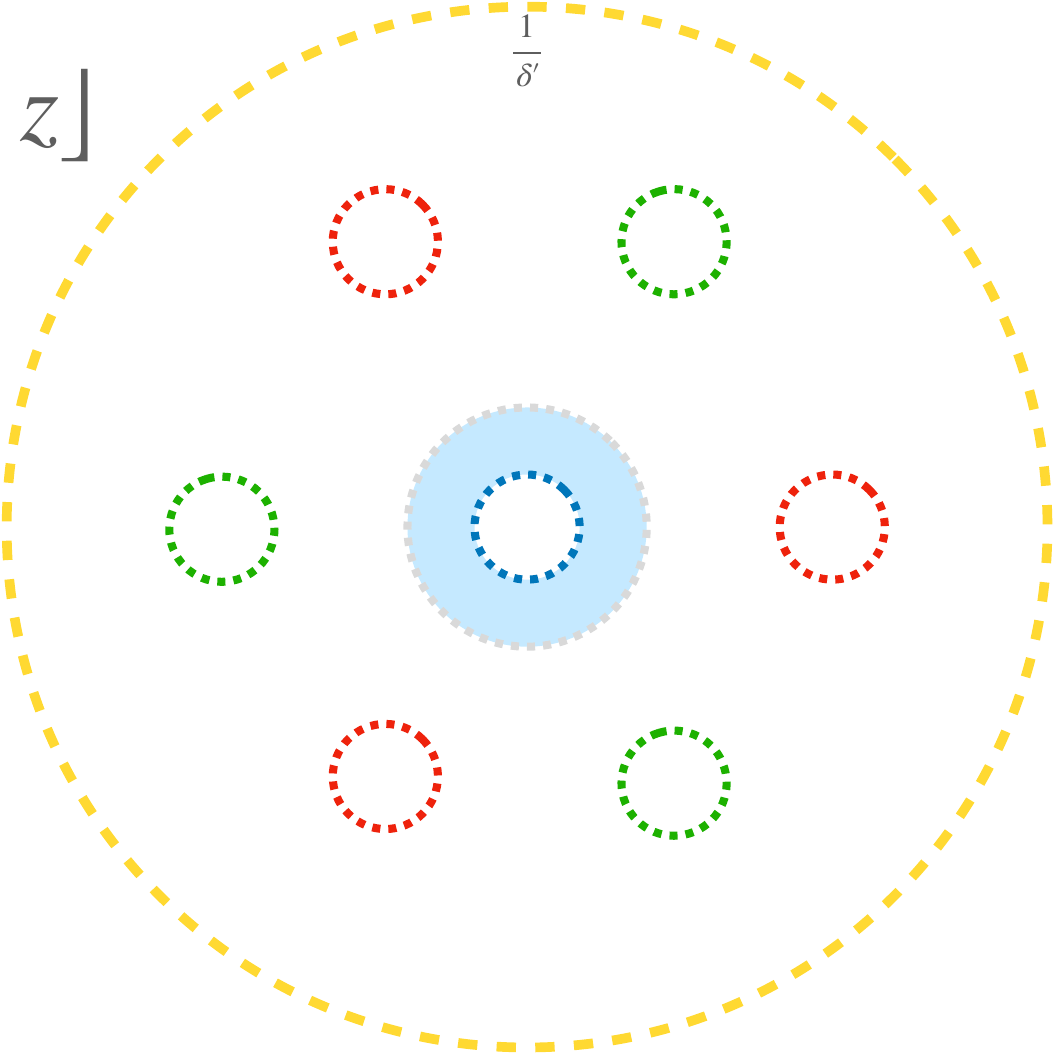}\quad
        \includegraphics[scale=0.20]{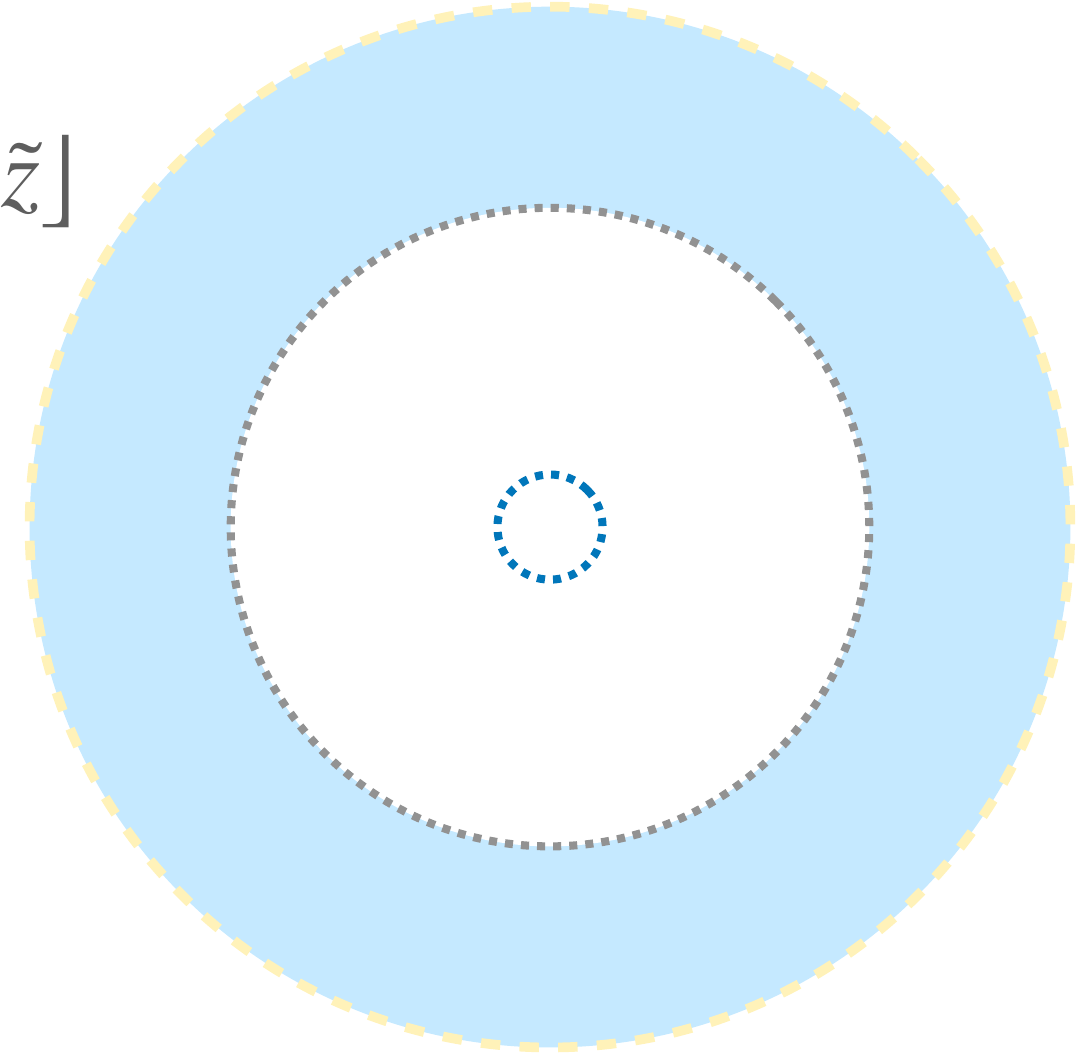}
    \end{center}
    \caption{This set up is shown for the dihedral three party measure (figure \ref{fin-cox-fig} top left) with covering map \eqref{eq:dihcovering}. Left two: We consider the insertion of three twist operators at 0,1,$\infty$ on the base space. We show the two coordinate charts with coordinates $x,\tilde x$ corresponding to each disc of radius $1/\delta$  as well as the cutoffs $\epsilon,\tilde{\epsilon}$. Right two: The covering space, also the Riemann sphere, given by two coordinate charts with coordinates $z,\tilde z$ corresponding to each disc of radius $1/\delta'$. We have shown the preimage of each boundary under the covering map and each of these will contribute to the evaluation of the Liouville action. The two blue regions are the inner and outer annuli these occur because crossing the yellow or gray contours necessitates evaluating the contribution with respect to a different set of coordinates.}
    \label{fig:charts}
\end{figure}
This set up is summarized in figure \ref{fig:charts}. The covering space $\CM_\CE^{\rm uni}$ is naturally divided into various regions. We evaluate the Liouville action on all these regions separately. 
\begin{itemize}
    \item Regular region: Pre-image of the first disc $x<1/\delta$. 
    \item Outer annulus: Region between the pre-image of $x=1/\delta$ and $z=1/\delta'$.
    \item Second half of $z$ space: Region between the pre-image of $\tilde x = \tilde \epsilon$ and $\tilde z= 1/\delta'$.
    \item Inner annuli surrounding finite pre-images of $\infty$:  Bounded by the pre-image of $\tilde x = \tilde \epsilon$ and that of $\tilde x = 1/\delta'$. 
\end{itemize}
We will not give details of this computation but simply summarize the results. 
\begin{itemize}
\item Regular region:
\begin{itemize}
\item For each pre-image $z_{i,a}, a\neq q$:
\be
S^{(i,a)}_L=-\frac{c}{12}(k_a-1)\left(\frac{1}{k_a}\log|\alpha_{i,a}|+\log(k_a)+\frac{k_a-1}{k_a}\log(\epsilon)\right)
\ee
\item The outer boundary of the regular region:
\be
S_L^{}=\frac{c}{12}(k_\infty-1)\left(\frac{1}{k_\infty}\log|\beta_0|+\log(k_\infty)-\frac{k_\infty-1}{k_\infty}\log(\delta)\right)
\ee
\item For each finite image of infinity there is a boundary with an inner annulus:
\be
S_{L}=-\frac{c}{12}(k_\infty+1)\left(\frac{1}{k_\infty}\log|\beta_j|-\log(k_\infty)+\frac{k_\infty+1}{k_\infty}\log(\delta)\right)
\ee
\end{itemize}
\item Outer annulus: The two boundaries together contribute
\be
S_L^{OA}= \frac{c}{12}(k_\infty+1)^2\left(\frac{1}{k_\infty}\log(|\beta_0|)+\frac{1}{k_\infty}\log(\delta)-\log(\delta')\right)
\ee
\item Inner annuli: For each finite image of infinity there is a inner annuli whose two boundaries together contribute
\be
S_L^{IA,j}= -\frac{c}{12}\frac{(k_\infty-1)^2}{k_\infty}\log(\tilde{\epsilon}\delta)
\ee
\item Second half of the sphere:  The two boundaries together contribute
\be
S^{SH}_{L}=-\frac{c}{12}\frac{(k_\infty-1)^2}{k_\infty}\left(\beta_0)+\log(\tilde{\epsilon})+2\log(\delta)-k_\infty\log(\delta')\right)
\ee
\item Curvature
\be
S_L^{cur}=\frac{c}{6}\phi|_{|z|=\frac{1}{\delta'}}=\frac{c}{3}\left(\log|\beta_0|+\log(k_\infty)-2\log(\delta)+(k_\infty+1)\log(\delta')\right)
\ee
\end{itemize}
Here the last ``curvature'' contribution is due to the $R\phi$ term in the Liouville action. This contribution comes only from the boundary $x=1/\delta'$ because that is where all the sphere curvature is concentrated. The rest of the contribution comes from the kinetic term in the Liouville action. After summing all the contributions, we get
\be\label{eq:master}
\begin{split}
S_L &= -\frac{c}{12}\Biggr\{
\sum_{a,i} \frac{k_a - 1}{k_{a}}\log |\alpha_{i,a}|
+\sum_j \frac{k_\infty +1}{k_\infty}\log |\beta_j|
-\frac{ k_\infty -1}{k_\infty} \log |\beta_0|\\
&\qquad+(n_r - m_a)\log k_a
- (n_r +m_\infty-2)\log k_\infty + \sum_a \frac{m_a(k_a -1)^2}{k_a}\log \epsilon\\
&\qquad 
  + \frac{m_\infty(k_\infty -1)^2}{k_\infty} 
             \log\tilde{\epsilon}+ 2\frac{m_\infty(k_\infty +1)^2}{k_\infty} 
             \log\delta -4\log\delta'\Biggr\}.
\end{split}
\ee
This is UV divergent. As explained in section \ref{cft-sec}, to get a finite answer, we must normalize the twist operators. We can use the above expression to compute the two point function of twist operators. Normalizing them as explained in section \ref{cft-sec}, we get equation \eqref{l-formula} for the twist operator correlation function. 

\subsection*{Extension to genus $1$}
In this section, we will see how to extend the above formula to the case when the covering space $\CM_\CE^{\rm uni}$ has genus $1$. This has a number of simplifications compared to the genus $0$ case. 
As the torus is flat, we can work with a single coordinate chart on the covering space. We will take all the pre-images of twist operators to be at finite points. 
As a result, we do not get any contribution from curvature, outer annulus and the second half of the disk. Removing these contributions and normalizing the twist operators appropriately we get the formula \eqref{l-formula-1}.

\section{Replica symmetry representations}\label{app:reps}
In this section we list a choice of explicit permutation representations for the replica symmetry as well as the group elements used for the construction of each of the multi-invariants  considered in this article. Finally, we show the resulting twist operators and their monodromies. In two cases $\mathbb{A}_5$ of order 60 and $GL(2,3)$ of order 48 we have omitted due to the lengthy expressions.

\paragraph{$\mathbb{Z}_{n}$}
\be
\langle a|a^n=e\rangle
\ee
\be
   a:\quad (1,2,3,\cdots, n)
\ee
Genus 0 2-party measure figure \ref{zn-orb-fig}:
\be\label{eq:zntau}
g_O=e,\quad g_A=a^{n-1}.
\ee
\be
\begin{split}
   \sigma_{a}:&\quad (1,2,\cdots, n)\\ 
   \sigma_{a^{n-1}}:&\quad  (1,n,n-1,\cdots, 2)
   \end{split}
\ee

\paragraph{$\mathbb{D}_{2n}$}
\be
\langle a,b|a^n=b^2=e;ba=a^{n-1}b\rangle
\ee
\be
\begin{split}
   &a:\quad (1,2,\cdots, n)(n+1,n+2,\cdots, 2n)\\
   &b:\quad  (1,2n)(2,2n-1)(3,2n-2)\cdots(n,n+1)\\.
\end{split}
\ee
Genus 0 3-party measure figure \ref{fin-cox-fig} top left:
\be
g_O=e,
\;g_A=a^{n-1},
\;g_B={b}
\ee
\be
\begin{split}
   \sigma_{a}:&\quad (1,2,\cdots, n)(n+1,n+2,\cdots, 2n)\\ 
   \sigma_{a^{n-1}b}:&\quad  (1,n+1)(2,2n)(3,2n-1)\cdots(n,n+2)\\
   \sigma_{b}:&\quad  (1,2n)(2,2n-1)(3,2n-2)\cdots(n,n+1).
\end{split}
\ee
Genus 1 3-party measure on four regions with $n=3$ figure \ref{schottky-quotient1} top left:
\be
g_O=e,\;g_A=ab,\;g_B=e,\;g_C=b
\ee
\be
\begin{split}
   \sigma_{ab}:&\quad (1,5)(2,4)(3,6) \\
   \sigma_{ab}:&\quad (1,5)(2,4)(3,6)  \\
   \sigma_{b}:&\quad (1,6)(2,5)(3,4) \\
     \sigma_{b}:&\quad (1,6)(2,5)(3,4)
\end{split}
\ee
Genus 1 4-party measure with $n=6$ figure \ref{schottky-quotient1} \emph{both} top right and bottom right:

\be
g_O=e,\;g_A=a^2b,\;g_B=a^3,\;g_C=b
\ee
\be
\begin{split}
   \sigma_{a^2b}:&\quad (1,10)(2,9)(3,8)(4,7)(5,12)(6,11) \\
   \sigma_{a^5b}:&\quad (1,7)(2,12)(3,11)(4,10)(5,9)(6,8)\\
   \sigma_{a^3b}:&\quad  (1,9)(2,8)(3,7)(4,12)(5,11)(6,10)\\
     \sigma_{b}:&\quad(1,12)(2,11)(3,10)(4,9)(5,8)(6,7)
\end{split}
\ee

\paragraph{$\mathbb{D}_{6}\times \mathbb{Z}_2$}\footnote{This is isomorphic to $\mathbb{D}_{12}$. Since the measure we are considering is genus 2 such a Riemann surface has a hyperelliptic involution. By writing the replica symmetry this way it makes this more naturally apparent. In particular note that in figure \ref{d4d6-g2} copies 1-6 are separated from copies 7-12 by the right circle which is the fixed point locus of the reflection $r_4$.}
\be
\langle a,b,c|a^3=b^2=c^2=e;ba=a^2b;ac=ca;bc=cb\rangle
\ee
\be
\begin{split}
   &a:\quad (1,2,3)(4,5,6)(7,8,9)(10,11,12)\\
   &b:\quad  (1,6)(2,5)(3,4)(7,12)(8,11)(9,10)\\
   &c:\quad  (1,7)(2,8)(3,9)(4,10)(5,11)(6,12)
\end{split}
\ee
Genus 2 4-party measure figure \ref{d4d6-g2}:
\be
g_O=e,
\;g_A=a^2,
\;g_B=b,\;g_C=cb
\ee
\be
\begin{split}
   \sigma_{a}:&\quad (1,2,3)(4,5,6)(7,8,9)(10,11,12)\\ 
   \sigma_{a^2b}:&\quad  (1,4)(2,6)(3,5)(7,10)(8,12)(9,11)\\
   \sigma_{c}:&\quad  (1,7)(2,8)(3,9)(4,10)(5,11)(6,12)\\
    \sigma_{cb}:&\quad  (1,12)(2,11)(3,10)(4,9)(5,8)(6,7).
\end{split}
\ee

\paragraph{$\mathbb{A}_{4}$}
\be
\langle a,b,c|a^2=b^2=c^3e;cac^2=ab=ba,cbc^2=a\rangle
\ee
\be
\begin{split}
   &a:\quad (1,4)(2,9)(3,12)(5,11)(6,7)(8,10)\\
   &b:\quad  (1,10)(2,5)(3,7)(4,8)(6,12)(9,11)\\
    &c:\quad  (1,2,3)(4,5,6)(7,8,9)(10,11,12).
\end{split}
\ee
Genus 0 3-party measure figure \ref{fin-cox-fig} top right:
\be\label{eq:tetratau}
g_O=e,\;g_A={a},\;g_B={c}
\ee
\be
\begin{split}
   \sigma_{a}:&\quad (1,4)(2,9)(3,12)(5,11)(6,7)(8,10)\\ 
   \sigma_{ac^2}:&\quad  (1,6,9)(2,8,12)(3,11,4)(5,10,7)\\
   \sigma_{c}:&\quad  (1,2,3)(4,5,6)(7,8,9)(10,11,12)
\end{split}
\ee

\paragraph{$\mathbb{S}_{4}$}
\be
\langle a,b,c|a^2=b^2=c^3=d^2=e;cac^2=ab=ba,cbc^2=a,bd=db,dcd=c^2\rangle
\ee
\be
\begin{split}
   &a:\quad (1, 15)(2, 12)(3, 21)(4, 24)(5, 9)(6, 16)(7, 23)(8, 17)(10, 14)(11, 19)(13, 20)(18, 22)\\
   &b:\quad  (1, 19)(2, 13)(3, 10)(4, 17)(5, 22)(6, 7)(8, 24)(9, 18)(11, 15)(12, 20)(14, 21)(16, 23)\\
    &c:\quad  (1, 2, 3)(4, 5, 6)(7, 8, 9)(10, 11, 12)(13, 14, 15)(16, 17, 18)(19, 20, 21)(22, 23, 24)\\
    &d:\quad (1, 6)(2, 5)(3, 4)(7, 19)(8, 21)(9, 20)(10, 17)(11, 16)(12, 18)(13, 22)(14, 24)(15, 23)\\
\end{split}
\ee
Genus 0 3-party measure figure \ref{fin-cox-fig} bottom left:
\be\label{eq:octtau}
g_O=e,\;g_A={(ad)^3},\;g_B={abcd}
\ee
\be
\begin{split}
   \sigma_{ad}:&\quad  (1,23,19,16)(2,18,13,9)(3,8,10,24)(4,14,17,21)(5,20,22,12)(6,11,7,15)\\
   \sigma_{bc^2}:&\quad  (1, 21, 13)(4, 15, 10)(3, 12, 19)(4, 16, 22)(5, 24, 7)(6, 9, 17)(8, 23, 18)(11, 14, 20)\\
   \sigma_{abcd}:&\quad  (1, 18)(2, 8)(3, 23)(4, 20)(5, 11)(6, 14)(7, 10)(9, 15)(12, 24)(13, 17)(16, 21)(19, 22)
\end{split}
\ee
Genus 3 4-party measure I figure \ref{d4d6-g3}:
\be
g_O=e,\;g_A=c^2,\;g_A=d,\;g_C=b
\ee
\be
\begin{split}
   \sigma_{c}:&\quad  (1, 2, 3)(4, 5, 6)(7, 8, 9)(10, 11, 12)(13, 14, 15)(16, 17, 18)(19, 20, 21)(22, 23, 24)  \\
   \sigma_{c^2d}:&\quad (1, 4)(2, 6)(3, 5)(7, 20)(8, 19)(9,21)(10, 18)(11, 17)(12, 16)(13, 23)(14, 22)(15, 24) \\
   \sigma_{db}:&\quad (1, 7)(2, 22)(3, 17)(4, 10)(5, 13)(6,19)(8, 14)(9, 12)(11, 23)(15, 16)(18, 20)(21, 24) \\
     \sigma_{b}:&\quad (1, 19)(2, 13)(3, 10)(4, 17)(5, 22)(6, 7)(8, 24)(9, 18)(11, 15)(12, 20)(14, 21)(16, 23)
\end{split}
\ee
Genus 3 4-party measure II figure \ref{d6a3-g3}:
\be
g_O=e,\;g_A=b,\;g_A=c^2\;g_C=d
\ee
\be
\begin{split}
   \sigma_{b}:&\quad (1, 19)(2, 13)(3, 10)(4, 17)(5, 22)(6, 7)(8, 24)(9, 18)(11, 15)(12, 20)(14, 21)(16, 23) \\
   \sigma_{bc}:&\quad (1, 20, 10)(2, 14, 19)(3, 11, 13)(4, 18, 7)(5, 23, 17)(6, 8, 22)(9, 16, 24)(12, 21, 15) \\
   \sigma_{c^2d}:&\quad (1, 4)(2, 6)(3, 5)(7, 20)(8, 19)(9,21)(10, 18)(11, 17)(12, 16)(13, 23)(14, 22)(15, 24)   \\
     \sigma_{d}:&\quad (1, 6)(2, 5)(3, 4)(7, 19)(8, 21)(9, 20)(10, 17)(11, 16)(12, 18)(13, 22)(14, 24)(15, 23)
\end{split}
\ee

\paragraph{$\mathbb{Z}_3\rtimes\mathbb{D}_{6}$}
\be
\langle a,b,c|a^3=b^3=c^2=e;ca=a^2c;cb=b^2c;ab=ba\rangle
\ee
\be
\begin{split}
   &a:\quad (1,2,3)(4,5,6)(7,8,9)(10,11,12)(13,14,15)(16,17,18)\\
   &b:\quad (1,5,8)(2,6,9)(3,4,7)(10,16,13)(11,17,14)(12,18,15)\\
   &c:\quad (1,10)(2,12)(3,11)(4,14)(5,13)(6,15)(7,17)(8,16)(9,18) \\
\end{split}
\ee
Genus 1 4-party measure figure \ref{schottky-quotient1} bottom left:
\be
g_O=e,\;g_A={cb^2},\;g_B={a^2},\;g_C={cb}
\ee
\be
\begin{split}
   \sigma_{cb^2}:&\quad (1,13)(2,15)(3,14)(4,17)(5,16)(6,18)(7,11)(8,10)(9,12) \\
   \sigma_{cab^2}:&\quad (1,14)(2,13)(3,15)(4,18)(5,17)(6,16)(7,12)(8,11)(9,10)  \\
   \sigma_{cab}:&\quad (1,17)(2,16)(3,18)(4,12)(5,11)(6,10)(7,15)(8,14)(9,13)  \\
     \sigma_{cb}:&\quad (1,16)(2,18)(3,17)(4,11)(5,10)(6,12)(7,14)(8,13)(9,15) 
\end{split}
\ee
\paragraph{$\mathbb{S}_3^2$}
\be
\langle a^3=b^2=c^3=d^2=e;(ab)^2=(cd)^2=e\rangle.
\ee
\be
\begin{split}
 a:\quad &(1,2,3)(4,5,6)(7,8,9)(10,11,12)(13,14,15)(16,17,18)\\
 &(19,20,21)(22,23,24)(25,26,27)(28,29,30)(31,32,33)(34,35,36)\\
b:\quad &(1,36)(2,35)(3,34)(4,32)(5,31)(6,33)(7,28)(8,30)(9,29)\\
&(10,27)(11,26)(12,25)(13,23)(14,22)(15,24)(16,19)(17,21)(18,20)\\
c:\quad &(1,19,22)(2,20,23)(3,21,24)(14,36,16)(13,35,18)(15,34,17)\\
&(4,25,28)(5,26,29)(6,27,30)(8,33,10)(7,32,12)(9,31,11)\\
d:\quad &(1,31)(2,32)(3,33)(4,35)(5,36)(6,34)(7,20)(8,21)(9,19)\\
&(10,24)(11,22)(12,23)(13,25)(14,26)(15,27)(16,29)(17,30)(18,28)
\end{split}
\ee
Genus 4 4-party measure figure \ref{d4d6-g4}:
\be
g_O=e,\;g_A=a^2c^2,\;g_B=bd,\;g_C=b
\ee
\be
\begin{split}
   \sigma_{ac}:\quad &(1,20,24)(2,21,22)(3,19,23)(4,26,30)(5,27,28)(6,25,29) \\
   &(7,33,11)(8,31,12)(9,32,10)(13,36,17)(14,34,18)(15,35,16)\\
   \sigma_{a^2c^2bd}:\quad &(1,27)(2,26)(3,25)(4,21)(5,20)(6,19)(7,14)(8,13)(9,15)\\
&(10,35)(11,34)(12,36)(16,32)(17,31)(18,33)(22,30)(23,29)(24,28)\\
   \sigma_{d}:\quad &(1,31)(2,32)(3,33)(4,35)(5,36)(6,34)(7,20)(8,21)(9,19)\\
&(10,24)(11,22)(12,23)(13,25)(14,26)(15,27)(16,29)(17,30)(18,28)\\
     \sigma_{b}:\quad &(1,36)(2,35)(3,34)(4,32)(5,31)(6,33)(7,28)(8,30)(9,29)\\
&(10,27)(11,26)(12,25)(13,23)(14,22)(15,24)(16,19)(17,21)(18,20)
\end{split}
\ee


\bibliography{LargeDCFT.bib}

\end{document}